\documentclass{article}

\def\foryap{01}

\def\dofigures{11}

\usepackage{calc}
\usepackage{amsmath}
\usepackage{amssymb}
\usepackage{amsthm}
\usepackage{color}
\usepackage[sans]{dsfont}

\usepackage{fancyhdr}

\if\dofigures%
\usepackage{graphicx}
\usepackage{texdraw}
\fi%

\newcommand{\sindex}[2]{}

\definecolor{maroon}{rgb}{0.3,0.0,0.0}
\newcommand{\cucon}[1]{}

\newcommand{\R}{\mathds{R}}
\newcommand{\N}{\mathds{N}}

\newcommand{\C}{\mathds{C}}
\newcommand{\Z}{\mathds{Z}}
\newcommand{\defm}[1]{\emph{#1}}
\newcommand{\subeq}[2]{\mathord{\underbrace{\mathop{#1}}_{#2}}}

\newcommand{\sgn}{\operatorname{sgn}}
\newcommand{\Det}{\operatorname{det}}

\newcommand{\graph}{\operatorname{graph}}
\newcommand{\sign}{\operatorname{sgn}}

\newcommand{\osc}{\operatorname{osc}}
\newcommand{\spC}{\mathcal{C}}

\newcommand{\tua}{\tau} 
\newcommand{\nva}{\beta}

\newcommand{\contemb}{\hookrightarrow}

\newcommand{\xo}{\chi}
\newcommand{\po}{\psi}
\newcommand{\so}{s}

\newcommand{\xn}{\hat\chi}
\newcommand{\pn}{\hat\psi}

\newcommand{\xnl}{\acute\chi}
\newcommand{\pnl}{\acute\psi}

\newcommand{\etat}{\eta^*}
\newcommand{\vxit}{\vec\xi^*}
\newcommand{\xia}{\xi}
\newcommand{\etaa}{\eta}
\newcommand{\vxia}{\vec\xi}

\theoremstyle{plain}
\newtheorem{theorem}{Theorem}
\newtheorem{lemma}{Lemma}

\newtheorem{proposition}[lemma]{Proposition}
\theoremstyle{definition}
\newtheorem{convention}[lemma]{Convention}

\newtheorem{definition}[lemma]{Definition}
\theoremstyle{remark}
\newtheorem{remark}[lemma]{Remark}

\definecolor{brown}{rgb}{0.5,0.4,0.0}
\definecolor{darkgreen}{rgb}{0.0,0.6,0.0}
\definecolor{orange}{rgb}{1.0,0.5,0.0}

\makeatletter\@addtoreset{equation}{subsection}\@addtoreset{equation}{section}\makeatother
\makeatletter\@addtoreset{lemma}{subsection}\@addtoreset{equation}{section}\makeatother

\newcommand{\myeqref}[1]{(\ref{#1})}
\newcommand{\myref}[1]{\ref{#1}}

\newcommand{\mylabel}[1]{\label{#1}}
\newcommand{\myeqlabel}[1]{\label{#1}}  

\begin{document}

\if\foryap\else%
\title{Supersonic flow onto a solid wedge}
\author{Volker Elling \and Tai-Ping Liu}
\date{}
\maketitle

\thispagestyle{empty}

\def\oldparindent{\parindent}%
\def\oldparskip{\parskip}%

\begin{abstract}\noindent%
\parindent=0cm%
\parskip=.7\baselineskip%
	We consider the problem of 2D supersonic flow onto a solid wedge, 
	or equivalently in a concave corner formed by two solid walls.
	For mild corners, there are two possible steady state solutions, 
	one with a strong and one with a weak shock emanating from the corner. 
	The weak shock is observed in supersonic flights.
	A long-standing natural conjecture is that the strong shock is unstable in some sense.

	We resolve this issue by showing that a sharp wedge will eventually 
	produce weak shocks 
	at the tip when accelerated to a supersonic speed. More precisely we prove that
	for upstream state as initial data in the entire domain, the time-dependent solution 
	is self-similar, with a weak shock at the tip of the wedge. 
	We construct analytic solutions for self-similar potential flow, both 
	isothermal and isentropic with arbitrary $\gamma\geq 1$.
       
	In the process of constructing the self-similar solution, we develop a large number of theoretical tools 
	for these elliptic regions. 
	These tools allow us to establish large-data results rather than a small perturbation. 
	We show that the wave pattern persists as long as the weak shock is supersonic-supersonic;
	when this is no longer true, numerics show a physical change of behaviour.
	In addition we obtain rather detailed information about the elliptic region, including analyticity as well as bounds for velocity components
	and shock tangents.
\end{abstract}

\fi%

\tableofcontents

\section{Introduction}

\subsection{Background}

\parindent=0cm
\parskip=.7\baselineskip

Gas flow onto a solid wedge, like forward edges of airplane wings or engine inlets, 
is a fundamental problem for aerodynamics (see Figure \myref{fig:wedgeflow}).
An equivalent problem is flow in a convex corner of an otherwise straight 
wall (see Figure \myref{fig:weakstrong}).
For \emph{supersonic} flow and sufficiently small $\tau$, 
it is well-known that this problem has steady solutions with a straight shock emanating from the corner, 
separating two constant-state regions (``upstream'' and ''downstream''). 
The shock is the consequence of compression of the gas by the downstream wall. 

\if\dofigures%
\begin{figure}
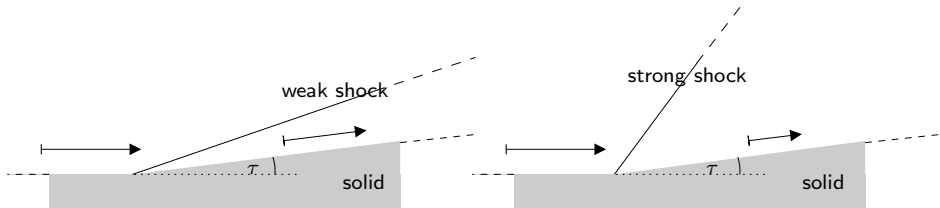

\parbox[b]{.49\linewidth}{\input{ws2.pstex_t}}\hfill\parbox[b]{.49\linewidth}{\input{ss2.pstex_t}}
\caption{Two steady solutions of supersonic flow along an infinite wall with a corner.}
\mylabel{fig:weakstrong}
\end{figure}
\begin{figure}
\center{\input{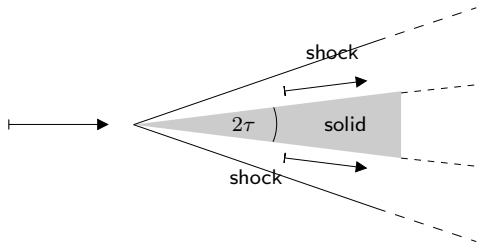}}
\caption{Reflect either part of Figure \myref{fig:weakstrong} across the upstream wall. There are four combinations; each is a steady solutions 
to supersonic flow onto a solid wedge. The one with two weak shocks is the one observed in nature and numerics.}
\mylabel{fig:wedgeflow}
\end{figure}
\fi%

A longstanding open and puzzling problem is that, for $\tau$ close to $0$ (corresponding to a sharp wedge resp.\ a mild corner), there are \emph{two}
possible steady solutions of the corner flow, one with a strong and one with a (comparatively) weak shock (see Figure \myref{fig:weakstrong}). 
Both shocks satisfy the entropy condition\footnote{There is a third shock that violates the entropy condition.}. 
However, only the \emph{weak} shocks are observed in nature.
To quote \cite{courant-friedrichs}: ``The question arises which of the two actually occurs. 
It has frequently been stated that the strong one is unstable and that, therefore, only the weak one could occur. 
A convincing proof of this instability has apparently never been given.''

The goal of the present paper is to understand this. 

For many purposes, in particular for many questions concerning flow around airplane wings, viscosity, heat conduction 
and kinetic effects can be neglected. 
It is natural to consider inviscid models, such as the full or isentropic compressible Euler equations or compressible potential flow. 
The appropriate boundary condition at solid surfaces is the \defm{slip condition}: the gas velocity is tangential.

In each model the shock and its upstream and downstream states satisfy the \defm{Rankine-Hugoniot relations}, a system of nonlinear 
algebraic equations. These relations determine the \defm{shock polar}: the curve of downstream velocities that result when 
varying the shock normal while holding the shock steady 
and the upstream velocity and density fixed (see Figure \myref{fig:spolar-lines}). 

\if\dofigures%
\begin{figure}
\center{\input{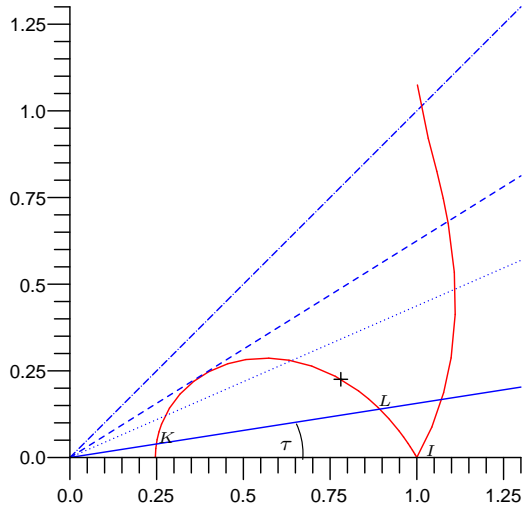}}
\caption{Curve: downstream velocities for all possible shock normals. 
The left branch corresponds to compression shocks, the right branch to unphysical expansion shocks.
The black $+$ is exactly sonic, points left of it are subsonic, right of it supersonic.
Solid line: there are three solutions; strong shock $K$, 
weak shock $L$, unphysical. 
Dotted line: still three solutions, but the weak shock is supersonic-subsonic now.
Dashed line: critical angle; weak and strong shock coincide. 
Dashed-dotted line: $\tau$ above critical, no physical solution.}
\mylabel{fig:spolar-lines}
\end{figure}
\fi%

In Figure \myref{fig:spolar-lines} the upstream velocity is $(1,0)$, labelled $I$. 
Possible downstream velocities are intersections of the shock polar with the ray at a counterclockwise angle $\tau$ from the positive
horizontal axis. Obviously for small $\tau$ there are three intersection points. The leftmost intersection, called $K$, corresponds to the strong shock.
The rightmost is an unphysical expansion shock which need not be considered. The middle point, called $L$, is the weak shock.
(The shock normals are parallel to the difference between downstream velocity (intersection point) and upstream velocity $(1,0)$.)

For $\tau\downarrow 0$, the strong shock approaches a \defm{normal shock}, whereas the weak shock vanishes ($L$ approaches $I$).

There is a \defm{critical angle} $\tau=\tau_*$ where $L$ and $K$ coincide; for larger $\tau$ no steady entropy-satisfying shock
can be attached to the wedge tip resp.\ corner.

The black $+$ on the shock polar indicates a downstream state that is exactly \defm{sonic} (Mach number $M=1$). 
Polar points left of it are \defm{subsonic} ($M<1$), polar points right of it are \defm{supersonic}. 
In numerical experiments, the weak shock detaches from the corner/wedge tip when its downstream changes from supersonic to subsonic
(e.g.\ by increasing $\tau$).

In \cite{elling-liu-rims05,elling-hyp2006} we have reported on numerical experiments: 
to our surprise and somewhat contrary to the aforementioned conjecture, 
both the strong and weak shocks are time-asymptotically stable under large, compactly supported perturbations. 
Instead, the strong shock is unstable under (generic) perturbations of the downstream state at infinity;
depending on the perturbation either the weak shock appears or the shock detaches from the wedge tip/corner entirely.
It may be possible to obtain a strong shock in very special cases, for example by placing a perfectly feedback-controlled nozzle somewhere downstream. 

The weak shock is stable under \emph{both} kinds of perturbation. 

While various conjectures and empirical observations have been made regarding weak vs.\ strong shock, 
previously no \emph{mathematical} arguments for either were known. 
To obtain one, we devise an ``unbiased'' test: at time $t=0$, fill the entire domain
with upstream data; check which shock appears for $t>0$. 

In numerics, the \emph{weak} shock appears spontaneously (see Figure \myref{fig:fullsol}). Motivated by this, we construct an \emph{analytical}
solution.

An equivalent experiment is to accelerate a solid wedge in motionless air
instantaneously to supersonic speed. More generally, if a finite wedge is accelerated from rest at time $0$ to a fixed supersonic speed at
time $\epsilon\ll L/c$ ($L$ wedge length), we may expect the solution to be a good approximation for times $t$ in the scale 
$\epsilon\ll t\ll L/c$.

\subsection{Numerical results}

\mylabel{section:numerics}

\if\dofigures%
\begin{figure}
\parbox[t]{.49\textwidth}{\includegraphics[width=\linewidth]{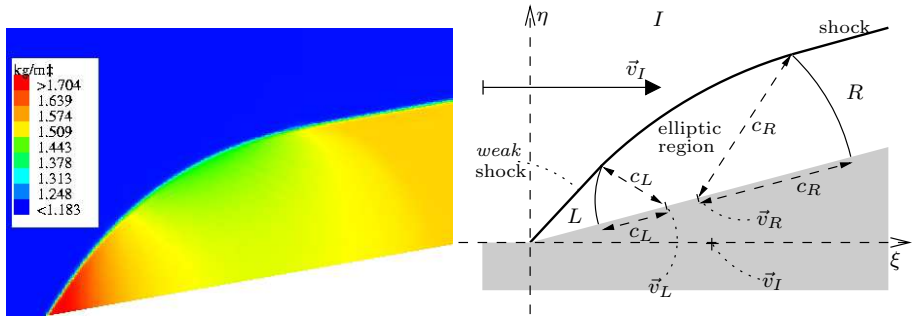}}\hfil
\input{frameorig.pstex_t}
\caption{Left: numerical solution of the wedge flow problem. Right: Structure. Three trivial hyperbolic regions $I$, $L$, $R$ 
are separated by straight shocks. They enclose a nontrivial elliptic region bounded by a curved shock and two parabolic circle arcs 
centered in $\vec v_L$ resp.\ $\vec v_R$ with radius $c_L$ resp.\ $c_R$. Density and velocity are functions of $\xi=x/t$, $\eta=y/t$ only.}
\mylabel{fig:fullsol}
\mylabel{fig:frameorig}
\end{figure}

Figure \myref{fig:fullsol} shows the flow pattern (density) solving our test problem, for some positive time. 
Here $\gamma=7/5$, $M_I=2.94$ and $\tua=10^\circ$.
$\vec v_I$ is horizontal from left to right. 
Blue is $\rho_I$; green, yellow and red are successively larger densities.

A straight shock (blue to red) emanates from the wedge tip. Calculation shows that
it is the weak shock. There is another straight shock on the right (blue to orange), parallel to the downstream wall. 
Below each straight shock lies a constant region. Both shocks are connected by a curved shock, with a nontrivial (elliptic) region below. 

The flow pattern is self-similar: density and velocity are constant along rays $x=\xi t,y=\eta t$ for fixed $\xi,\eta$. 
This can be visualized as $t$ being a ``zoom'' parameter, with $t=0$ corresponding to ``infinitely far away''
and $t\uparrow\infty$ to ``infinitely close to the origin'' (wedge tip resp.\ wall corner).
In particular the flow structure is the same for all times.

\begin{figure}
\parbox[t]{\textwidth}{\includegraphics[width=\linewidth]{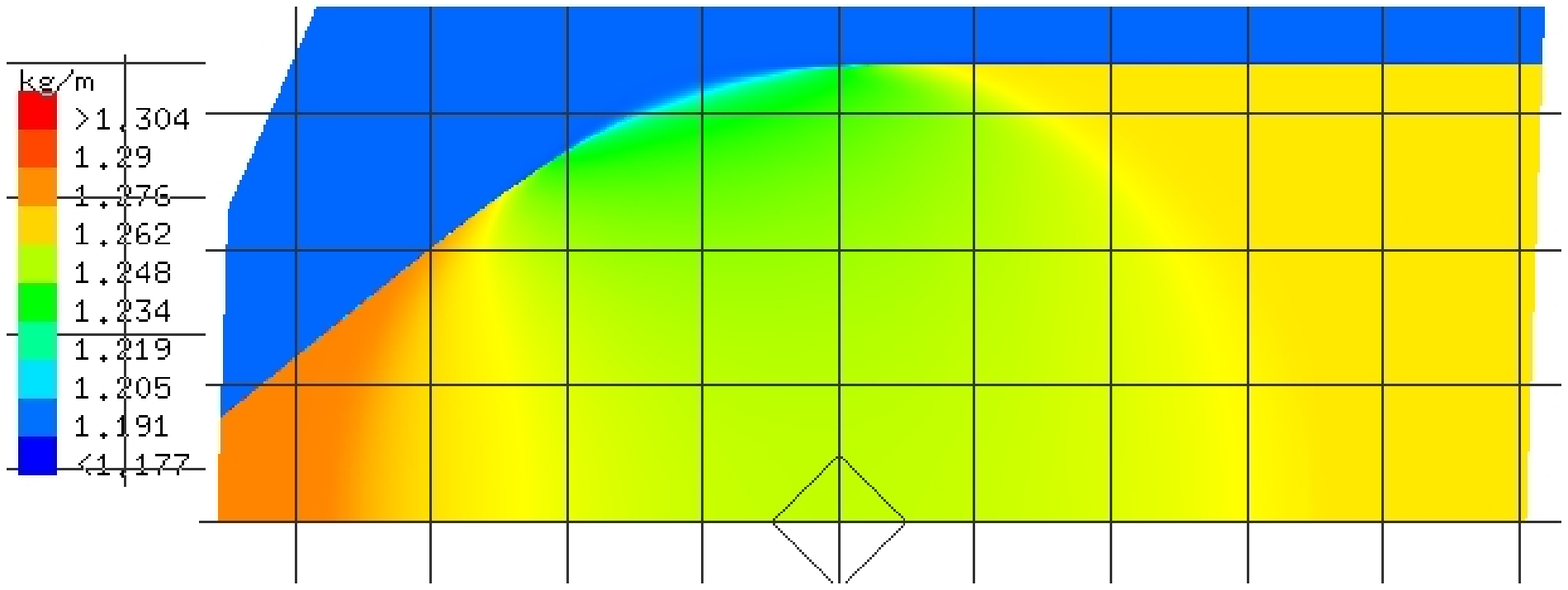}}
\parbox[t]{\textwidth}{\includegraphics[width=\linewidth]{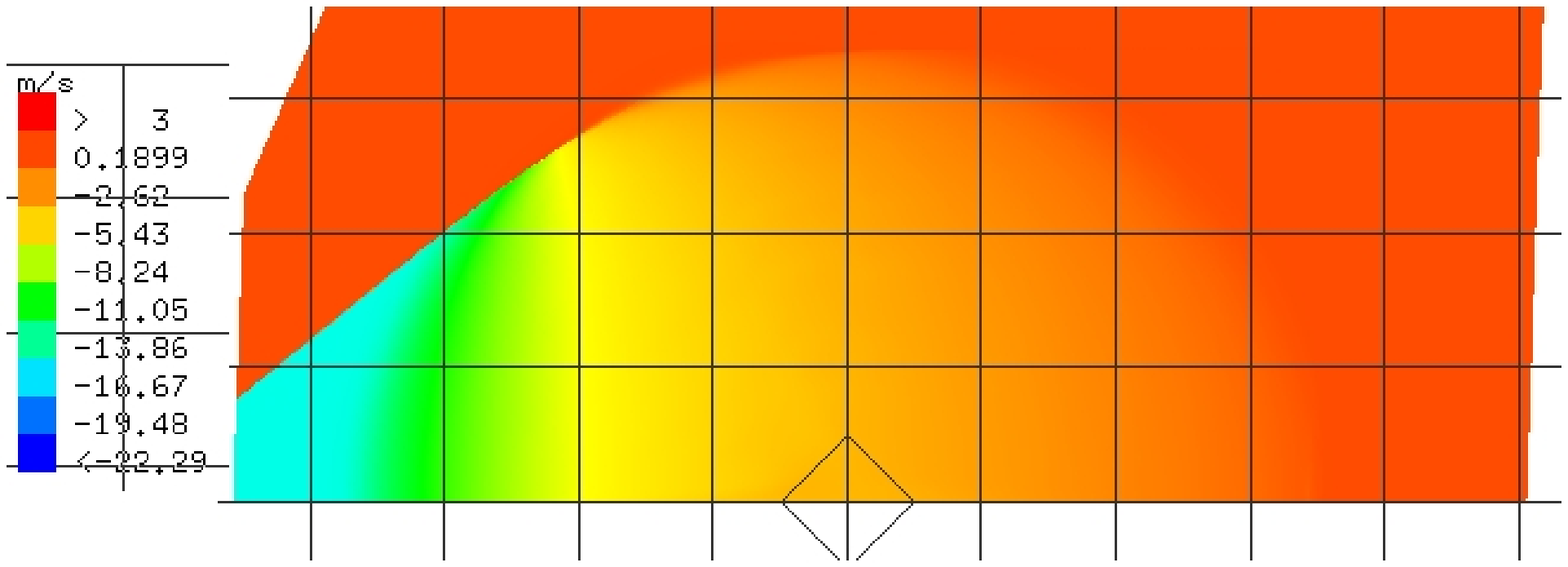}}
\parbox[t]{\textwidth}{\includegraphics[width=\linewidth]{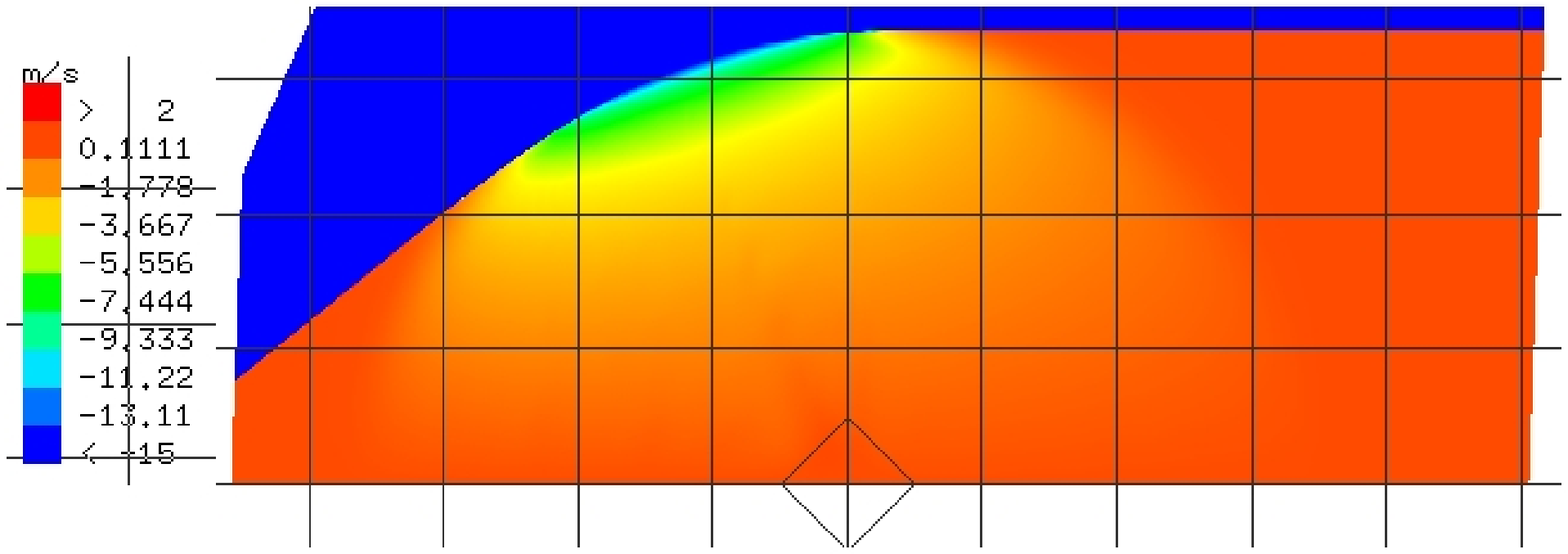}}
\caption{Elliptic region. From top to bottom: $\rho$, $\vec v$ tangential, and $\vec v$ normal to the wedge surface.
Corner/wedge tip is intersection of left shock and bottom domain boundary (left outside diagrams). Diamond indicates
origin in Figure \ref{fig:frameR} coordinates; velocities are relative to it.}
\mylabel{fig:numerics}
\end{figure}
\fi%

Figure \myref{fig:numerics} shows the elliptic region in more detail. In the top picture density is shown; its minimum
in the elliptic region is attained at the shock (Proposition \myref{prop:density-shock} will show that the minimum is a ``pseudo-normal'' point).
In the middle picture the velocity tangential to the downstream wall is shown; the bottom picture shows velocity normal to the downstream wall. 

The diamond in the center of the bottom domain boundaries indicates the origin in self-similar coordinates, where we use standard
coordinates (Figure \myref{fig:frameR}).

It should be emphasized that numerical computations only suggest the structure of the solution. For instance, 
it is not clear that the constant states $L$ and $R$ extend to the pseudo-sonic circles $P_L$ and $P_R$. 
Although in one dimension with viscosity, some techniques can convert a numerical solution with sufficiently small residual into an existence proof for
an exact solution (see e.g.\ \cite{jiang-yu}), only partial results are available in multiple dimensions without viscosity 
(see \cite{kuznetsov,lax-wendroff,elling-lax-wendroff-journal}); it is not clear whether a general result is even true (see \cite{elling-nuq-journal}).

\subsection{Main result}

To obtain a mathematical argument, we construct the self-similar solution \emph{exactly} rather than numerically. 
We use compressible potential flow as model:
\begin{theorem}
	\mylabel{th:elling-liu}%
	Let $\tua\in(0,\frac\pi2)$, $M_I,\rho_I,c_I\in(0,\infty)$, $\gamma\in[1,\infty)$;
	set $\vec v_I:=(M_Ic_I,0)$.
	Define the wedge
	\begin{alignat}{1}
		W &:= \{(x,y)\in\R^2:y<x\tan\tua\}\notag
	\end{alignat}
	and
	\begin{alignat}{1}
		\Omega &:= (0,\infty)\times\complement W.\notag
	\end{alignat}

	Assume the following conditions are satisfied:
	\begin{enumerate}
	\item Unsteady potential flow with $\gamma$-law pressure admits a steady straight shock with upstream data $\rho_I$ and $\vec v_I$ and 
		downstream velocity $\vec v_L$ and sound speed $c_L$ so that\footnote{$\measuredangle$ is the counterclockwise angle
		from first to second vector, ranging from $0$ to $2\pi$.}
		$$\measuredangle(\vec v_I,\vec v_L)=\tua.$$
	\item The shock is supersonic-supersonic:
		$$M_L:=\frac{|\vec v_L|}{c_L}>1.$$
	\item Of the two intersection points of the shock with the circle $\partial B_{c_L}(\vec v_L)$,
		let $\vxit_L$ be the one closer to the corner (origin). Let the \defm{$R$ shock} be the unique shock 
		parallel to $\vec v_L$, 
		with upstream data $\rho_I$ and $\vec v_I$, downstream sound speed $c_R$ and downstream velocity $\vec v_R$ 
		parallel to $\vec v_L$ as well. Of the two intersection points with $\partial B_{c_R}(\vec v_R)$,
		let $\vxit_R$ be the one farther from the corner (see Figure \myref{fig:techcond}). We require that
		\begin{alignat}{1}
			\{\text{Line segment from $\vxit_L$ to $\vxit_R$}\} \cap\overline B_{c_I}(\vec v_I) &= \emptyset.
			\myeqlabel{eq:techcond}
		\end{alignat}
	\end{enumerate}

	Then there is a weak solution (see Remark \myref{rem:weaksol}) $\phi\in C^{0,1}(\overline\Omega)$ of 
	\begin{alignat}{3}
		& \text{unsteady potential flow} \qquad && \text{in $\Omega$,} \myeqlabel{eq:prob1} \\
		& \nabla\phi\cdot\vec n = 0 \qquad && \text{on $\partial W$,} \myeqlabel{eq:prob2} \\
		& \rho = \rho_I, \quad \nabla\phi = \vec v_I \qquad && \text{for $t=0$} \myeqlabel{eq:prob3} 
	\end{alignat}
\end{theorem}

\if\dofigures%
\begin{figure}
	\center{\input{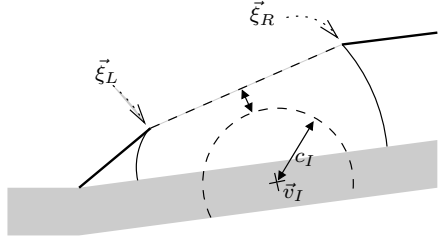}}
	\caption{Solutions are constructed for all cases that satisfy the condition \myeqref{eq:techcond}:
	the dashed circle and line must be separated. A shock has positive strength if and only if it does not
	touch the circle.}
	\mylabel{fig:techcond}
\end{figure}
\fi%

In addition to existence, detailed results about the structure of the weak solution can be obtained (see Remark \myref{rem:structure}).
At this point we emphasize only that each solution consists, in some neighbourhood of the origin, of the \emph{weak} shock separating
two constant-state regions.

\begin{remark}
	\mylabel{rem:weaksol}%
	See Section \myref{section:pf} for introduction and precise definition of potential flow.
	By weak solution we mean that 
	\begin{alignat}{1}
		\nabla\phi(0,\vec x) &= \vec v_I \qquad\text{for a.e.\ $\vec x\in\complement W$} \myeqlabel{eq:weak-ini}
	\end{alignat}
	and
	\begin{alignat}{1}
		\int_\Omega
		\rho\vartheta_t
		+\rho\nabla\phi\cdot\nabla\vartheta~d\vec x~dt
		+\int_{\complement W}\vartheta(0,x,y)\rho_I~d\vec x 
		&= 0  \notag
	\end{alignat}
	for all test functions $\vartheta\in C_c^\infty(\overline\Omega)$.

	(For $\phi\in C^{0,1}(\overline\Omega)$, the velocity $\nabla\phi$ is a.e.\ well-defined on $\{0\}\times\complement W$, but
	$\phi_t$ and hence $\rho$ may not be well-defined.)
\end{remark}

\begin{remark}
	As Remark \myref{rem:notvoid} shows, there is a large set of tip shocks and parameters that satisfy the conditions of
	Theorem \myref{th:elling-liu}.

	The first and second condition are physically necessary, not technical limitations. If the first is violated
	(for $M_I<1$ or large $\tau$), there is no straight steady physical shock attached to the corner at all.
	If either of them is violated, numerical experiments show 
	a flow pattern with a shock \emph{detached} from the corner, moving upstream (left). 

	The third condition is technical.
	It is needed in some cases to prove the shock does not vanish (which is never observed in numerics);
	none of the other estimates requires this condition.
	We expect that the condition will be removed with some additional analysis.
	
	It should be emphasized that the theorem and its proof are global in nature: large parameter changes are possible.
\end{remark}

\begin{remark}
	Incidentally we also solve the problem for \emph{asymmetric} wedges, as long as both sides allow a supersonic-supersonic weak
	shock and as long as \myeqref{eq:techcond} is satisfied on both sides.
\end{remark}

\subsection{Related work}

\cite[Section 117, 122 and 123]{courant-friedrichs} explain in detail shock polars and the corner flow problem. 
Despite its age, its discussion of weak vs.\ strong shock is still a good reflection of the state of prior research.
\cite{ferrari-tricomi} is another useful reference.

\cite{canic-keyfitz-kim} consider the classical problem of regular reflection of a shock by a symmetric wedge;
this problem, like ours, has a self-similar solution.
They consider the unsteady transonic small disturbance equation as model.
\cite{yuxi-zheng-rref} studies the same problem for the pressure-gradient system.
The monographs \cite{yuxi-zheng-book,li-zhang-yang} compute various self-similar flows numerically and present some analysis 
and simplified models.

\cite{gqchen-feldman} prove existence of small perturbations of a plane shock in 
steady potential flow.

\cite{shuxing-chen-cone3d} constructs steady solutions for 3D cones rather than 2D wedges.
\cite{lien-liu} 
discuss stability of 3D flow past a perturbed cone; \cite{shuxing-dening} show existence and linear stability in the case
of the isentropic Euler equations.
\cite{chen-zhang-zhu} study existence and stability of supersonic flows onto perturbed wedges with attached shocks; the introduction gives a detailed
discussion of previous work.

So far the only other paper that proves global existence of some nontrivial time-dependent solution of potential flow is
\cite{chen-feldman-selfsim-journal}: they construct exact solutions for regular reflection, assuming sufficiently sharp wedges.

\subsection{Overview}

In Section \myref{section:ppf} we give an introduction to unsteady potential flow. We derive self-similar potential flow, discuss its shock
conditions and analyze the properties of shocks in detail. 

In Section \myref{section:apriori} we discuss a collection of maximum principles for elliptic regions of self-similar potential flow.
Some of these identify circumstances in which certain quantities (density, ...) can or cannot have maxima or minima in the interior.
Other results discuss local extrema at solid (slip condition) walls and finally shocks with a constant-state hyperbolic region on the other side.

Since the hyperbolic regions are trivial (see Figure \myref{fig:fullsol}), the heart of the problem
is the construction of the elliptic region. This is accomplished in Section \myref{section:ellreg}. 
Readers interested in more overview should go to Section \myref{section:approach}, where all proof steps are surveyed.

A crucial ingredient are the maximum principles from Section \myref{section:apriori}, combined with ODE-type arguments at the parabolic
arcs in Sections \myref{section:L-control} to \myref{section:lowerbounds}, 
and techniques to control shock location and normals (Sections \myref{section:densitycontrol} and \myref{section:v-control}). 
Section \myref{section:entireflow} combines the elliptic region with its hyperbolic counterparts to construct the full flow pattern.
The remaining sections are standard but delicate applications of nonlinear elliptic theory.
Some literature results, such as regularity in corners and at free boundaries, need extension which is done in the Appendix.

\subsection{Notation}

For the most part we use standard notation. Subscripts and superscripts may denote tensor indices, partial derivatives or powers,
depending on the context.

$\measuredangle(\vec x,\vec y)$ is the \emph{counterclockwise} angle from $\vec x$ to $\vec y$.
For $\vec x=(x_1,x_2)$, $\vec y=(y_1,y_2)$, 
$$\vec x^\perp := (-x_2,x_1)$$
(\emph{counterclockwise} rotation by $90^\circ$), 
$$\vec x\times\vec y:=x_1y_2-x_2y_1.$$
$\vec x^2$ is the rank-one matrix $\vec x\vec x^T$ whereas $|\vec x|^2$ is the norm.
Correspondingly, $\nabla^2=\nabla\nabla^T$ is the Hessian (\emph{not} the Laplacian).

Normals $\vec n$ are outer normals to a domain, except on the shock $S$ (defined later) where they are downstream, so usually inner.
Tangents $\vec t$ are always defined as $\vec t:=\vec n^\perp$.

\section{Potential flow}

\mylabel{section:ppf}%

\subsection{Unsteady potential flow}
\mylabel{section:pf}%

We consider the isentropic Euler equations of compressible gas dynamics in $d$ space dimensions:
\begin{alignat}{1}
    \rho_t + \nabla\cdot(\rho\vec v) &= 0 \myeqlabel{eq:rhodiv} \\
    (\rho\vec v)_t + \sum_{i=1}^d(\rho v^i\vec v)_{x^i} + \nabla(p(\rho)) &= 0, \myeqlabel{eq:mom}
\end{alignat}
Hereafter, $\nabla$ denotes the gradient with respect either to the space coordinates ${\vec x}=(x^1,x^2,\cdots,x^d)$ or
the similarity coordinates $t^{-1}\vec x$.
${\vec v}=(v^1,v^2,\cdots,v^d)$ is the velocity of the gas, $\rho$ the density, $p(\rho)$ pressure.
In this article we consider only polytropic pressure laws ($\gamma$-laws) with $\gamma\geq 1$:
\begin{alignat}{1}
    p(\rho) &= \frac{c_0^2\rho_0}{\gamma}\left(\frac{\rho}{\rho_0}\right)^\gamma \myeqlabel{eq:p-polytropic}
\end{alignat}
(here $c_0$ is the sound speed at density $\rho_0$). Many subsequent results extend with little or no change to 
$\gamma<1$ or to general pressure laws, but in special cases some steps require more work or break down entirely. 
To keep the presentation simple we don't strive for generality with respect to pressure laws.

For smooth solutions, substituting \myeqref{eq:rhodiv} into \myeqref{eq:mom} yields the simpler form
\begin{alignat}{1}
    \vec v_t+\vec v\cdot\nabla^T\vec v + \nabla(\pi(\rho)) &= 0. \myeqlabel{eq:v}
\end{alignat}
Here $\pi$ is defined as
\begin{alignat}{1}
    \pi(\rho) &= c_0^2\cdot\begin{cases}
	\frac{(\rho/\rho_0)^{\gamma-1}-1}{\gamma-1}, & \gamma> 1 \\
	\log(\rho/\rho_0), & \gamma=1.
    \end{cases}\notag
\end{alignat}
This $\pi$ is $C^\infty$ in $\rho\in(0,\infty)$ and $\gamma\in[1,\infty)$
and has the property
$$\pi_\rho=\frac{p_\rho}{\rho}.$$

If we assume \defm{irrotationality}
$$v^i_j=v^j_i$$
(where $i,j=1,\dotsc,d$), then the Euler equations are reduced to potential flow:
\begin{alignat*}{1}
    \vec v &= \nabla_{\vec x}\phi
\end{alignat*}
for some scalar \defm{potential}\footnote{We consider simply connected domains; otherwise $\phi$ might be multivalued.} 
function $\phi$. For smooth flows, substituting this into \myeqref{eq:v} yields, for $i=1,\dotsc,d$,
\begin{alignat}{1}
    0 
    &= \phi_{it} + \nabla\phi_i\cdot\nabla\phi + \pi(\rho)_i = \big(\phi_t + \frac{|\nabla\phi|^2}{2} + \pi(\rho)\big)_i. \notag
\end{alignat}
Thus, for some constant $A$, 
\begin{alignat}{1}
    \rho &= \pi^{-1}(A-\phi_t-\frac{|\nabla\phi|^2}{2}). \myeqlabel{eq:rhoA}
\end{alignat}
Substituting this into \myeqref{eq:rhodiv} yields a single second-order quasilinear hyperbolic equation, the \defm{potential flow} equation, for a scalar field $\phi$:
\begin{alignat}{1}
	\big(\rho(\phi_t,|\nabla\phi|)\big)_t+\nabla\cdot\big(\rho(\phi_t,|\nabla\phi|)\nabla\phi\big) &= 0. \myeqlabel{eq:potflow-divform}
\end{alignat}
Henceforth we omit the arguments of $\rho$. Moreover we eliminate $A$ with the substitution 
\begin{alignat}{1}
	&A\leftarrow 0,\qquad\phi(t,\vec x)\leftarrow\phi(t,\vec x)-tA\notag
\end{alignat}
(so that $\phi_t\leftarrow\phi_t-A$). Hence we use
\begin{alignat}{1}
	\rho &= \pi^{-1}(-\phi_t-\frac{1}{2}|\nabla\phi|^2)\myeqlabel{eq:rho}
\end{alignat}
from now on.

Using $c^2=p_\rho$ and 
\begin{alignat}{1}
  (\pi^{-1})' &= (\pi_\rho)^{-1} = (\frac{p_\rho}{\rho})^{-1}=\frac{\rho}{c^2} \myeqlabel{eq:Dpiinv}
\end{alignat}
the equation can also be written in nondivergence form:
\begin{alignat}{1}
    \phi_{tt} + 2\nabla\phi_t\cdot\nabla\phi + \sum_{i,j=1}^d\phi_i\phi_j\phi_{ij} - c^2\Delta\phi &= 0 \myeqlabel{eq:potential-flow}
\end{alignat}
\myeqref{eq:potential-flow} is hyperbolic (as long as $c>0$).
For polytropic pressure law the local sound speed $c$ is given by
\begin{alignat}{1}
	c^2 &= c_0^2 + (\gamma-1)(-\phi_t-\frac{1}{2}|\nabla\phi|^2). \myeqlabel{eq:cs-uspf}
\end{alignat}

\subsection{Self-similar potential flow}
\mylabel{section:sspf}%

Our initial data is self-similar: it is constant along rays emanating from $\vec x=(0,0)$.
Our domain $\complement W$ is self-similar too: it is a union of rays emanating from $(t,x,y)=(0,0,0)$.
In any such situation it is expected --- and confirmed by numerical results --- that the solution is self-similar as well, i.e.\ that
$\rho,\vec v$ are constant along rays $\vec x=t\vec\xi$ emanating from the origin. 
Self-similarity corresponds to the ansatz
\begin{alignat}{1}
	\phi(t,\vec x) &:= t\psi(\vec\xi),\qquad\vec\xi := t^{-1}\vec x \myeqlabel{eq:psi-phi}.
\end{alignat}
Clearly, $\phi\in C^{0,1}(\Omega)$ if and only if $\psi\in C^{0,1}(\complement W)$.
This choice yields
\begin{alignat}{1}
	\vec v(t,\vec x) &= \nabla\phi(t,\vec x) = \nabla\psi(t^{-1}\vec x), \notag\\
	\rho(t,\vec x) &= \pi^{-1}(-\phi_t-\frac{1}{2}|\nabla\phi|^2) 
	= \pi^{-1}(-\psi+\vec\xi\cdot\nabla\psi-\frac{1}{2}|\nabla\psi|^2).\notag
\end{alignat}
The expression for $\rho$ can be made more pleasant (and independent of $\vec\xi$) by using
\begin{alignat}{1}
	\chi(\vec\xi) &:= \psi(\vec\xi)-\frac{1}{2}|\vec\xi|^2;\notag
\end{alignat}
this yields
\begin{alignat}{1}
	\rho &= \pi^{-1}(-\chi-\frac{1}{2}|\nabla\chi|^2). \myeqlabel{eq:rhoeq}
\end{alignat}
$\nabla\chi=\nabla\psi-\vec\xi$ is called \defm{pseudo-velocity}.

\myeqref{eq:potflow-divform} then reduces to
\begin{alignat}{1}
	\nabla\cdot(\rho\nabla\chi)+2\rho &= 0 \myeqlabel{eq:chi-divform}
\end{alignat}
(or $+d\rho$, in $d$ dimensions)
which holds in a distributional sense.
For smooth solutions we obtain the non-divergence form
\begin{alignat}{1}
	(c^2I-\nabla\chi\nabla\chi^T):\nabla^2\chi
	= (c^2-\chi_\xi^2)\chi_{\xi\xi}-2\chi_\xi\chi_\eta\chi_{\xi\eta}+(c^2-\chi_\eta^2)\chi_{\eta\eta}
	&= |\nabla\chi|^2-2c^2 \myeqlabel{eq:chi} 
\end{alignat}
Another convenient form is
\begin{alignat}{1}
	(c^2I-\nabla\chi\nabla\chi^T):\nabla^2\psi 
	&= (c^2-\chi_\xi^2)\psi_{\xi\xi}-2\chi_\xi\chi_\eta\psi_{\xi\eta}+(c^2-\chi_\eta^2)\psi_{\eta\eta} 
	= 0. \myeqlabel{eq:psi}
\end{alignat}
Here, \myeqref{eq:cs-uspf} for polytropic pressure law yields 
\begin{alignat}{1}
	c^2 &= c_0^2 + (\gamma-1)(-\chi-\frac{1}{2}|\nabla\chi|^2) \myeqlabel{eq:css}
\end{alignat}

\begin{remark}
	\mylabel{rem:symmetries}%
	\myeqref{eq:chi-divform} inherits a number of symmetries from \myeqref{eq:rhodiv}, \myeqref{eq:mom}: 
	\begin{enumerate}
	\item It is invariant under rotation.
	\item It is invariant under reflection.
	\item It is invariant under translation in $\vec\xi$, which is not as trivial as translation in $\vec x$:
	it corresponds to the Galilean transformation $\vec v\leftarrow \vec v+\vec v_0$, 
	$\vec x\leftarrow \vec x-\vec v_0t$ (with constant $\vec v_0\in\R^d$) in $(t,\vec x)$ coordinates.
	This is sometimes called \defm{change of inertial frame}.
	\end{enumerate}
\end{remark}

\myeqref{eq:chi} is a PDE of mixed type. The type
is determined by the \defm{(local) pseudo-Mach number}
\begin{alignat}{1}
    L &:= \frac{|\nabla\chi|}{c}, \myeqlabel{eq:L}
\end{alignat}
with $0\leq L<1$ for elliptic (pseudo-subsonic), $L=1$ for parabolic (pseudo-sonic), $L>1$ for hyperbolic (pseudo-supersonic) regions.

The pseudo-Mach number $L$ can be interpreted in a way analogous to the Mach number $M$: 
consider a \emph{steady} solution of the \emph{unsteady} potential flow equation. 
Loosely speaking, in an $M<1$ (subsonic) region a small localized disturbance will be propagated in all directions, 
whereas in an $M>1$ region it is propagated only in the \defm{Mach cone}. 
$L<1$ and $L>1$ are analogous, except that we study the propagation of disturbances in 
the unsteady potential flow equation written in $(t,\vec x/t)$ coordinates, rather than $(t,\vec x)$. 

There is no strong relation between 
$M<1$ and $L<1$: consider two constant-state (hence steady and selfsimilar) flows with zero velocity ($M=0$ constant) 
resp.\ supersonic velocity $\vec v$ ($M>1$ constant). Each flow has $L=0$ in the point $\vec\xi=0$ 
resp.\ $\vec\xi=\vec v$ and $L\uparrow\infty$ as $\vec\xi\uparrow\infty$, so
there are examples for each of the four cases $M,L<1$, $M<1<L$, $L<1<M$ and $1<L,M$.

While velocity $\vec v$ is motion
relative to space coordinates $\vec x$, pseudo-velocity 
$$\vec z:=\nabla\chi$$
is motion relative to similarity coordinates
$\vec\xi$ \emph{at time $t=1$}.

The simplest class of solutions of \myeqref{eq:chi} are the \defm{constant-state solutions}: $\psi$ affine in $\vec\xi$, 
hence $\vec v$, $\rho$ and $c$ constant. They are elliptic in a circle centered in $\vec\xi=\vec v$ with radius $c$, parabolic
on the boundary of that circle and hyperbolic outside.

\begin{convention}
	If we study a function called (e.g.) $\tilde\chi$, then $\tilde\psi$, $\tilde\rho$, $\tilde L$ etc.
	will refer to the quantities computed from it as $\psi$, $\rho$, $L$ are computed from $\chi$
	(e.g.\ $\tilde\psi=\tilde\chi+\frac{1}{2}|\vec\xi|^2$). We will
	tacitly use this notation from now on.
\end{convention}

\subsection{Shock conditions}

Consider a ball $U$ and a simple smooth curve $S$ so that $U=U^u\cup S\cup U^d$ where
$U^u,U^d$ are open, connected, and $S,U^u,U^d$ disjoint.
Consider $\chi:U\rightarrow\R$ so that $\chi=\chi^{u,d}$ in $U^{u,d}$ where 
$\chi^{u,d}\in \spC^2(\overline{U^{u,d}})$. 

$\chi$ is a weak solution of \myeqref{eq:chi-divform} if and only if it is a strong solution in each point of $U_-$ and $U_+$ and 
if it satisfies the following conditions in each point of $S$: 
\begin{alignat}{1}
	\chi^u &= \chi^d, \myeqlabel{eq:chijump} \\
	\vec n\cdot(\rho^u\nabla\chi^u-\rho^d\nabla\chi^d) &= 0 \myeqlabel{eq:momjump}
\end{alignat}
Here $\vec n$ is a normal to $S$.

\myeqref{eq:chijump} and \myeqref{eq:momjump} are the \defm{Rankine-Hugoniot} conditions for self-similar potential flow
shocks. They do not depend on $\vec\xi$ or on the shock speed explicitly; these quantities
are hidden by the use of $\chi$ rather than $\psi$. The Rankine-Hugoniot conditions are derived in the same way as
those for the full Euler equations (see \cite[Section 3.4.1]{evans}).

Note that \myeqref{eq:chijump} is equivalent to 
\begin{alignat}{1}
	\psi^u &= \psi^d. \myeqlabel{eq:psijump}
\end{alignat}
Taking the tangential derivative of \myeqref{eq:chijump} resp.\ \myeqref{eq:psijump} yields
\begin{alignat}{1}
	\frac{\partial\chi^u}{\partial t} &= \frac{\partial\chi^d}{\partial t} \myeqlabel{eq:chitan}, \\
	\frac{\partial\psi^u}{\partial t} &= \frac{\partial\psi^d}{\partial t} \myeqlabel{eq:psitan}.
\end{alignat}
The shock relations imply that the tangential velocity is continuous across shocks.

Define $(z^x_u,z^y_u):=\vec z_u:=\nabla\chi^u$ and $(v^x_u,v^y_u):=\vec v_u:=\nabla\psi^u$.
Abbreviate $z^t_u:=\vec z_u\cdot\vec t$, $z^n_u:=\vec z_u\cdot\vec n$, and same for $v$ instead of $z$.
Same definitions for $d$ instead of $u$.
We can restate the shock relations as
\begin{alignat}{1}
	\rho_uz^n_u &= \rho_dz^n_d, \myeqlabel{eq:steady-continuity} \\
	z_u^t &= z_d^t. \myeqlabel{eq:chitan-z}
\end{alignat}
Using the last relation, we often write $z^t$ without distinction.

The \defm{shock speed} is $\sigma=\vec\xi\cdot\vec n$, where $\vec\xi$ is any point on the shock. 
A shock is \defm{steady} in a point if its tangent passes through the origin.
We can restate \myeqref{eq:steady-continuity} as
\begin{alignat}{1}
	\rho_uv^n_u-\rho_dv^n_d &= \sigma(\rho_u-\rho_d)\notag
\end{alignat}
which is a more familiar form.

We focus on $\rho_u,\rho_d>0$ from now on, which will be the case in all circumstances. 
If $\rho_u=\rho_d$ in a point, we say the shock \defm{vanishes}; in this case $z^n_d=z^n_u$ in that point, by \myeqref{eq:chitan-z}. 
In all other cases $z^n_d,z^n_u$ must have equal sign by \myeqref{eq:chitan-z}; we fix $\vec n$ so that $z^n_d,z^n_u>0$. 
This means the normal points \defm{downstream}. 
The shock is \defm{admissible} if and only if $\rho_u\leq\rho_d$ which is equivalent to $z^n_u\geq z^n_d$.

A shock is called \defm{pseudo-normal} in a point $\vec\xi$ if $z^t=0$ there.
For $\vec\xi=0$, this means that the shock is \defm{normal} ($v^t=0$), but for $\vec\xi\neq 0$ normal and pseudo-normal are not always equivalent.

It is good to keep in mind that for a \emph{straight} shock, $\rho_d$ and $\vec v_d$ are constant if $\rho_u$ and $\vec v_u$ are.
Obviously $\vec z_d$ may vary in this case.

\begin{remark}
	The Rankine-Hugoniot conditions for the original isentropic Euler equations cannot be used for potential flow: 
	even if the flow on one side of a shock is irrotational, the flow on the other side has nonzero vorticity 
	for curved shocks.
\end{remark}

\subsection{Pseudo-normal shocks}

\mylabel{section:shocks}

Here we study the consequences and solutions of the shock relations in a pseudo-normal point.
We state all results for pseudo-velocities $\vec z$ and for pseudo-Mach numbers $L$ because moving shocks 
are ubiquitous in this article. For better intuition the reader may bear in mind that $\vec z=\vec v$
and $L=M$ if the shock is steady, i.e.\ passes through the origin. In fact 
by Remark \myref{rem:symmetries}, weak and entropy solutions of self-similar potential flow are invariant
under translation and rotation, so we may always consider translating the shock so that it becomes steady,
which does not change $\vec z,\rho,L$ whereas $\vec v$ is changed only by a constant vector.
Hence the behaviour of arbitrary shocks is entirely determined by those of steady shocks.

$L^n:=z^n/c$ and $L^t:=z^t/c$ will be referred to as normal resp.\ tangential pseudo-Mach number.
\myeqref{eq:chijump} and \myeqref{eq:rhoeq} imply
\begin{alignat}{1}
	\rho_d &= \pi^{-1}(\pi(\rho_u)+\frac{|z_u|^2}{2}-\frac{|z_d|^2}{2}). \myeqlabel{eq:steady-rho}
\end{alignat}
It is apparent that \myeqref{eq:steady-rho} reduces to
\begin{alignat}{1}
	\rho_d &= \pi^{-1}(\pi(\rho_u)+\frac{(z^n_u)^2}{2}-\frac{(z^n_d)^2}{2}). \myeqlabel{eq:normal-rho}
\end{alignat}
Combined with \myeqref{eq:chitan-z}
there is a direct relation between normal velocities, independent of the tangential velocities.

For polytropic pressure laws we may use a rather convenient simplification: there is an explicit relation connecting
$L^n_u,L^n_d$, independent of $\rho_u,\vec z_u$.

\begin{lemma}
	\mylabel{lemma:srel-M}%
	For $L^n_u,L^n_d>0$, \myeqref{eq:normal-rho} and \myeqref{eq:steady-continuity} are equivalent to
	\begin{alignat}{1}
		g(L^n_u) &= g(L^n_d), \myeqlabel{eq:srel-M} \\
		g(x) &= \begin{cases}
			\left(x^2+\frac{2}{\gamma-1}\right)x^\frac{2(1-\gamma)}{\gamma+1}, & \gamma>1, \\
			x^2-2\log x, & \gamma=1.
		\end{cases}
		\myeqlabel{eq:g}
	\end{alignat}
	Note that
	\begin{alignat}{1}
		\frac{\partial g}{\partial x} &= \frac{4}{\gamma+1}(x-x^{-1})x^{-2\frac{\gamma-1}{\gamma+1}}, \myeqlabel{eq:dgdM} \\
		\frac{\partial^2g}{\partial x^2} &= \frac{4}{(\gamma+1)^2}\big((3-\gamma)+(3\gamma-1)x^{-2}\big)x^{-2\frac{\gamma-1}{\gamma+1}}. \myeqlabel{eq:ddgddM} \\
		g(x) &\sim \begin{cases}
			x^{-2\frac{\gamma-1}{\gamma+1}},&\gamma>1 \\
			-\log x,&\gamma=1 
			\end{cases} \qquad\text{as $x\downarrow 0$,} \myeqlabel{eq:gM0} \\
		g(x) &\sim x^\frac{4}{\gamma+1}\qquad\text{as $x\uparrow\infty$,} \myeqlabel{eq:gMinf}  
	\end{alignat}
	Moreover
	\begin{alignat}{1}
		\frac{c_u}{c_d} &= \left(\frac{L^n_d}{L^n_u}\right)^\frac{\gamma-1}{\gamma+1}.\myeqlabel{eq:cM}
	\end{alignat}
	as well as
	\begin{alignat}{1}
		\frac{\rho_u}{\rho_d} &= \left(\frac{L^n_d}{L^n_u}\right)^\frac{2}{\gamma+1}.\myeqlabel{eq:rhoM}
	\end{alignat}	
\end{lemma}
\begin{proof}
	\myeqref{eq:steady-continuity} can be written
	\begin{alignat}{1}
		\rho_uL^n_uc_u &= \rho_dL^n_dc_d   \notag\\
		\Rightarrow\qquad (\rho_u)^{(\gamma+1)/2}L^n_u &= \rho_d^{(\gamma+1)/2}L^n_d; \myeqlabel{eq:cMpre}
	\end{alignat}
	this yields \myeqref{eq:rhoM} which yields \myeqref{eq:cM}.

	Consider $\gamma>1$, so that $c^2=c_0^2+(\gamma-1)\pi(\rho)$.
	\myeqref{eq:normal-rho} can be transformed to
	\begin{alignat}{1}
		\frac{(z^n_u)^2}{2}+\frac{c_u^2}{\gamma-1} &= \frac{(z^n_d)^2}{2}+\frac{c_d^2}{\gamma-1}  \notag\\
		\Leftrightarrow\quad \left((L^n_u)^2+\frac{2}{\gamma-1}\right)c_u^2 &= \left((L^n_d)^2+\frac{2}{\gamma-1}\right)c_d^2 \notag
	\end{alignat}
	Substitute \myeqref{eq:cM} to obtain \myeqref{eq:srel-M}.

	For $\gamma=1$, \myeqref{eq:normal-rho} is
	\begin{alignat}{1}
		c_0^2\log\frac{\rho_d}{\rho_0}+\frac{(z^n_d)^2}{2} &= c_0^2\log\frac{\rho_u}{\rho_0}+\frac{(z^n_u)^2}{2} \notag\\
		\Rightarrow\qquad \frac{\rho_d}{\rho_u} &= \exp\frac{(L^n_u)^2-(L^n_d)^2}{2}; \notag
	\end{alignat}
	using \myeqref{eq:rhoM} we obtain \myeqref{eq:srel-M}.
\end{proof}

\begin{proposition}
	\mylabel{lemma:srel}%
	There is an analytic strictly decreasing function $L^n_d=L^n_d(L^n_u)$, which is its own inverse,
	so that the shock relation \myeqref{eq:srel-M} is solved for all $L^n_u\in(0,\infty)$. 
	For $L^n_u\neq 1$, the only other solution of \myeqref{eq:srel-M} is the trivial one: $L^n_d=L^n_u$. 
	For $L^n_u=1$, both coincide.
	
	\begin{alignat}{1}
		L^n_d & \begin{cases}
			\uparrow\infty, & L^n_u\downarrow 0, \\
			= 1, & L^n_u=1, \\
			\sim (L^n_u)^\frac{-2}{\gamma-1}, & L^n_u\uparrow\infty,\ \gamma>1, \\
			\sim \exp(-\frac{(L^n_u)^2}{2}), & L^n_u\uparrow\infty,\ \gamma=1.
		\end{cases} \myeqlabel{eq:Mnd-asym}
	\end{alignat}
	The resulting shock is admissible if and only if $L^n_u\geq 1$.
	\begin{alignat}{1}
		\frac{\partial L^n_d}{\partial L^n_u} &= 
		\frac{L^n_u-1/L^n_u}{L^n_d-1/L^n_d}\left(\frac{L^n_u}{L^n_d}\right)^{-2\frac{\gamma-1}{\gamma+1}} < 0,
		\myeqlabel{eq:dMRdMLgen} \\
		\frac{\partial L^n_d}{\partial L^n_u}_{|L^n_u=L^n_d(=1)} &=-1, \myeqlabel{eq:dMRdML} \\
	\end{alignat}
	For $L^n_u>1>L^n_d$, we have $\rho_d>\rho_u$ and $z^n_d<z^n_u$. $\rho_d,c_d$ are strictly increasing in $L^n_u$ for $\rho_u$ fixed. 
	For $\gamma>1$, $c_d>c_u$ as well, and
	$c_d$ is strictly increasing in $L^n_u$ for $\rho_u$ fixed.
\end{proposition}
\begin{proof}
	Assume $L^n_u>1$. From \myeqref{eq:dgdM} it is obvious that $\frac{\partial g}{\partial x}(x)>0$ for $x>1$. 
	Hence $g(L^n_u)>g(1)$; moreover \myeqref{eq:srel-M} cannot have more than one solution $L^n_d$ in $[1,\infty]$ for fixed $L^n_u\geq1$; 
	in fact $L^n_d=L^n_u$ is the unique solution. 	
	For $x<1$, $\frac{\partial g}{\partial x}(x)<0$, so \myeqref{eq:srel-M} cannot have more than one solution $L^n_d$ in $(0,1)$.
	It must have one, though, because $g(0+)=+\infty>g(L^n_u)>g(1)$.

	For $L^n_u<1$, the existence of a nontrivial solution $L^n_d\in(1,\infty)$ is obtained from the previous case by analogous arguments. 
	For $L^n_u=1$, the sign of $\partial g/\partial x$ rules out any other solutions. 
	Since \myeqref{eq:srel-M} is symmetric in $L^n_u,L^n_d$, it is obvious that $L^n_u\mapsto L^n_d$ is its own inverse. 

	The trivial solution branch $L^n_d=L^n_u$ is obviously smooth; by the implicit function theorem the the nontrivial branch
	is analytic away from $L^n_u=1$. In $L^n_u=1$ there is a degeneracy which has to be analyzed by inspecting the Hessian of 
	$h(L^n_u,L^n_d):=g(L^n_u)-g(L^n_d)=0$: 
	\begin{alignat}{1}
		A:=\begin{bmatrix}
			\frac{\partial^2h}{(\partial L^n_u)^2} & \frac{\partial^2h}{\partial L^n_u\partial L^n_d} \\
			\frac{\partial^2h}{\partial L^n_u\partial L^n_d} & \frac{\partial^2h}{(\partial L^n_d)^2}
		\end{bmatrix}
		&= \begin{bmatrix}
			\frac{\partial^2g}{\partial x^2}(L^n_u) & 0 \\ 0 & \frac{-\partial^2g}{\partial x^2}(L^n_d)
		\end{bmatrix}_{|L^n_u=L^n_d=1} \overset{\text{\myeqref{eq:ddgddM}}}{=} \begin{bmatrix}
			\frac{8}{\gamma+1} & 0 \\ 0 & \frac{-8}{\gamma+1}
		\end{bmatrix}.\notag
	\end{alignat}
	$A$ is an invertible indefinite matrix; the solutions of $\vec w^TA\vec w=0$ are $\vec w=(1,1)$ and $\vec w=(1,-1)$. 
	The classical Morse lemma (see e.g.\ \cite[Lemma 12.19]{smoller}) shows that in a small neighbourhood of $(L^n_u,L^n_d)=(1,1)$, 
	the solution of \myeqref{eq:srel-M} form two analytic curves that intersect in $(1,1)$ with tangents $(1,1)$ (trivial branch) and $(1,-1)$ 
	(nontrivial branch). The latter yields \myeqref{eq:dMRdML}.

	The remainder of \myeqref{eq:Mnd-asym} follows from the asymptotics of $g$ (see \myeqref{eq:gM0}, \myeqref{eq:gMinf}).

	From \myeqref{eq:rhoM} and \myeqref{eq:dMRdMLgen} we see that $\rho_d>\rho_u$ for $L^n_d<1<L^n_u$ and that $\rho_d$ is an increasing function of
	$L^n_u$ for $\rho_u$ held fixed. 
	Clearly the same applies to $c_d$ if $\gamma>1$, 
	and (because of \myeqref{eq:steady-continuity}) $z^n_d<z^n_u$. This means the shock is admissible for $L^n_u>1$.
\end{proof}

\begin{remark}
	In the remaining arguments we always assume that the shock is admissible (or vanishing) and ignore the branch $L^n_d=L^n_u$.
\end{remark}

\begin{proposition}
	\mylabel{prop:shockv}%
	If we hold $c_u,\rho_u$ fixed:
	\begin{alignat}{1}
		\frac{z^n_u}{z^n_d}\cdot\frac{\partial z^n_d}{\partial z^n_u}
		< \frac{\gamma-1}{\gamma+1} < 1.\myeqlabel{eq:DDspecial}
	\end{alignat}
	In particular
	\begin{alignat}{1}
		\frac{\partial z^n_d}{\partial z^n_u}
		< \frac{\gamma-1}{\gamma+1}.\myeqlabel{eq:DvndDvnu}
	\end{alignat}
	Therefore
	\begin{alignat}{1}
		\frac{z^n_d}{c_u}-1<\frac{\gamma-1}{\gamma+1}\Big(\frac{z^n_u}{c_u}-1\Big) \myeqlabel{eq:zndznucomp}
	\end{alignat}
\end{proposition}
\begin{remark}
\myeqref{eq:DvndDvnu} is not very tight (we can show $<0$ for $\gamma<3$), but sufficient for our purposes.
\end{remark}
\begin{proof}
	We use \myeqref{eq:cM}:
	\begin{alignat*}{1}
		\frac{\partial c_d}{\partial L^n_u} 
		&= c_u\frac{\partial}{\partial L^n_u}\left(\frac{L^n_u}{L^n_d}\right)^\frac{\gamma-1}{\gamma+1} 
		= \frac{c_u}{L^n_d}\left(\frac{L^n_u}{L^n_d}\right)^\frac{-2}{\gamma+1}\frac{\gamma-1}{\gamma+1}
		\left(1-\frac{L^n_u}{L^n_d}\frac{\partial L^n_d}{\partial L^n_u}\right)
	\end{alignat*}
	so
	\begin{alignat*}{1}
		\frac{\partial z^n_d}{\partial z^n_u} 
		&= c_u^{-1}\frac{\partial z^n_d}{\partial L^n_u} 
		= \frac{L^n_d}{c_u}\frac{\partial c_d}{\partial L^n_u}+\frac{c_d}{c_u}\frac{\partial L^n_d}{\partial L^n_u} \\
		&\overset{\myeqref{eq:cM}}{=} \left(\frac{L^n_u}{L^n_d}\right)^\frac{-2}{\gamma+1}
		\left(\frac{2}{\gamma+1}\cdot\frac{L^n_u}{L^n_d}
		\cdot\frac{\partial L^n_d}{\partial L^n_u}
		+\frac{\gamma-1}{\gamma+1}\right)
	\end{alignat*}
	Then
	\begin{alignat*}{1}
		\frac{z^n_u}{z^n_d}\frac{\partial z^n_d}{\partial z^n_u} 
		&= \frac{\rho_d}{\rho_u}\left(\frac{L^n_u}{L^n_d}\right)^\frac{-2}{\gamma+1}
		\left(\frac{2}{\gamma+1}\cdot\frac{L^n_u}{L^n_d}
		\cdot\frac{\partial L^n_d}{\partial L^n_u}
		+\frac{\gamma-1}{\gamma+1}\right) \\
		&\overset{\myeqref{eq:rhoM}}{=} \subeq{\frac{2}{\gamma+1}}{>0}\cdot\subeq{\frac{L^n_u}{L^n_d}}{>0}
		\cdot\subeq{\frac{\partial L^n_d}{\partial L^n_u}}{<0}
		+\frac{\gamma-1}{\gamma+1} < \frac{\gamma-1}{\gamma+1}.
	\end{alignat*}
	Integrating \myeqref{eq:DvndDvnu} from $z^n_d=z^n_u=c_u$ for a vanishing shock, we obtain \myeqref{eq:zndznucomp}.
\end{proof}

\begin{proposition}
	\mylabel{lemma:movingnormal}%
	Consider a shock with velocity $\sigma:=\vec\xi\cdot\vec n$. Our convention $z^n_u>0$ requires $\sigma<v^n_u$. 
	Vary $\sigma$ while holding $\vec n$ and $\vec v_u$ fixed. Then:
	\begin{alignat}{1}
		\frac{\partial v^n_d}{\partial\sigma} &\geq 1-\frac{\partial z^n_d}{\partial z^n_u} > 0; \myeqlabel{eq:DvndDsigma} \\
		\qquad \frac{\partial\rho_d}{\partial\sigma} &<0. \myeqlabel{eq:DrhoDsigma}
	\end{alignat}
\end{proposition}
\begin{proof}
	For moving shocks, $v^n_d=z^n_d+\sigma$ and $z^n_u=v^n_u-\sigma$, so
	\begin{alignat*}{1}
		\frac{\partial v^n_d}{\partial\sigma} &= 1+\frac{\partial z^n_d}{\partial\sigma}
		= 1-\frac{\partial z^n_d}{\partial z^n_u} \overset{\text{\myeqref{eq:DvndDvnu}}}{>} 
		1-\frac{\gamma-1}{\gamma+1} = \frac{2}{\gamma+1} > 0
	\end{alignat*}
	and
	\begin{alignat*}{1}
		\frac{\partial\rho_d}{\partial\sigma} &= -\frac{\partial\rho_d}{\partial z^n_u} 
		= -\frac{\partial\rho_d}{\partial L^n_u}c_u^{-1} \overset{\text{\myeqref{eq:rhoM}}}{\underset{\text{\myeqref{eq:dMRdMLgen}}}{<}} 0
	\end{alignat*}
\end{proof}

\subsection{Shock polar}

\mylabel{section:shockpolar}

Here we prove only the results needed for our purposes.

\begin{proposition}
	\mylabel{prop:shockpolar}%
	Consider a fixed point on a shock with upstream density $\rho_u$ and pseudo-velocity $\vec z_u$ held fixed
	while we vary the normal. Define $\nva:=\measuredangle(\vec z_u,\vec n)$.
	$\rho_d$ is strictly decreasing in $|\nva|$, whereas
	$L_d,|\vec z_d|$ are strictly increasing. $c_d$ is strictly decreasing for $\gamma>1$, constant otherwise.
	Moreover
	\begin{alignat}{1}
		(\partial_\nva\vec v_d)\cdot\vec n = (\partial_\nva\vec z_d)\cdot\vec n &= z^t\Big(\frac{\partial z^n_d}{\partial z^n_u}-1\Big), \myeqlabel{eq:DvdxDnva} \\
		(\partial_\nva\vec v_d)\cdot\vec t = (\partial_\nva\vec z_d)\cdot\vec t &= z^n_d-z^n_u. \myeqlabel{eq:DvdyDnva} 
	\end{alignat}

	If $\vec z_u=(z^x_u,0)$ with $z^x_u>0$, then $z^x_d$ is increasing in $|\nva|$.
\end{proposition}
\begin{proof}
	\begin{alignat}{1}
		\partial_\nva\vec z_d 
		&= \partial_\nva(z^n_d\vec n+z^t\vec t) 
		= \frac{\partial z^n_d}{\partial z^n_u}\partial_\nva z^n_u\vec n+z^n_d\vec t-z^t\vec n+\partial_\nva z^t\vec t \notag\\
		&= z^t\Big(\frac{\partial z^n_d}{\partial z^n_u}-1\Big)\vec n+(z^n_d-z^n_u)\vec t \myeqlabel{eq:blark}
	\end{alignat}
	This is \myeqref{eq:DvdxDnva}, \myeqref{eq:DvdyDnva}, using that $\vec\xi$ is fixed.
	For $\nva>0$, 
	$z^t=-|\vec z_u|\sin\nva$ is strictly decreasing (and negative).
	\begin{alignat*}{1}
		\frac12\partial_\nva(|\vec z_d|^2)
		&= \frac12\partial_\nva((z^n_d)^2+(z^t)^2)
		= z^n_d\frac{\partial z^n_d}{\partial z^n_u}\partial_\nva z^n_u + z^t\partial_\nva z^t \\
		&= z^t\Big(z^n_d\frac{\partial z^n_d}{\partial z^n_u} - \subeq{z^n_u}{\geq z^n_d}\Big) 
		\geq \subeq{z^t}{<0}\subeq{z^n_d}{>0}\Big(\subeq{\frac{\partial z^n_d}{\partial z^n_u}}{\leq(\gamma-1)/(\gamma+1)} - 1\Big)
		> 0
	\end{alignat*}
	by \myeqref{eq:DvndDvnu}, so $|\vec z_d|$ is strictly increasing. 
	Then by \myeqref{eq:steady-rho}
	$$\rho_d=\pi^{-1}\big(\pi(\rho_u)+\frac12(|\vec z_u|^2-|\vec z_d|^2)\big),$$
	is strictly decreasing, as is $c_d$ (except constant for $\gamma=1$), so $L_d=|\vec z_d|/c_d$ is increasing.

	For $\nva>0$, \myeqref{eq:blark} yields
	\begin{alignat*}{1}
		\partial_\nva(z^x_d) 
		&= \subeq{z^t}{<0}\subeq{\Big(\frac{\partial z^n_d}{\partial z^n_u}-1\Big)}{<0}\subeq{\cos\nva}{>0}-\subeq{(z^n_d-z^n_u)}{<0}\subeq{\sin\beta}{>0} > 0
	\end{alignat*}
	by \myeqref{eq:DvndDvnu}.

	$\nva<0$ is analogous by symmetry.
\end{proof}

\subsection{Shock-parabolic corners with fixed vertical downstream velocity}

The corners of our elliptic region (see Figure \myref{fig:frameR}) are points on shocks where $L_d=1$ (or $L_d=\sqrt{1-\epsilon}$ if regularized);
on the other hand $v^y_d=0$ (if the $x$ direction is tangential to the wall).
We study such shocks in detail.

\begin{proposition}
	\mylabel{prop:steady-shock-circle}%
	Consider a straight shock (see Figure \myref{fig:shockL}) passing through a point $\vec\xi$.
	The shock is pseudo-normal in the point $\vec\xi_M=\vec\xi+\vec t\cdot(\vec v_d-\vec\xi)\vec t$ and pseudo-oblique in
	every other point. $\vec\xi_M$ is the closest point on the shock both to $\vec v_u$ and to $\vec v_d$.
	The circle with center $\vec v_u$ and radius $c_u$ does not intersect the shock. 
	The downstream flow has $L<\sqrt{1-\epsilon}$ inside the circle with radius
	$c_d\sqrt{1-\epsilon}$ and center $\vec v_d$, $L=\sqrt{1-\epsilon}$ on the circle and $L>\sqrt{1-\epsilon}$ outside.
	If $L^n_d<\sqrt{1-\epsilon}$, then the circle intersects the shock in the two points $\vec\xi_M\pm c_d\sqrt{1-\epsilon-(L^n_d)^2}\cdot\vec t$.
\end{proposition}
\begin{proof}
	Straightforward to check. $\vec\xi_M\cdot\vec t=\vec v_d\cdot\vec t=\vec v_u\cdot t$, so 
	$|\vec v_u-\vec\xi_M|=|(\vec v_u-\vec\xi_M)\cdot\vec n|=|z^n_u|>c_u$.
\end{proof}

\if\dofigures%
\begin{figure}
\input{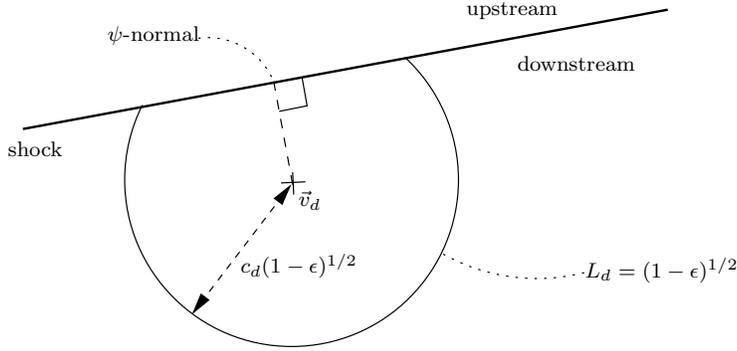}
\caption{$L$ values downstream of a shock.}
\mylabel{fig:shockL}
\end{figure}
\fi

\if\dofigures%
\begin{figure}
\center{\input{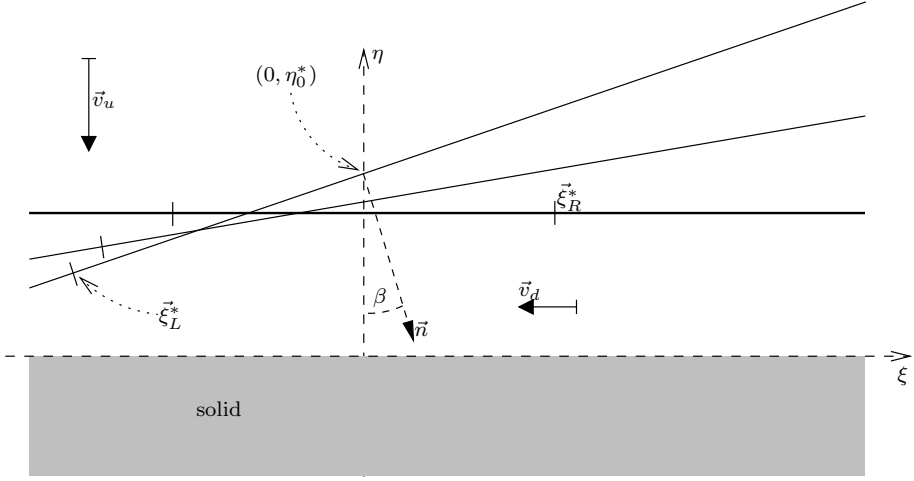}}
\caption{Changing shock normals while keeping $\vec v_d$ horizontal. $\vec\xi_L^*$ is the left $L_d^2=1-\epsilon$ point for each
shock.}
\mylabel{fig:horvzero}
\end{figure}
\fi%

\begin{proposition}
	\mylabel{prop:vdzero}%
	Consider a straight shock with $v^x_u=0$, $v^y_u<0$ and downstream normal $\vec n=(\sin\nva,-\cos\nva)$
	through $\vec\xi=(0,\eta)$ (see Figure \myref{fig:horvzero}).
	For every $\nva\in(-\frac{\pi}{2},\frac{\pi}{2})$ there is a unique $\eta=\eta^*_0\in\R$ so that $v^y_d=0$. $\eta^*_0$ and the corresponding
	downstream data are analytic functions of $\nva$. $\eta^*_0$ is strictly increasing in $|\nva|$.

	For the shock passing through $(0,\eta^*_0)$, 
	let $\vxit_L$ and $\vxit_R$ be the two points with $L_d=\sqrt{1-\epsilon}$, as given by Proposition \myref{prop:steady-shock-circle}.
	These points are analytic functions of $\nva$.
	$L^n_u$, $\rho_d$ and $z^n_u$ are increasing functions\footnote{All of these are independent of the location along the (straight) shock.} 
	of $\nva$; 
	$v^x_d$ and $L^n_d$ are decreasing functions of $\nva$.
	For $\nva\in[0,\frac{\pi}{2})$, $\etat_L$ is a strictly decreasing function of $\nva$ with range $(\underline\eta^*_L,\overline\eta^*_0]$,
	where $\overline\eta^*_0$ is the $\eta^*_0$ for $\nva=0$, and $\underline\eta^*_L$ is some \emph{negative} constant.
\end{proposition}
\begin{proof}
	First regard everything as a function of $\nva$ and $\eta$, with $\vec v_u$, $\rho_u$ held fixed and the shock held
	as passing through $(0,\eta)$. 
	In $(0,\eta)$:
	\begin{alignat}{1}
		z^n_u 
		&= z^x_un^x+z^y_un^y 
		= (\subeq{v^x_u}{=0}-\subeq{\xi}{=0})\sin\nva-(v^y_u-\eta)\cos\nva
		= (\eta-v^y_u)\cos\nva, \mylabel{eq:bluaaa}\\
		\partial_\eta z^n_u 
		&= \cos\nva, \notag 
	\end{alignat}
	\begin{alignat}{1}
		v^y_d 
		&= v^y_u+(v^y_d-v^y_u) 
		= v^y_u+(v^n_d-v^n_u)n^y 
		= v^y_u+(z^n_d-z^n_u)n^y
		= v^y_u+(z^n_u-z^n_d)\cos\nva, \notag \\
		\partial_\eta v^y_d 
		&= (1-\frac{\partial z^n_d}{\partial z^n_u})\partial_\eta z^n_u\cos\nva 
		= (1-\frac{\partial z^n_d}{\partial z^n_u})\cos^2\nva 
		\overset{\text{\myeqref{eq:DvndDvnu}}}{>}\frac{2}{\gamma+1}\cos^2\nva > 0 \myeqlabel{eq:vydeta} 
	\end{alignat}
	$v^y_d$ is an increasing function of $\eta$, so for fixed $\nva$ there can be at most one $\eta$ with $v^y_d=0$.
	For $\eta=v^y_u+c_u/\cos\nva$ we have $L^n_u=1$ in the point $(0,\eta)$, so $v^y_d=v^y_u<0$ there;
	on the other hand \myeqref{eq:vydeta} has uniformly lower-bounded right-hand side (for every fixed $\nva\in(-\frac\pi2,\frac\pi2$),
	so $v^y_d\uparrow+\infty$ if we take $\eta\uparrow+\infty$. Therefore there is exactly one solution $\eta=\eta^*_0$
	for each $\nva$. 

	Now consider $\nva\in(0,\frac\pi2)$ first (so $z^t=(v^y_u-\eta)\sin\nva<0$); the case $\nva<0$ is symmetric.
	\begin{alignat}{1}
		\partial_\nva v^y_d 
		&= (1-\frac{\partial z^n_d}{\partial z^n_u})\partial_\nva z^n_u\cos\nva - (z^n_u-z^n_d)\sin\nva 
		= (1-\frac{\partial z^n_d}{\partial z^n_u})z^t\cos\nva - (z^n_u-z^n_d)\sin\nva \notag \\
		\partial_\nva\eta^*_0 
		&= -\partial_\nva v^y_d/\partial_\eta v^y_d 
		= \subeq{\frac{1}{\cos\beta}}{>0}(\subeq{-z^t}{>0}+\subeq{\frac{z^n_u-z^n_d}{1-\partial z^n_d/\partial z^n_u}}{>0}\subeq{\tan\nva}{>0}) > 0 \notag
	\end{alignat}

	For the remainder of the proof, fix $\eta=\eta_0^*$ and consider everything a function of $\nva$. Consider $\nva\in(0,\frac\pi2)$ 
	increasing (the other case is symmetric). Then
	\begin{alignat}{1}
		z^n_u &= (\vec v_u-\vec\xi)\cdot\vec n \notag\\
		\partial_\nva z^n_u
		&= \subeq{(\vec v_u-\vec\xi)\cdot\vec t}{=z^t} - \partial_\nva\eta_0^*\subeq{n^y}{=-\cos\beta} 
		= \frac{z^n_u-z^n_d}{1-\partial z^n_d/\partial z^n_u}\tan\nva > 0 \notag
	\end{alignat}
	Therefore $L^n_u$ and $\rho_d$ are also strictly increasing, whereas $L^n_d$ is strictly decreasing.
	\begin{alignat}{1}
		v^x_d &=\subeq{v^x_u}{=0}+v^x_d-v^x_u\overset{v^t_d=v^t_u}{=}(v^n_d-v^n_u)n^x
		\overset{v^t_d=v^t_u}{=}(\subeq{v^y_d}{=0}-v^y_u)n^x/n^y=\subeq{v^y_u}{<0}\tan\nva<0 \notag\\
		\partial_\nva v^x_d &= \frac{v^y_u}{\cos^2\nva} < 0 \notag
	\end{alignat}
	Obviously $v^x_d$ is strictly decreasing.
	\begin{alignat}{1}
		\eta_M &= v^y_u - z^n_un^y = v^y_u + z^n_u\cos\nva \notag\\
		\partial_\nva\eta_M 
		&= -z^n_u\sin\nva + \partial_\nva z^n_u\cos\nva 
		= \Big( -z^n_u + \frac{z^n_u-z^n_d}{1-\partial z^n_d/\partial z^n_u} \Big)\sin\nva \notag\\
		&= \frac{z^n_u\partial z^n_d/\partial z^n_u-z^n_d}{1-\partial z^n_d/\partial z^n_u} \sin\nva  
		\overset{\myeqref{eq:DDspecial}}{\underset{\myeqref{eq:DvndDvnu}}{<}}0
		\mylabel{eq:etaMbeta}
	\end{alignat}
	So $\eta_M$ is strictly decreasing.
	\begin{alignat}{1}
		\etat_L
		&=\eta_M-c_d\sin\nva\sqrt{1-\epsilon-(L^n_d)^2}\notag
	\end{alignat}
	$c_d$ is increasing and $L^n_d$ strictly decreasing, so $\etat_L$ is strictly decreasing. Moreover
	\begin{alignat}{1}
		\etat_L &=\subeq{v^y_d}{=0}+z^n_d\cos\nva-c_d\sin\nva\sqrt{1-\epsilon-(L^n_d)^2} 
		=\big(L^n_d\cos\nva-\sin\nva\sqrt{1-\epsilon-(L^n_d)^2}\big)c_d\notag
	\end{alignat}
	$L^n_d$ is $<1-\epsilon$ for $\nva=0$ and decreasing in $\nva>0$, so it is uniformly bounded above	 away from $1-\epsilon$ 
	as $\nva\uparrow\frac{\pi}{2}$. For $\nva$ large enough the $\etat_L$ expression is negative. 
	Thus $\etat_L$ covers an interval $(\underline\eta^*_L,\overline\eta_0^*]$
	for some $\underline\eta^*_L<0$.
\end{proof}

\section{Maximum principles}

\mylabel{section:apriori}

In this section we derive many a priori estimates for smooth elliptic regions
and for smooth shocks separating elliptic and constant-state hyperbolic regions.

\newcommand{\vxz}{\vec\xi_0}
\newcommand{\vxzp}{(\vxz)}
\newcommand{\pteq}{\overset{\bullet}{=}}
\newcommand{\ptneq}{\overset{\bullet}{\neq}}
\newcommand{\ptleq}{\overset{\bullet}{\leq}}
\newcommand{\ptgeq}{\overset{\bullet}{\geq}}
\newcommand{\ptless}{\overset{\bullet}{<}}
\newcommand{\ptgtr}{\overset{\bullet}{>}}
\newcommand{\ptlessgtr}{\overset{\bullet}{\lessgtr}}

\subsection{Common techniques for extremum principles}

\begin{lemma}
	\mylabel{lemma:alemma2}%
	Let $m\geq 1$.
	If $a_0,\dotsc,a_m$ define a positive semidefinite tensor, i.e.
	\begin{alignat}{1}
		\sum_{j=0}^m\binom{m}{j}a_j\xi^j\eta^{m-j} &\geq 0 \qquad\forall \xi,\eta\in\R,\notag
	\end{alignat}
	and if 
	\begin{alignat}{1}
		Aa_k + Ba_{k+1} + Ca_{k+2} &= 0 \qquad\forall k\in\{0,\dotsc,m-2\}\myeqlabel{eq:alemma2}
	\end{alignat}
	with constants $A,B,C\in\R$ so that $4AC>B^2$, then
	$a_0=\dotsb=a_m=0$.
\end{lemma}
\begin{proof}
	$4AC>B^2$ means that $A+Bz+Cz^2=0$ has two roots $z,\overline z\in\C-\R$.
	Then
	\begin{alignat*}{1}
		a_k &= \Re(az^k) \qquad(k=0,\dotsc,m)\notag
	\end{alignat*}
	where $a\in\C$ is some linear combination of $a_0,a_1$,
	so
	\begin{alignat}{1}
		0 &\leq \sum_{k=0}^m\binom{m}{k}\Re(az^k)\xi^k\eta^{m-k} 
		= \Re\Big(a\sum_{k=0}^m\binom{m}{k}z^k\xi^k\eta^{m-k}\Big) 
		= \Re(a(z\xi+\eta)^m) \notag
	\end{alignat}
	for all $\vec\xi\in\R^2$.
	Since $\Im z\neq 0$, $(\xi,\eta)\mapsto(z\xi+\eta)^m$ is onto $\C$. The inequality cannot be true
	unless $a=0$, so $a_0=\dotsb=a_m=0$.
\end{proof}

\begin{lemma}
	\mylabel{lemma:alemma}%
	Let $m\geq 3$ and $\vec b=(b_1,b_2)\in\R^2-\{0\}$, $\vec n\in\R^2-\{0\}$. If $a_0,\dotsc,a_m\in\R$ satisfy
	\begin{alignat}{1}
		\sum_{k=0}^{m-1}\binom{m-1}{k}(b_1a_k+b_2a_{k+1})\xi^k\eta^{m-1-k} &\geq 0 \qquad\forall\vec\xi\in\R^2,~\vec\xi\cdot\vec n>0\myeqlabel{eq:alemma}
	\end{alignat}
	as well as \myeqref{eq:alemma2} 
	with constants $A,B,C\in\R$ so that $4AC>B^2$, then
	$$a_k=0\qquad\forall k\in\{0,\dotsc,m\}.$$
\end{lemma}
\begin{proof}
	$4AC>B^2$ means that $A+Bz+Cz^2=0$ has two roots $z,\overline z\in\C-\R$.
	The general solution of \myeqref{eq:alemma2} is 
	\begin{alignat}{1}
		a_k &= \Re(az^k) \qquad(k=0,\dotsc,m).\notag
	\end{alignat}
	where $a\in\C$ is some linear combination of $a_0$ and $a_1$. 
	Substitute this into \myeqref{eq:alemma}:
	\begin{alignat}{1}
		0 &\leq \sum_{k=0}^{m-1}\binom{m-1}{k}\big(b_1\Re(az^k)+b_2\Re(az^{k+1})\big)\xi^k\eta^{m-1-k} \notag\\
		&= \Re\Big(a(b_1+b_2z)\sum_{k=0}^{m-1}\binom{m-1}{k}z^k\xi^k\eta^{m-1-k}\Big) \notag\\
		&= \Re(a(b_1+b_2z)(z\xi+\eta)^{m-1}) \qquad\forall \vec\xi\in\R^2,~\vec\xi\cdot\vec n\geq 0\myeqlabel{eq:aineq}
	\end{alignat}
	We may use $\vec\xi\cdot\vec n\geq 0$ by continuity, which defines a closed halfplane of $\R^2$.
	$z\in\C-\R$, so the range of $(\xi,\eta)\mapsto\xi+z\eta$ is a closed halfplane of $\C$. 
	The range of $(\xi,\eta)\mapsto(\xi+z\eta)^{m-1}$ is all of $\C$ because $m-1\geq 2$. 
	Moreover $b_1+b_2z\neq 0$ because either $b_2=0$, then $b_1+b_2z=b_1\neq 0$,
	or $b_2\neq 0$, then $\Im(b_1+b_2z)=b_2\Im(z)\neq 0$.
	Thus the range of $(\xi,\eta)\mapsto a(b_1+b_2z)(z\xi+\eta)^{m-1}$ is all of $\C$, contradicting \myeqref{eq:aineq}, 
	unless $a=0$ which means $a_k=0$ for all $k=0,\dotsc,m$.
\end{proof}

\begin{lemma}
	\mylabel{lemma:interior-higher}%
	Consider an open set $U$ and a point $\vxz\in\overline U$. Assume that there is
	a $\vec n\neq 0$ (quasi an inner normal) so that for every $\vec\xi$ with $\vec n\cdot\vec\xi>0$
	there is a $\delta>0$ with 
	$$\{\vxz+t\vec\xi:t\in(0,\delta)\}\subset U.$$
	Let $\psi$ be an analytic solution of \myeqref{eq:psi} in $\overline{U}$,
	so that $L<1$, $\rho>0$ and $D^2\psi=0$ in $\vxz$. 
	Then 
	$$f=f(\psi,\nabla\psi,\subeq{\chi+\frac{1}{2}|\nabla\chi|^2}{a})$$
	(with $f$ a $C^\infty$ function of its arguments)
	cannot have an extremum in $\vxz$,
	unless $\psi$ is linear or
	\begin{alignat}{1}
		\frac{\partial f}{\partial(\nabla\psi)}+\frac{\partial f}{\partial a}\nabla\chi &= 0\qquad\text{in $\vec\xi_0$.} \myeqlabel{eq:lihcond}
	\end{alignat}
\end{lemma}
\begin{proof}
	We use dot notation ($\pteq$ etc.) to indicate relations holding only in $\vec\xi_0$.
	We show by complete induction over $k=3,4,\dotsc$ that $D^k\psi\pteq 0$.
	Induction step ($3,\dotsc,k-1\rightarrow k\geq 3$):
	for $j=0,\dotsc,k-2$ take $\partial_1^j\partial_2^{k-2-j}$ of the equation. This yields
	\begin{alignat}{1}
		(c^2I-\nabla\chi^2):\nabla^2\partial_1^j\partial_2^{k-2-j}\psi \pteq 0 \myeqlabel{eq:genDpsi}
	\end{alignat}
	because all other terms contain at least one component of $D^2\psi,\dotsc,D^{k-1}\psi$ as factor,
	hence vanish.

	We may exploit that $D^j\chi=D^j\psi$ for $j\geq 3$.
	$$\nabla^2(\chi+\frac{1}{2}|\nabla\chi|^2)=\nabla^2\chi\nabla^2\psi+\sum_{i=1}^2\partial_i\chi\nabla^2\partial_i\chi
	\pteq\sum_{i=1}^2\partial_i\chi\nabla^2\partial_i\psi,$$
	and for multiindices $\alpha$ with $3\leq|\alpha|<k$,
	$$\partial^\alpha(\chi+\frac{1}{2}|\nabla\chi|^2)\pteq\sum_{i=1}^2\partial_i\chi\partial_i\partial^\alpha\psi,$$
	so for all $2\leq|\alpha|<k$
	$$\partial^\alpha(\chi+\frac{1}{2}|\nabla\chi|^2)\pteq\nabla\chi\cdot\nabla\partial^\alpha\psi,$$
	Thus for $j=0,\dotsc,k-1$:
	\begin{alignat*}{1}
		\partial_1^j\partial_2^{k-1-j}(f)
		&=
		\frac{\partial f}{\partial(\nabla\psi)}\cdot\nabla\partial_1^j\partial_2^{k-1-j}\psi 
		+\frac{\partial f}{\partial a}\nabla\chi\cdot\nabla\partial_1^j\partial_2^{k-1-j}\psi \\
		&+ \text{terms with $D^2\psi,\dotsc,D^{k-1}\psi$ components as factor} \\
		&= 
		\Big(\frac{\partial f}{\partial(\nabla\psi)}+\frac{\partial f}{\partial a}\nabla\chi\Big)
		\cdot\nabla\partial_1^j\partial_2^{k-1-j}\psi
	\end{alignat*}
	because $D^2\psi,\dotsc,D^{k-1}\psi\pteq0$.

	In a similar way we obtain $D^\ell(f)\pteq 0$ for $\ell=2,\dotsc,k-2$, and $D(f)=0$ already by assumption. Therefore
	the $k-1$st order minimum conditions for $f$ apply: for all $\vec\xi=(\xi,\eta)\in\R^2$ with
	$\vec\xi\cdot\vec n>0$,
	\begin{alignat}{1}
		0 &\leq \sum_{j=0}^{k-1}\binom{k-1}{j}\partial_1^j\partial_2^{k-1-j}(f)\xi^j\eta^{k-1-j} \notag \\
		&= \sum_{j=0}^{k-1}\binom{k-1}{j}
		\subeq{\left(\frac{\partial f}{\partial(\nabla\psi)}+\frac{\partial f}{\partial a}\nabla\chi\right)}{=:\vec b}
		\cdot
		\nabla\partial_1^j\partial_2^{k-1-j}\psi\cdot\xi^j\eta^{k-1-j}. \myeqlabel{eq:genDf}
	\end{alignat}
	Applying Lemma \myref{lemma:alemma} to \myeqref{eq:genDpsi} and \myeqref{eq:genDf},
	with $a_j=\partial^j\partial^{k-1-j}\psi$ and using $L<1$, yields $D^k\psi\pteq 0$.
	Note that $\vec b\neq 0$ iff \myeqref{eq:lihcond} is not satisfied.
	The induction step is complete.

	We have shown that $D^k\psi\pteq 0$ for all $k\geq 2$. Since $\psi$ is analytic, it must be linear
	which represents constant density and velocity.
\end{proof}

\begin{remark}
	Lemma \myref{lemma:interior-higher} applies trivially to the interior case $\vxz\in U$: any $\vec n\neq 0$ will do.
\end{remark}

\subsection{Density in the interior}

\begin{proposition}
	\mylabel{prop:c-principle}%
	Let $\chi$ be an analytic solution of \myeqref{eq:chi} in an open connected domain $\Omega$. 
	Assume that $L<1$ in $\Omega$ and that $\rho$ is positive and not constant. Then $\rho$ does not have maxima
	in points where $\nabla\chi\neq 0$, and it does not have minima anywhere.
\end{proposition}
\begin{remark}
	Proposition \myref{prop:c-principle} trivially implies corresponding results for variables like $p$ and $c$
	that are strictly monotone functions $\rho$ (except for $c$ in the isothermal case where it is constant).
\end{remark}
\begin{proof}[Proof of Proposition \myref{prop:c-principle}]
	The first-order condition for a critical point is 
	$$0\pteq\nabla(\rho)=\nabla(\pi^{-1}(-\chi-\frac{1}{2}|\nabla\chi|^2))=-\frac{\rho}{c^2}\nabla^2\psi\nabla\chi.$$

	If $\nabla\chi\ptneq 0$, then combined with the PDE \myeqref{eq:psi} we obtain $D^2\psi\pteq 0$.
	Now we can apply Lemma \myref{lemma:interior-higher} to show that $\psi$ is actually a constant-state solution.
	In applying the lemma we choose $f=f(\psi,\nabla\psi,a)=f(a)$ only, using $f_a\neq 0$ and $\nabla\chi\neq 0$
	so that \myeqref{eq:lihcond} is false.

	If $\nabla\chi\pteq0$, then $\nabla(\rho)\pteq 0$ is trivially satisfied. We need to study the second-order
	condition for a minimum, which implies in particular
	\begin{alignat*}{1}
		0 &\ptleq \Delta(\rho)
		=(\pi^{-1})'(-\Delta\chi-|\nabla^2\chi|^2-\subeq{\nabla\chi}{\pteq 0}\cdot\nabla\Delta\chi)
		+(\pi^{-1})''|\subeq{\nabla\chi}{\pteq 0}+\nabla^2\chi\subeq{\nabla\chi}{\pteq 0}|^2 \\
		&\pteq \rho c^{-2}(\Delta\psi-|\nabla^2\psi|^2)
	\end{alignat*}
	The equation \myeqref{eq:psi} reduces to $\Delta\psi\pteq 0$, so
	\begin{alignat}{1}
		0&\ptleq -\rho c^{-2}|\nabla^2\psi|^2. \myeqlabel{eq:rhomin2}
	\end{alignat}
	Since $\rho,c>0$ this implies $\nabla^2\psi\pteq 0$. Then $\nabla^2(\rho)\pteq 0$ as well.

	We show for $k=3,4,5,\dotsc$ by induction that $D^k(\rho)\pteq 0$ and $D^k\psi\pteq 0$ as well. Induction step 
	($2,\dotsc,k-1\rightarrow k\geq 3$): a minimum of $\rho=\pi^{-1}(-\chi-\frac{1}{2}|\nabla\chi|^2)$ 
	is the same as a maximum of $\chi+\frac{1}{2}|\nabla\chi|^2$. For $j=0,\dotsc,k$:
	\begin{alignat*}{1}
		\partial_1^j\partial_2^{k-j}(\chi+\frac{1}{2}|\nabla\chi|^2)
		&= \partial_1^j\partial_2^{k-j}\chi
		+ \subeq{\nabla\chi}{\pteq0}\cdot\nabla\partial_1^j\partial_2^{k-j}\chi\\
		&+ j\partial_1\nabla\chi\cdot\partial_1^{j-1}\partial_2^{k-j}\nabla\chi 
		+(k-j)\partial_2\nabla\chi\cdot\partial_1^j\partial_2^{k-j-1}\nabla\chi \\
		&+\text{terms with components of $D^3\psi,\dotsc,D^{k-1}\psi$ as factor} \\
		&\pteq \partial_1^j\partial_2^{k-j}\chi
		- j\partial_1^{j-1}\partial_2^{k-j}\partial_1\chi
		-(k-j)\partial_1^j\partial_2^{k-1-j}\partial_2\chi \\
		&= (1-k)\partial_1^j\partial_2^{k-j}\psi
	\end{alignat*}
	Here we used that $\partial_1\nabla\chi\pteq(-1,0)$, $\partial_2\nabla\chi\pteq(0,-1)$ because 
	$\nabla^2\psi\pteq 0$. The induction assumption, $D^j\psi\pteq 0$ for $j=2,\dotsc,k-1$, eliminates
	the other terms.

	Since $D^j(\chi+\frac12|\nabla\chi|^2)\pteq 0$ for $j=1,\dotsc,k-1$, a maximum requires that 
	$D^k(\chi+\frac12|\nabla\chi|^2)\ptgeq 0$ (i.e.\ is a negative semidefinite
	tensor), so $0\ptgeq D^k(\chi+\frac{1}{2}|\nabla\chi|^2)\pteq(1-k)D^k\psi$, so $D^k\psi\ptgeq 0$.
	Taking $k-2$ derivatives of the equation yields (for $\nabla\chi\pteq 0$)
	$$\partial_1^j\partial_2^{k-2-j}\Delta\psi=0\qquad(j=0,\dotsc,k-2).$$
	Lemma \myref{lemma:alemma2} implies that $D^k\psi\pteq 0$. 
	The induction step is complete.

	Again we have shown that $D^k\psi\pteq 0$ for all $k\geq 2$. Therefore $\psi$, which is analytic, must be linear.
\end{proof}
\begin{remark}
	Note that the proof fails for \emph{maxima} in $\nabla\chi$: in that case $\leq$ in \myeqref{eq:rhomin2}
	turns into $\geq$ which does not yield sufficient information. Indeed there are counterexamples.
\end{remark}

\subsection{Velocity components in the interior}

\begin{proposition}
	\mylabel{prop:interior-velocity}%
	Let $\chi$ be an analytic solution of \myeqref{eq:chi} in an open connected 
	domain $\Omega$, with $\rho>0$ and $L<1$ in $\Omega$.
	For any $\vec w\in\R^2-\{0\}$, the velocity component $\vec w\cdot\nabla\psi$ does not have a maximum or a minimum in $\Omega$,
	unless $\chi$ is a constant-state solution in $\Omega$.
\end{proposition}
\begin{proof}
	Assume that $\psi_1$ has a minimum in some point. Then $\nabla\psi_1\pteq 0$;
	using the equation \myeqref{eq:psi} yields $D^2\psi\pteq 0$ because $L<1$ implies $c^2-\chi_2^2>0$.
	Now we apply Lemma \myref{lemma:interior-higher} to obtain that $\psi$ must be a constant-state solution.

	Any other $\vec w$ can be treated by rotating around the origin so that $\vec w\cdot\nabla\psi$
	becomes $\psi_1$ (see Remark \myref{rem:symmetries}).
\end{proof}

\subsection{Velocity components on the wall}

\begin{proposition}
	\mylabel{prop:v-wall}%
	Consider a point $\vec\xi_0$ on a straight line $I$, let $r>0$, $U:=B_r(\vec\xi_0)$, $\Gamma:=I\cap U$ 
	and $U^+$ one of the two connected components of $U-I$. 
	Consider a solution $\po$ of \myeqref{eq:chi} that is analytic in $\overline{U^+}$ and satisfies the slip condition
	$\xo_n=0$ on $\Gamma$.
	Assume that $L<1$ in $\vec\xi_0$.
	
	For any $\vec w$, 
	$$\nabla\psi(\vec\xi_0)\cdot\vec w=\inf_{\overline U^+}\nabla\psi\cdot\vec w$$
	is not possible 
	unless $\psi$ is a constant-state solution, or unless $\vec w$ is normal to $\Gamma$.
\end{proposition}
\begin{proof}
	Assume there is an extremum point on $\Gamma$. 
	We may assume (by rotation and translation) that $\Gamma$ is a piece of the horizontal axis, 
	that the extremum point is the origin, and that $U_+$ is contained in the upper halfplane;
	then $\vec w=(w^1,w^2)$ with $w^1\neq 0$ (not normal).
	A tangential derivative of the boundary condition $\psi_2=0$ implies $\psi_{12}=0$ on $\Gamma$. A minimum requires
	$$0\pteq (\vec w\nabla\psi)_1=w^1\psi_{11}+w^2\psi_{12}=w^1\psi_{11}\quad\Rightarrow\quad\psi_{11}\pteq 0.$$
	The equation \myeqref{eq:psi} yields that $D^2\psi\pteq 0$.
	Now the result is delivered by Lemma \myref{lemma:interior-higher}.
\end{proof}

\subsection{Velocity components at shocks}

\begin{proposition}
	\mylabel{prop:vshock}%
	Consider disjoint open connected domains $\Omega$ and $\Omega^h$ 
	and a simple analytic curve $S\subset\overline\Omega\cap\overline\Omega^h$ (excluding the endpoints).
	Consider a constant-state (linear) potential $\psi^h$ of \myeqref{eq:chijump} in $\Omega^h$.
	Let $\chi$ be an analytic solution of \myeqref{eq:chi} in $\Omega\cup S$.
	Let the shock relations \myeqref{eq:chijump} and \myeqref{eq:momjump} be satisfied on $S$
	and assume the shock is admissible. 
	Assume that $\chi$ satisfies $\rho>0$ and $L<1$ in $\Omega\cup S$.

	Let $\vec w\neq 0$. 
	Assume that $\vec w\cdot\nabla\psi$ has a local maximum (with respect to $\Omega\cup S$) in $\vec\xi\in S$.
	Then either $S$ is straight and $\psi$ is constant-state in $\Omega$,
	or
	\begin{alignat}{1}
		\vec w\cdot \Big((1-c^{-2}\chi_n^2)\vec t+\chi_t(\frac{1}{\chi_n^h}+c^{-2}\chi_n)\vec n\Big) &=0.
		\myeqlabel{eq:propvx1}
	\end{alignat}
	and 
	\begin{alignat}{1}
		\sgn\kappa &= \sgn w^n \neq 0
		\mylabel{eq:propvx-s11}
	\end{alignat}
	where $\kappa$ is the curvature of $S$ in $\vec\xi$ ($\kappa>0$ for $\Omega$ locally convex).
\end{proposition}
\begin{proof}
	We use a dot to indicate relations that hold only in the hypothetical extremum point.
	By $L<1$, $\Omega$ must be downstream and $\Omega^h$ is upstream.
	Without loss of generality, rotate around $\vec\xi_0$ until 
	$\vec n\pteq(0,-1)$. 
	In this setting $\partial_1\pteq\partial_t$ and $-\partial_2\pteq\partial_n$,
	and $\chi_1\pteq\chi_t=\chi^h_t\pteq\chi_1^h$ by \myeqref{eq:chitan}.
	Use horizontal translation (Remark \myref{rem:symmetries}) so that $\psi^h_1\pteq0$ and therefore $\psi_1\pteq0$;
	this adds a constant vector to velocities, while leaving $\rho$ and $L$ unchanged, so no generality is lost.
	Let the shock be parametrized by $\xi\mapsto(\xi,s(\xi))$ locally; then $s_1\pteq 0$. 

	Take $\partial_{tt}$ of \myeqref{eq:psijump}:
	\begin{alignat}{1}
		\psi_{11} + (\psi_2-\psi_2^h)s_{11} &\pteq 0. \myeqlabel{eq:blurb}
	\end{alignat}
	Take $\partial_t$ of \myeqref{eq:momjump}:
	\begin{alignat}{1}
		0 &= \partial_t\Big((\pi^{-1}(-\chi-\frac{1}{2}|\nabla\chi|^2)\nabla\chi-\rho^h\nabla\chi^h)\cdot\vec n\Big) \notag\\
		&= 
		\vec n\cdot\partial_t\Big(\pi^{-1}(-\chi-\frac{1}{2}|\nabla\chi|^2)\nabla\chi-\rho^h\nabla\chi^h\Big) 
		+\Big(\pi^{-1}(-\chi-\frac{1}{2}|\nabla\chi|^2)\nabla\chi-\rho^h\nabla\chi^h\Big)\cdot(\vec n)_t \notag\\
		&\pteq
		\begin{bmatrix}0\\-1\end{bmatrix}\cdot\partial_1\Big(\pi^{-1}(-\chi-\frac{1}{2}|\nabla\chi|^2)\nabla\chi-\rho^h\nabla\chi^h\Big) 
		+\Big(\pi^{-1}(-\chi-\frac{1}{2}|\nabla\chi|^2)\nabla\chi-\rho^h\nabla\chi^h\Big)\cdot\begin{bmatrix}s_{11}\\0\end{bmatrix} \notag\\
		&\pteq
		\rho c^{-2}(\chi_1+\chi_1\chi_{11}+\chi_2\chi_{12})\chi_2-\rho\chi_{12}+\rho^h\chi_{12}^h
		+(\rho\chi_1-\rho^h\chi_1^h)s_{11} \notag\\
		&\pteq
		\rho\Big(-(1-c^{-2}\chi_2^2)\psi_{12}+c^{-2}\chi_1\chi_2\psi_{11}+\chi_1(1-\frac{\chi_2}{\chi_2^h})s_{11}\Big)
		\myeqlabel{eq:rhos-psi12}
	\end{alignat}
	Combining these results with the equation and $(\vec w\cdot\nabla\psi)_t=0$ we get the system
	\begin{alignat}{1}
		\begin{bmatrix}
			w^1 & w^2 & 0 & 0 \\
			1 & 0 & 0 & \chi_2-\chi_2^h \\
			c^2-\chi_1^2 & -2\chi_1\chi_2 & c^2-\chi_2^2 & 0 \\
			-c^{-2}\chi_1\chi_2 & 1-c^{-2}\chi_2^2 & 0 & (\frac{\chi_2}{\chi_2^h}-1)\chi_1
		\end{bmatrix}
		\begin{bmatrix}
			\psi_{11} \\ \psi_{12} \\ \psi_{22} \\ s_{11}
		\end{bmatrix}
		&= 0. \myeqlabel{eq:wsys}
	\end{alignat}
	Determinant of the system matrix:
	\begin{alignat}{1}
		\Det &= -(c^2-\chi_2^2)\Big(w^2\frac{\chi_1}{\chi_2^h}(\chi_2-\chi_2^h)
		+(\chi_2-\chi_2^h)\big(w^1(1-c^{-2}\chi_2^2)+w^2c^{-2}\chi_1\chi_2\big)\Big) \notag \\
		&= -\subeq{(c^2-\chi_2^2)}{>0}\subeq{(\chi_2-\chi_2^h)}{>0}\vec w\cdot
		\begin{bmatrix}
			1-c^{-2}\chi_2^2 \\
			\chi_1(1/\chi_2^h+c^{-2}\chi_2)
		\end{bmatrix} \myeqlabel{eq:wdet}
	\end{alignat}
	The determinant is zero iff the final scalar product is zero. 
	The latter condition can be written \myeqref{eq:propvx1}, if we return to original coordinates.

	If the determinant is nonzero, then $D^2\psi\pteq 0$ and $s_{11}\pteq 0$ is the only solution.
	If the determinant is zero, but $s_{11}\pteq 0$, still $D^2\psi\pteq 0$. In either case
	we can invoke Lemma \myref{lemma:interior-higher} to get that $\psi$ is a constant-state solution;
	then \myeqref{eq:psijump} shows that the shock is straight.

	Now assume the determinant is zero and $s_{11}\ptneq 0$. 
	By row 2 of the system this implies $\psi_{11}\ptneq 0$. Then by row 1 and $\vec w\neq 0$, necessarily $w^2\ptneq 0$.

	By solving rows 1,2,3 of the system for $\nabla^2\psi$ as a function of $s_{11}$, 
	then substituting the result into $(\vec w\cdot\nabla\psi)_2$, 
	we obtain \cucon{verified in maple/vvshock.txt}
	\begin{alignat}{1}
		(\vec w\cdot\nabla\psi)_2 
		&= 
		\frac{(\chi_2-\chi_2^h)\big((c^2-\chi_2^2)w_1^2+2\chi_1\chi_2w^1w^2+(c^2-\chi_1^2)w_2^2\big)}{w^2(c^2-\chi_2^2)}s_{11} \notag\\
		&= 
		\frac{(\chi_2-\chi_2^h)(\vec w^\perp)^T(c^2I-\nabla\chi^2)\vec w^\perp}{c^2-\chi_2^2}\cdot\frac{s_{11}}{w_2}\notag
	\end{alignat}
	All factors in the coefficient of $s_{11}/w_2$ are positive; $s_{11},w_2\neq 0$ is already known, so $(\vec w\cdot\nabla\psi)_2\ptneq 0$.
	A maximum requires $(\vec w\cdot\nabla\psi)_2\ptgtr 0$, so $\sgn\kappa = -\sgn s_{11}=-\sgn w^2=\sgn w^n\neq 0$.
\end{proof}

\subsection{Pseudo-Mach number at shocks}

\begin{proposition}
	\mylabel{prop:L-minmax}%
	Consider the setting\footnote{We do not need analyticity here.} in the first paragraph of the statement of Proposition \myref{prop:vshock}.

	Let $\delta_{L\rho}>0$ be such that
	\begin{alignat}{1}
		\frac{\rho}{\rho^h} \in[\delta_{L\rho},1-\delta_{L\rho}]\qquad\text{on $S$} \myeqlabel{eq:rhoLb}
	\end{alignat}
	Let $b\in\spC^1(\Omega\cup S)$.
	There is a $\delta_{LS}>0$ (depending continuously and only on $\delta_{L\rho},\gamma,c^h,\rho^h$) with the following property:

	$L^2+b$ cannot attain a local (with respect to $\Omega\cup S$) maximum in a point on $S$ where $L^2\in[1-\delta_{LS},1)$ and 
	$|\nabla b|\leq\delta_{LS}$.
\end{proposition}
\begin{proof}
	We use the same notation and simplifications as explained at the start of the proof of Proposition \myref{prop:vshock}.

	From $\rho/\rho^h\leq 1-\delta_{L\rho}<1$ we obtain
	\begin{alignat}{1}
		\frac{\chi_2^2}{c^2},\frac{\chi_2}{\chi^h_2} &\leq 1-C_s\myeqlabel{eq:rho-sstr}
	\end{alignat}
	for some constant $C_s=C_s(\delta_{L\rho},\gamma)>0$. 

	\begin{alignat}{1}
		\partial_1(L^2+b) \\
		&= \partial_1(\frac{|\nabla\chi|^2}{c^2})+b_1 \notag\\
		&= \partial_1(\frac{|\nabla\chi|^2}{c_0^2+(1-\gamma)(\chi+\frac{1}{2}|\nabla\chi|^2)})+b_1 \notag\\
		&= c^{-2}\Big((2+(\gamma-1)L^2)(\chi_1\chi_{11}+\chi_2\chi_{12})+(\gamma-1)L^2\chi_1\Big)+b_1 \notag\\
		&= c^{-2}\Big((2+(\gamma-1)L^2)(\chi_1\psi_{11}+\chi_2\psi_{12})-2\chi_1\Big)+b_1 \notag
	\end{alignat}
	and analogously
	\begin{alignat}{1}
		\partial_2(L^2+b)
		&=c^{-2}\Big((2+(\gamma-1)L^2)(\chi_1\psi_{12}+\chi_2\psi_{22})-2\chi_2\Big)+b_2 \myeqlabel{eq:L2}
	\end{alignat}

	Combining $\partial_1(L^2)\pteq 0$ with the equation \myeqref{eq:psi}, and \myeqref{eq:blurb} and 
	\myeqref{eq:rhos-psi12} we have a linear system
	\begin{alignat}{1}
		\begin{bmatrix}
			\chi_1 & \chi_2 & 0 & 0 \\
			1 & 0 & 0 & \chi_2-\chi_2^h \\
			c^2-\chi_1^2 & -2\chi_1\chi_2 & c^2-\chi_2^2 & 0 \\
			-\frac{\chi_1\chi_2}{c^2} & 1-\frac{\chi_2^2}{c^2} & 0 & (\frac{\chi_2}{\chi_2^h}-1)\chi_1
		\end{bmatrix}
		\begin{bmatrix}
			\psi_{11} \\ \psi_{12} \\ \psi_{22} \\ s_{11}
		\end{bmatrix}
		&\pteq\begin{bmatrix}
			\frac{2\chi_1-c^2b_1}{2+(\gamma-1)L^2} \\ 0 \\ 0 \\ 0
		\end{bmatrix}\notag
	\end{alignat}

	First consider $\nabla b=0$ and $L=1$, i.e.\ $\chi_1=\sgn\chi_1\cdot\sqrt{c^2-\chi_2^2}$. 
	Then the inverse of the system matrix has entries polynomial in $c,\chi_2,\chi_2^h,\sgn\chi_1$
	divided by a common denominator
	$$c^2(c^2-\chi_2^2)^{3/2}\big((\chi_2^h)^2-\chi_2^2\big).$$
	This denominator is bounded below away from zero by $c^7\delta_D$ where $\delta_D$ depends continuously and only
	on $\delta_{L\rho}$ and $\gamma$. The rest of the inverse matrix is bounded by some constant depending only on
	$\rho^h,c^h,\gamma,\delta_{L\rho}$.

	Solving for $D^2\psi$ and substituting the result yields
	$$\partial_2(L^2+b) = \frac{2}{\chi_2^h+\chi_2} < 0,$$
	so clearly a maximum of $L^2+b$ is not possible.

	For $\nabla b\neq 0$ and $L<1$ we use that the inverse has been bounded away from $0$, so that small perturbations are possible.
	Thus, there is a $\delta_{LS}$, depending only on $\delta_{L\rho}$, $\gamma$, $\rho^h$, $c^h$,
	so that no maximum is possible if $|\nabla b|\leq\delta_{LS}$ and $L^2\geq1-\delta_{LS}$.
\end{proof}

\begin{remark}
	We could assume $L\leq\overline L<1$ which by itself would imply $\rho/\rho^h\leq C(\overline L)<1$. 
	However, this is not sufficient as $\delta_{LS}$ would depend on $\overline L$, 
	so Proposition \myref{prop:L-minmax} would be void.
	It is necessary to obtain uniform shock strength bounds separately; only then can $L$ be controlled.
\end{remark}

\subsection{Density at shocks}

\begin{proposition}
	\mylabel{prop:density-shock}%
	Consider the setting in the first paragraph of Proposition \myref{prop:vshock}.

	If $\rho$ has a local extremum with respect to $\Omega\cup S$ in $\vec\xi_0$,
	then one of the following alternatives must hold:
	\begin{enumerate}
	\item $\chi$ is a constant-state solution in $\Omega\cup S$, and $S$ is straight.
	\item The shock is pseudo-normal in $\vec\xi_0$. In a local \emph{minimum}, $S$ has curvature $\kappa>0$ 
		(as before, $\kappa>0$ for $\Omega$ locally strictly convex in that point).
		\cucon{and $\geq c^{-1}$; for max $>0$ or $\leq c^{-1}$.
		$=c^{-1}$ is unclear.}
	\end{enumerate}

	In the latter case, assume stronger that $\rho$ has a \emph{global minimum} with respect to $\overline\Omega$ in $\vec\xi_0$.
	Assume for technical convenience that the shock tangents differ by no more than an angle $<\pi/2$ from the one in $\vec\xi_0$. 
	Let $S^*$ be the tangent to $S$ in $(\xi_0,s(\xi_0))$. Then $S^*$ does not meet $S$ anywhere else.
\end{proposition}
\begin{proof}
	We use the same notation and simplifications as explained at the start of the proof of Proposition \myref{prop:vshock}.

	A $\rho$ extremum requires
	\begin{alignat}{1}
		0 &\pteq \partial_t(\rho) \pteq \partial_1(\pi^{-1}(-\chi-\frac{1}{2}|\nabla\chi|^2))
		= -\frac{\rho}{c^2}(\chi_1+\chi_1\chi_{11}+\chi_2\chi_{12}) \notag\\
		\Rightarrow\qquad \chi_1\psi_{11}+\chi_2\psi_{12} &\pteq 0 \notag
	\end{alignat}
	We combine this with the now-familiar equations \myeqref{eq:psi}, \myeqref{eq:blurb} and \myeqref{eq:rhos-psi12}. The resulting linear system is
	\begin{alignat}{1}
		\begin{bmatrix}
			\chi_1 & \chi_2 & 0 & 0 \\
			1 & 0 & 0 & \chi_2-\chi_2^h \\
			c^2-\chi_1^2 & -2\chi_1\chi_2 & c^2-\chi_2^2 & 0 \\
			-c^{-2}\chi_1\chi_2 & 1-c^{-2}\chi_2^2 & 0 & (\frac{\chi_2}{\chi_2^h}-1)\chi_1
		\end{bmatrix}
		\begin{bmatrix}
			\psi_{11} \\ \psi_{12} \\ \psi_{22} \\ s_{11}
		\end{bmatrix}
		&= 0.\notag
	\end{alignat}
	The determinant is 
	\begin{alignat}{1}
		&= -(c^2-\chi_2^2)(\chi_2-\chi_2^h)\nabla\chi\cdot
		\begin{bmatrix}
			1-c^{-2}\chi_2^2 \\
			\frac{\chi_1}{\chi_2^h}+c^{-2}\chi_1\chi_2
		\end{bmatrix} \notag\\
		&= -\subeq{(c^2-\chi_2^2)}{>0}\subeq{(\chi_2-\chi_2^h)}{>0}
			\subeq{(1+\frac{\chi_2}{\chi_2^h})}{>0}\cdot\chi_1 \notag
	\end{alignat}
	It is nonzero if and only if $\chi_1\ptneq0$. In that case, $\nabla^2\psi\pteq 0$ and $s_{11}\pteq 0$.
	Now Lemma \myref{lemma:interior-higher} yields that the shock is straight and the solution constant-state.
	(The Lemma applies because $\rho$ is a strictly decreasing 
	function of $\chi+\frac{1}{2}|\nabla\chi|^2$ alone, and \myeqref{eq:lihcond} is satisfied
	because $\chi_n\neq 0$, hence $\nabla\chi\neq 0$, at any shock).

	If $\chi_1\pteq 0$ but $s_{11}\pteq 0$, then the equations imply that $\nabla^2\psi\pteq 0$, so the Lemma still applies.
	This concludes the proof of the first part.

	For part two we may assume that all of $S$ is parametrized by $s$, because the shock tangents cannot become vertical, by assumption.
	Consider the remaining case $\chi_1\pteq 0$ and $s_{11}\ptneq 0$. $\chi_1\pteq0$ and $s_1\pteq0$ imply $\xi\pteq 0$.
	Here it is sufficient to argue without the interior. So we may exploit that at the shock downstream, $\rho_d$ is an increasing function of 
	$z^n_u$ (by \myeqref{eq:rhoM} and \myeqref{eq:dMRdMLgen}, for $\rho_u$ held fixed).
	Geometrically, $z^n_u$ in a point $\vec\xi$ on the shock is the distance of $\vec v_u$ to the shock tangent through $\vec\xi$.
	A pseudo-normal point is actually the closest point on the tangent to $\vec v_u$. 
	Obviously the tangents through nearby points are \emph{closer} to $\vec v_u$ if $s_{11}\ptgtr 0$ 
	(see Figure \myref{fig:rhos11}), which contradicts
	the assumption that $\rho$ is a local (in fact global) minimum. 
	Therefore $s_{11}\ptleq 0$. $s_{11}\pteq0$ has already been excluded, so $s_{11}\ptless 0$.
	\if\dofigures%
	\begin{figure}
	\parbox{.4\textwidth}{%
	\input{rhosarg.pstex_t}
	\caption{Tangent distance argument}
	\mylabel{fig:rhos11}
	} \parbox{.59\textwidth}{%
	\input{rhotan.pstex_t}
	\caption{Shock below global density minimum tangent}
	\mylabel{fig:rhotangent}
	}
	\end{figure}
	\begin{figure}
	\center{\input{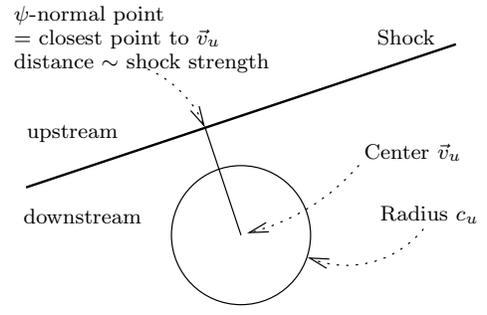}}
	\caption{Shock strength: distance of tangent from upstream velocity}
	\mylabel{fig:shockstrength}
	\end{figure}
	\fi

	Let $\xi\mapsto s^*(\xi)$ parametrize $S$. Assume that $S^*$ meets $S$ somewhere else, e.g.\ 
	that $s(\xi_1)=s^*(\xi_1)$ for some $\xi_1<\xi_0$ (see Figure \myref{fig:rhotangent}).
	By the mean value theorem there must be a $\xi_2\in(\xi_1,\xi_0)$ so that $s_1(\xi_2) = (s(\xi_1)-s(\xi_0))/(\xi_1-\xi_0) = s_1(\xi_0)$.
	Clearly we may choose $\xi_2$ maximal with this property, so that
	$s_1(\xi)\neq s_1(\xi_0)$ for $\xi\in(\xi_2,\xi_0)$. But $s_{11}\ptless 0$, so
	$s_1(\xi)>s_1(\xi_0)$ for $\xi\in(\xi_2,\xi_0)$. This implies $s(\xi_2) < s^*(\xi_0)$:  
	the shock tangent through $(\xi_2,s(\xi_2))$ is parallel to the one in $\vec\xi_0$, but \emph{lower}.
	That means the shock strength is smaller, so the downstream density is lower (by \myeqref{eq:DrhoDsigma}) 
	--- contradiction, because we assumed that $\rho$ has a global minimum
	in $\vec\xi_0$. 

	Therefore $s(\xi)>s(\xi_0)$ for all $\xi<\xi_0$ (on $S$).
	For $\xi>\xi_0$ the arguments are symmetric.
\end{proof}

\section{Construction of the flow}
\mylabel{section:ellreg}

\subsection{Problems}

Our solution, as observed in numerics (see Figure \myref{fig:fullsol}), has the structure in Figure \myref{fig:frameR}.
The upstream region, labeled ``$I$'', is a constant-state hyperbolic region. The shock has three parts: a straight shock emanating from the tip,
with a constant-state hyperbolic region labeled ``$L$'' below; a curved shock $S$ with a nontrivial elliptic region below; and a straight shock parallel
to the wall, with another constant-state hyperbolic region (``$R$'') below. The $L$ and $R$ regions are separated from the elliptic region 
by parabolic arcs $P_L$ resp.\ $P_R$ with radius $c_L$ resp.\ $c_R$, centered in $\vec v_L$ resp.\ $\vec v_R$. 

Several difficulties complicate the problem: 
the equation is nonlinear (quasilinear, divergence form), with coefficients depending on $\chi$ \emph{and}
$\nabla\chi$.
The boundary conditions (except on the wall) are fully nonlinear, i.e.\ they are nonlinear in $\nabla\chi$ as
well, which makes compactness hard to obtain.
Moreover, the boundary conditions linearize to oblique derivative conditions where the $\chi$
and $\chi_n$ coefficients have \emph{opposite} signs --- the most complicated case.
The equation and boundary conditions have singularities: for example we have to avoid
the vacuum ($\rho=0$).
The shock is a \emph{free} boundary, forming two\footnote{The wall-arc corners can be removed by reflection across the wall.} corners with the arcs.

Most significantly, the equation is \emph{mixed-type}: if $\chi$ and $\nabla\chi$ are not sufficiently controlled, points in
a supposedly elliptic region could be parabolic or hyperbolic. 
While the elliptic region is uniformly elliptic at the shock, it is \emph{degenerate} elliptic at the
parabolic arcs $P_L$ and $P_R$. 
Moreover it appears from numerics that $\nabla\chi$ is normal on the parabolic arcs, meaning they are \defm{characteristic} 
(in the Cauchy-Kovalevskaya sense),
so loss of regularity has to be expected.
The linearization of the degenerate problem is useless in this case. E.g.\ although we expect finite gradients (= velocities) in
the nonlinear problem (and observe them in numerics), it can be checked for simple examples 
that the linear equation has solutions with gradient that is infinite at parabolic arcs.

For many of these obstacles there are theoretical tools in the literature; many have been addressed 
in other contexts. However, the nonlinear characteristic degeneracy seems unprecedented.
There is little theory on degenerate elliptic equations; most of it can be found in \cite{oleinik-radkevic}. 
A large part of the theory was motivated by the linear \defm{Tricomi equation} which arises from steady 
potential flow via the \defm{hodograph transform} (see \cite[Section 4.4.3.a]{evans}).
The hodograph transform 
applies to quasilinear equations whose non-divergence form coefficients depend 
only on the gradient of the solution; the original equation is converted into a linear equation.
Unfortunately the coefficients of selfsimilar potential flow also depend on the solution itself.
It is not clear whether a modified hodograph transform can be developed for selfsimilar potential flow.

\emph{Nonlinear} degenerate elliptic equations, other than steady potential flow, have
not been explored much, apart from steady potential flow which can often be reduced to a linear equation.
The combination of nonlinearity and degeneracy is particularly
difficult: the solution can be characteristic degenerate, noncharacteristic degenerate, elliptic or hyperbolic
in each point of the arcs; each case is qualitatively very different. 
Precise bounds on the solution have to be established before we even know
which case occurs in which point.

\newcommand{\fusp}{\mathcal{F}}
\newcommand{\cfusp}{\overline\fusp}
\newcommand{\IT}{\mathcal{K}}
\newcommand{\BL}{\mathcal{L}}

\subsection{Approach}
\mylabel{section:approach}

In this paper we construct a weak solution, so only $\chi$ continuity 
and conservation are shown across the parabolic arcs. Most of the effort is concerned with the elliptic region.

\paragraph{Symmetries}

Due to Remark \myref{rem:symmetries}, we can change our coordinate system in several ways.

Starting in original coordinates (see Figure \myref{fig:frameorig}), 
we translate $\vec\xi\leftarrow\vec\xi-\vec v_R$ and $\vec v\leftarrow\vec v-\vec v_R$, then rotate so that the right shock is horizontal (and in the upper half-plane);
see Figure \myref{fig:frameR}.
Now $\vec v_R=0$, $\vec v_L=(v^x_L,0)$ with $v^x_L\leq0$, and $\vec v_I=(0,v^y_I)$ with $v^y_I<0$. 
Moreover, $\psi$ is \emph{constant} in the $R$ region, and $P_R$ is centered in the origin, which will be rather convenient. 
We use this as the standard picture, as it is the most convenient for discussing the elliptic region, which consumes most of our efforts.

For another choice, which we call ``L picture'', we start in original coordinates, translate $\vec\xi\leftarrow\vec\xi-\vec v_L$ and $\vec v\leftarrow\vec v-\vec v_L$, then rotate so that the left
shock is horizontal (and in the upper half-plane), then reflect everything across the vertical coordinate axis; see Figure \myref{fig:frameL}. In this picture $\vec v_L=0$ whereas
$\vec v_R=(v^x_R,0)$ with $v^x_R\leq 0$, and $\vec v_I=(0,v^y_I)$ with $v^y_I<0$ (which need not have the same value as before). 
Now $\psi$ is constant in the $L$ region, and $P_L$ is centered in the origin.

\if\dofigures%
\begin{figure}
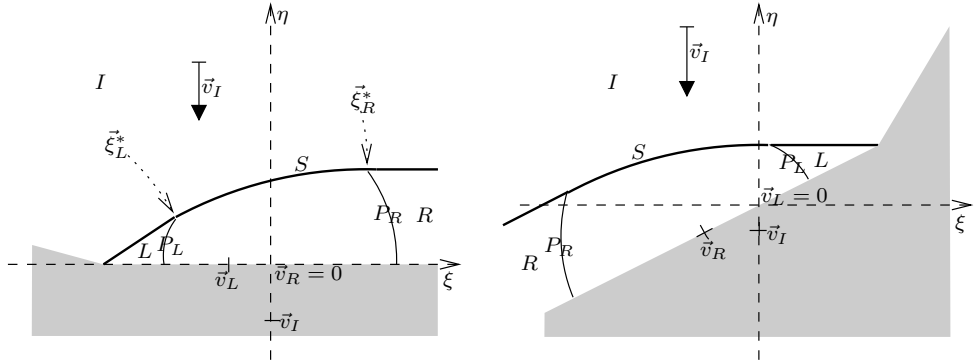

\input{frameR.pstex_t}\hspace{4mm}
\input{frameL.pstex_t}
\caption{Left: standard coordinates; right: L picture}
\mylabel{fig:frameR}
\mylabel{fig:frameL}
\end{figure}
\fi

\paragraph{Extremum principles}

As for most nonlinear elliptic problems, maximum principles are the key technique.
In Section \myref{section:apriori} we have developed many that apply to selfsimilar potential flow
in general. Their interior versions are comparable
to the classical strong maximum principle \cite[Theorem 3.5]{gilbarg-trudinger} or to maximum principles
for gradients of special quasilinear equations \cite[Section 15.1]{gilbarg-trudinger}.
In addition we have several extremum principles at the shock (Propositions \myref{prop:density-shock}, \myref{prop:vshock} and \myref{prop:L-minmax}), 
ruling out local (with respect to the domain) extrema of certain variables at the shock.

The general proof technique for extremum principles is to combine the equation 
with the first and second order conditions for an extremum to obtain a contradiction. At the shock we also 
include the boundary conditions. In many cases it is necessary to include first and higher derivatives of
equation and boundary conditions as well as higher-order extremum conditions. In some borderline cases (notably density and velocity)
it is necessary to consider \emph{all} derivatives and to assume that the solution is analytic
(see the proofs of Propositions \myref{prop:c-principle}, \myref{prop:density-shock}, \myref{prop:interior-velocity} and \myref{prop:vshock}).

\paragraph{Regularization and ellipticity}

\if\dofigures%
\begin{figure}
\input{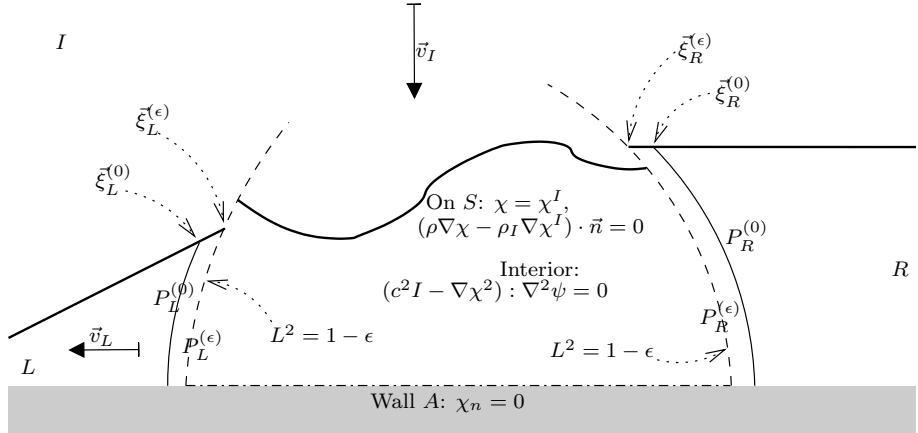}
\caption{The solution in the elliptic region and its delimiting shock $S$ are unknown. The parabolic arcs (solid) are modified
to quasi-parabolic arcs (dashed) with boundary condition $L=\sqrt{1-\epsilon}$ which is slightly elliptic.}
\mylabel{fig:regularized}
\end{figure}
\fi%

As we have mentioned, our elliptic region is nonlinear and characteristic degenerate; there are very few results for such problems.
The type of our PDE is governed by the pseudo-Mach number $L$.
We avoid the degeneracy by \emph{regularization}: we consider slightly different elliptic regions, with $L=\sqrt{1-\epsilon}$ on
the parabolic arcs rather than $L=1$ (see Figure \myref{fig:regularized}). 
We need $L$ uniformly bounded above away from $1$, so that our regularized
problems are uniformly elliptic and standard theory can be applied.

A basic problem of our constructive approach is that it is infeasible if the flow pattern has a complicated structure.
For example if the elliptic region could, upon perturbation, develop a (potentially infinite) number of parabolic and hyperbolic bubbles
in its interior, only abstract functional-analytic methods have a chance to succeed. 
Fortunately we have shown in
\cite{elling-liu-ellipticity-journal} that the pseudo-Mach number $L$ cannot have maxima in the interior of an elliptic
region or at a straight wall. On the arcs we have \emph{imposed} the value of $L$. It remains to control $L$ at the shock.
This result (Proposition \myref{prop:L-minmax}) is less general: we can rule out $L$ maxima close to $1$ at the shock \emph{if}
the shock has uniform strength. This condition needs to be verified with separate methods (see below). 

All combined we can be sure that our elliptic region stays uniformly elliptic under perturbation. We obtain, 
after many additional steps described below, one solution $\chi^{(\epsilon)}$ for each sufficiently small $\epsilon>0$. 
To obtain a solution of the original degenerate problem it is necessary to obtain estimates that are uniform in $\epsilon\downarrow 0$.

As discussed in \cite{elling-liu-ellipticity-journal}, the maximum principle for $L$ can be strengthened to a maximum principle for $L+b$
where $b$ is a small smooth positive function which is zero on the arcs. As a consequence, $L$ is \emph{uniformly in $\epsilon\downarrow 0$}
bounded above away from $1$, in subsets of the elliptic region that have positive distance from the arcs. We obtain uniform regularity
in all $C^{k,\alpha}$ norms (in fact analyticity) in each such subset. Hence we have compactness in each subset, so we can find a 
converging subsequence. By a diagonalization argument we have a subsequence that converges everywhere away
from the arcs, in any norm (see Section \myref{section:entireflow}).

Another of the many benefits of the ellipticity principle: $L\leq 1-\epsilon\leq 1$ implies a uniform bound on the gradient $|\nabla\chi|$
(see \myeqref{eq:lipode} etc), which is a basic ingredient of all higher regularity estimates for nonlinear elliptic equations 
(see the discussion in \cite[Section 11.3]{gilbarg-trudinger}). Moreover it shows that $\chi^{(\epsilon)}$ is uniformly Lipschitz, so
$\chi^{(0)}$ is Lipschitz as well, including the arcs. This is not quite sufficient for the boundary condition $L=1$ to be well-defined; but
we only seek a weak solution.

\paragraph{Iteration}

For any nonlinear elliptic problem, there are several ways of turning a priori estimates into an existence proof
(cf.\ \cite[Chapter 11, Section 17.2]{gilbarg-trudinger}). The \defm{method of continuity} is straightforward:
verify that the linearizations of the problem are isomorphisms between suitable spaces, apply the
inverse function theorem to obtain small perturbations, then use the a priori estimates to show that we can
repeat small perturbations indefinitely. 

A major drawback is that the linearizations are entirely determined by the nonlinear problem.
In comparison, \defm{fixed-point iteration methods} are more flexible: there is a large variety of
maps whose fixed points solve the nonlinear problem. In our case this flexibility is vital: the Fr\'echet
derivative of $L^2=1-\epsilon$ is an oblique derivative condition where $\delta\chi$ and $\delta\chi_n$
($\delta\chi$ being the variation of $\chi$) have opposite sign. In this case, maximum principles fail, but
they are needed to verify uniqueness (the other option, energy methods, seems unpractical for our complicated problem).

Another (non-fatal) drawback is that linearizations of a free-boundary problem involve a complicated coordinate
transform. Iteration methods can alternate solving a fixed-boundary problem with adjusting the free boundary
(see e.g.\ \cite{canic-keyfitz-lieberman} and \cite{gqchen-feldman}), which is easier.

Fixed-point methods can be subdivided into applications of the Schauder fixed point theorem (or its generalizations),
and of Leray-Schauder degree theory (see \cite{deimling}, \cite{zeidler-one} or \cite{smoller}).

The Schauder fixed point theorem requires showing
that the iteration is maps a closed ball (or homeomorphic image thereof) into itself. This corresponds to 
certain estimates for all elements of the ball, most of which are not fixed points (only approximately perhaps).
These new estimates are particular to the chosen iteration, which 
is unattractive and perhaps difficult\footnote{However, \cite{gqchen-feldman} have successfully applied the Schauder 
fixed point theorem to solve a steady potential flow problem.}. Moreover, every time details of the iteration are changed, all the a priori estimates
have to be checked and changed; this takes an excessive amount of time for complicated problems like the present one. 

In addition, while many function sets in nonlinear elliptic problems are defined by a single constraint 
$\|u\|_{C^{2,\alpha}}\leq M<\infty$, so that the set is obviously a closed ball, 
our function set $\cfusp$ is defined by more than twenty different constraints (see Definition \myref{def:fusp}). 
For many choices $\cfusp$ has nontrivial topology; it is rather difficult to check
whether an intersection of many sets is homeomorphic to a ball (unless convexity can be used).

Ultimately the Schauder fixed point theorem is a means of showing that a particular iteration has nonzero Leray-Schauder degree. 
However, there are other ways of computing a degree which turn out to be simpler in our case (see Section \myref{section:ls}).

The iteration has two steps. The first step solves a fixed-boundary elliptic problem for a function $\xn$;
in this step we use a modification of the second shock condition \myeqref{eq:momjump};
the data of this problem depends on $\xo$ (argument of the iteration map). In the second step the
shock is adjusted (and $\xn$ mapped to a new domain) to satisfy the first shock condition \myeqref{eq:chijump}. 

As mentioned, we have some flexibility in choosing the first-step elliptic problem. An obvious constraint is that fixed points
of the iteration, $\xo=\xn$, must solve the original problem. This is satisfied by 
replacing every occurence of $\chi$ in the original equation and boundary conditions
either by $\xo$ (old iterate) or by $\xn$ (new iterate after step 1).

For the arc boundary condition \myeqref{eq:Lsimple} we choose
$$\frac{1}{2}|\nabla\xn|^2 + \frac{(1-\epsilon)\big((\gamma-1)\xo-c_0^2\big)}{2+(\gamma-1)(1-\epsilon)}=0$$
for the iteration, where $\xo$ is the old and $\xn$ the new solution. Linearization with respect to $\xn$
no longer has a $0$th order term! Hence the opposite-sign problem does not apply, and we can apply the Hopf lemma
to show that the linearizations of the nonlinear elliptic problems arising in the iteration are isomorphisms.
We use the same technique for the interior equation, but keep $\xn$ in the $\rho$ function in the shock condition
\myeqref{eq:momjump}: we need at least one occurence of $\xn$ to be able to use maximum principles to achieve uniqueness
in various contexts. Obviously if $\xn$ does not appear anywhere in the elliptic problem defining the iteration, then
the next iterate is not uniquely determined: any constant can be added.

To apply Leray-Schauder degree theory, we argue that we are solving a continuum of problems, the simplest one
being the ``unperturbed'' problem (see Figure \myref{fig:unperturbed}) 
with $\gamma=1$. This problem is simple enough to verify that its Leray-Schauder
degree is $\pm 1$, in particular nonzero (see Section \myref{section:ls}). All other problems
have the same degree, due to the combined a priori estimates (no fixed points on the boundary of the iteration domain)
and continuity.

The opposite-sign difficulty resurfaces in two aspects: first, to prove degree $\neq 0$ we have to show
uniqueness of the unperturbed problem (see Proposition \myref{prop:unperturbed-unique}). For $\gamma>1$ this 
seems difficult, but for $\gamma=1$ the $\chi$ coefficient is zero (see above). Moreover, 
it is necessary to show that the linearization of the iteration does not have eigenvalue $1$. Again
$\gamma>1$ has unpleasant boundary conditions, but $\gamma=1$ is amenable. Thus solving the isothermal
and isentropic problems in a single paper is \emph{necessity} rather than choice.

Unfortunately it is necessary to show that the iteration is compact, which requires somewhat stronger regularity
estimates than the method of continuity. In fact
\cite[page 482]{gilbarg-trudinger} state: ``Even for the quasilinear [equation] case 
[with fully nonlinear boundary conditions], the fixed point methods [...] are not appropriate, 
since it is not in general possible to construct a \emph{compact} operator [...].'' 

Recent results, many of which are due to Gary Lieberman, have alleviated this problem, at least for our case.
Some unpleasant choices are necessary, however: by replacing the occurences of $\nabla\chi$
by a clever balance of $\nabla\xo$ and $\nabla\xn$ we could obtain a \emph{linear} elliptic problem as part the iteration.
But if $\nabla\xo$ occurs in any of the boundary conditions, then the new solution $\xn$ 
cannot be expected to be smoother than $\xo$ at the boundary. While we may use $\xo,\nabla\xo$ in the interior
coefficients, we must use $\nabla\xn$ for the boundary conditions, which stay nonlinear.

Now our general approach has a hierarchy of three elliptic problems: the original nonlinear problem, a modified nonlinear
problem as part of the iteration, and its linearization. It appears that we have to start all over showing
a priori estimates for the modified nonlinear problem. A trick\footnote{Another, less elegant,
	trick is to use the a priori estimates of the original problem to cut off the coefficients
	of the second problem so that they grow linearly and avoid all singularities like vacuum or 
	parabolicity. The cutoff terms and their derivatives would make the linearization unnecessarily complicated.} 
avoids this:
if the linearizations evaluated at $\xo=\xn$ are isomorphisms, then the nonlinear problems are local isomorphisms. 
Thus we restrict the set of $\xo$ so that the next iterate $\xn=\IT(\xo)$ is \emph{close} to $\xo$. 
Existence and uniqueness of $\xn$ as well as its continuous dependence on $\xo$ then follow 
by linearization around $\xo$; large perturbations are not necessary.

\paragraph{Boundary of the function set}

Both method of continuity and degree theory have a common feature: the bulk of the effort goes in showing that the problem
does not have solutions on the boundary of the function set. For function sets defined by inequality constraints, we have to show that 
for a solution of the original problem (a fixed point of the iteration), each inequality is in fact a strict ($<,>$) inequality. In
this step we are allowed to use the nonstrict ($\leq,\geq$) forms: we are moving from one well-behaved (smooth, elliptic, no vacuum, ...) 
solution to another, without having to show any results for arbitrary weak solutions with no prior information. In particular
we do not have to verify the strict inequalities in any particular order.

\paragraph{Parabolic arcs and corners}

The core of the solution --- and its most difficult part --- is the treatment of the parabolic arcs: 
to obtain a weak solution in the limit, we have to verify the shock conditions are satisfied across the arcs.
To this end we show that $\chi^{(\epsilon)}$ and $\nabla\chi^{(\epsilon)}$ are almost (up to $O(\epsilon^{1/2})$ jumps)
continuous across the arcs; in the $\epsilon\downarrow 0$ limit continuity and mass conservation for $\chi^{(0)}$ are implied.

$\chi,\nabla\chi$ continuity corresponds to two boundary conditions,
$$\lim_{\text{hyperbolic}}\chi=\lim_{\text{elliptic}}\chi,\qquad
\lim_{\text{hyperbolic}}\chi_n=\lim_{\text{elliptic}}\chi_n$$
(continuity for $\chi_t$ is implied by continuity for $\chi$). However, with two boundary conditions
the problem is \defm{overdetermined}, at least in the regularized case.
Hence we can impose only one boundary condition. A trick is needed to verify the second condition 
--- approximately for the regularized solutions; exactly (albeit weakly) in the degenerate limit:

In each point on the arcs, there are three components of $\nabla^2\chi$, but only two relations constraining them ---
the equation \myeqref{eq:chi} and the tangential derivative of the boundary condition, $L^2=1-\epsilon$
(see \myeqref{eq:Ltchi}).
Our key observation is that $L^2<1-\epsilon$ inside the elliptic region (as we have already shown previously),
whereas $L^2=1-\epsilon$ on the arcs: $L$ has a global maximum in each arc point.
Therefore $(L^2)_n\geq 0$ on the arcs, which provides an additional inequality. Thus we obtain 
``$2\frac{1}{2}$'' relations; solving them yields one explicit inequality for each second derivative, 
with right-hand side depending only on $\chi$ and $\nabla\chi$. Only the twice tangential derivative
matters (see \myeqref{eq:refp}).

In the shock-arc corners $\chi$ is $\spC^1$ (Proposition \myref{prop:fp-regularity}, using Section \myref{section:corner}), 
so we may combine the boundary conditions (one on $P$, two on the shock,
but with unknown shock tangent) to represent $\chi$ and $\nabla\chi$ as functions of the corner location. 
If we assume that the corners have precisely the ``expected'' location, then 
we know the corner values of $\chi$ and $\nabla\chi$; in particular $\chi_t=0$, where $\partial_t$ is the 
counterclockwise tangential derivative along $P$, 
and $c=c_L$ (upper left corner) resp.\ $c=c_R$ (upper right corner). However, our analysis is greatly complicated by having \emph{free} corners.
In the other endpoint of each arc the wall boundary condition $\chi_\eta=0$ fixes $\chi_t=0$.
Then the ordinary differential inequalities mentioned above yield \emph{tight}\footnote{It 
is worth noting that if, for whatever reason, we had $(L^2)_n\geq a$
for some bounded (smooth) function $a$, the trick would
still work. The tightness does not stem from $a=0$ (in fact $0$ is far from optimal), but rather from
the characteristic degeneracy.}
bounds for $\chi$ and $\nabla\chi$ on the arcs: they have the desired values, 
up to $O(\epsilon)$. This fortunate circumstance is the key to our solution.

We first discuss isothermal flow which is comparatively simple. In this case the inequality for
$\chi_t$ on $P_R$ has the form
\begin{alignat}{1}
	\chi_{tt} &\geq f(\chi_t) = -\epsilon+O(\chi_t^2), \mylabel{eq:chitt-exp}
\end{alignat}
(see \myeqref{eq:pphi-isen}), with $O$ term constant independent of $\epsilon$.
$\chi_t=0$ in the wall-arc corner, so integration to the arc-shock corner yields 
\begin{alignat}{1}
	\chi_t &\geq -O(\epsilon) \mylabel{eq:chitt-exp2}
\end{alignat}
along the entire arc.

In Section \myref{section:cornersmoving} we compute the values of $\chi,\nabla\chi$ and dependent quantities in the right shock-arc corner
as the corner location varies along the arc. \myeqref{eq:pc-zphi} shows that $\chi_t$ decreases uniformly as the corner moves down from
its expected location.
This contradicts $\chi_t\geq-O(\epsilon)$ --- the second-derivative inequalities already imply (see Proposition \myref{prop:etaa-lowerbound})
a lower bound for each corner, to within $O(\epsilon^{1/2})$ of their expected location.

However, an entirely different argument (see Proposition \myref{prop:etaa-upperbound}) is needed to bound the right shock-arc corner \emph{above}: 
$\psi$ must attain a global minimum in the domain. It can never attain a local minimum
in the interior, on the wall, or on the left arc. For \emph{isothermal} flow we have $\psi_n\geq 0$ on the right arc,
so the Hopf lemma excludes a minimum there as well. Hence there has to be a global minimum on the shock or
in the right shock-arc corner. Essential observation: if that corner is above its expected location,
then $\psi_2>0$ in it, as well as in any hypothetical global minimum on the shock, so we obtain a contradiction. 

Having obtained upper and lower bounds on the corner, we know that $\chi_t=O(\epsilon^{1/2})$ in it. Integrating \myeqref{eq:chitt-exp}
again, but in opposite direction, we obtain
\begin{alignat}{1}
	\chi_t &\leq -O(\epsilon^{1/2}) \mylabel{eq:chitt-exp3}
\end{alignat}
Combined with \myeqref{eq:chitt-exp2} we have $\chi_t=O(\epsilon^{1/2})$ on the arcs. By integration we also control $\chi$;
the boundary condition $|\nabla\chi|^2=c^2$ yields control over $\chi_n$. 

The arguments for the left arc are analogous, with a few modifications.

For \emph{isentropic} flow the arguments are similar, but much more complicated, because the sound speed $c$ can vary.
Instead of considering a single-variable ordinary differential inequality we have to combine it with an ODE for $c$ 
(see \myeqref{eq:cc-arc}). To make the system more tractable it is linearized (\myeqref{eq:pphi}, \myeqref{eq:kphi}) and
restated in polar variables (\myeqref{eq:qphi}, \myeqref{eq:thetaphi}). A delicate analysis (see the proofs of Propositions \myref{prop:pararc} and 
\myref{prop:etaa-lowerbound}) again establishes a lower bound for the right shock-arc corner, as well as a lower bound for $\chi_t$.
For an upper bound on the corner we adapt the isothermal argument, considering minima of $\psi+a\xi$ instead of $\psi$,
for some small $a>0$. This is necessary because for isentropic flow $\psi_n\geq-O(\epsilon^{1/2})$ rather than $\geq 0$ on the right arc. 
More delicate analysis (see the proof of Proposition \myref{prop:etaa-upperbound} and its references)
shows that $a$ can be chosen so large that $(\psi+a\xi)_n\geq 0$, but so small that still $(\psi+a\xi)_\eta>0$ in the right shock-arc corner
and in any hypothetical minimum point on the shock. Having bounded the corner above, an upper bound for $\chi_t$ on the arc is obtained as well.

An entirely different approach to parabolic arcs can be found in \cite{chen-feldman-selfsim-journal}.

\paragraph{Shock control}

It is clear that we have to expect degeneracy at the parabolic arcs. However, there can be degeneracy on the elliptic side of a hyperbolic-elliptic
shock, too, which we have to rule out. Moreover we need some control over the shock location and shape.

To prove that the shock has uniform strength, we show that the density on the elliptic side is uniformly bounded below away from $\rho_I$. 
Proposition \myref{prop:c-principle} rules out density minima in the interior of the elliptic region or at the wall (Remark \myref{rem:symmetries}).
The analysis described above controls the density on the arc up to $O(\epsilon^{1/2})$ (see \myeqref{eq:rhoP}). Finally, Proposition \myref{prop:density-shock}
shows that density cannot have local minima at the shock \emph{except} in a pseudo-normal point. The shock curvature must be positive 
(elliptic region locally convex) in such a point. Moreover if we have a \emph{global} minimum in this point, then the rest of the shock,
including the shock-arc corners, must be below the shock tangent in that point: otherwise there would be another shock tangent parallel to this one,
but lower, meaning lower density, which is a contradiction (see Proposition \myref{prop:rho}, Figure \myref{fig:shocktanarg} for the detailed argument).

Hence we can have density minima at the shock, but only in pseudo-normal points and \emph{above} the line connecting the shock-arc corners.
In such a point, the density is $>\rho_I$ if and only if the connecting line does not meet the circle with center $\vec v_I$ and radius $c_I$.
(Otherwise the shock vanishes or is entropy-violating.) This is precisely \myeqref{eq:techcond}. Unfortunately the cases where \myeqref{eq:techcond}
is violated are not covered in the present paper.

Having established uniform shock strength, Proposition \myref{prop:L-minmax} shows that $L$ cannot have local maxima (with respect to the domain) 
close to $1$ at the shock. This is a necessary ingredient for $L$ control in the elliptic region; the ellipticity must be uniform at the shock
(away from the corners). Hence the shock is analytic, except perhaps in the corners.

To control the shock tangents, $\rho$ and $L$ arguments are not sufficient. It is necessary to control some components of the velocity vector.
More precisely we control the horizontal (in standard coordinates) velocity $v^x$, as well as horizontal velocity in the $L$ picture which 
is $v^x+\alpha v^y$ in
the standard picture, for some $\alpha>0$. We show that $v^x$ cannot have maxima in the interior (Proposition \myref{prop:interior-velocity}),
at the wall (Proposition \myref{prop:v-wall}), or at the shock (Proposition \myref{prop:vshock}). In the latter case, like for density, there
are exceptional cases (see \myeqref{eq:propvx1}) where the shock curvature (see \myeqref{eq:propvx-s11}) is needed to rule out maxima. 
On the arcs we control velocity (like everything else) up to $O(\epsilon^{1/2})$ (see \myeqref{eq:vP}). 

All combined we have that $v^x$ must be between $v^x_L<0$ and $v^x_R=0$, up to $O(\epsilon^{1/2})$,
and an analogous result in $L$ coordinates (see \myeqref{eq:horvel}, \myeqref{eq:leftvel}). 
This yields a slew of additional information (Proposition \myref{prop:ny}): 
bounds on the shock normals \myeqref{eq:shocknormal}, the distance of shock to wall \myeqref{eq:shockwall}, the shock-arc corner angles
\myeqref{eq:cornerregion}, and the vertical velocity \myeqref{eq:vertvel}.

Note that the vertical velocity \emph{can} have minima at the shock, as can be observed in numerics (Figure \myref{fig:numerics} bottom). 
Other velocity directions can have extrema as well; we would be able to cover all cases of the theorem, even those violating \myeqref{eq:techcond},
if we had perfect velocity control. 

As a convenient sideeffect we have ruled out vacuum or negative densities, which is another type of singularity affecting self-similar
potential flow.

\subsection{Parameter set}

\mylabel{section:parmset}

We consider $\gamma\in[1,\infty)$, $\rho_I\in(0,\infty)$, $c_I\in(0,\infty)$.
In addition we use $M^y_I\in(-\infty,0)$ which defines $\vec v_I=(0,M^y_Ic_I)$ (in \emph{standard} coordinates; see Figure \myref{fig:frameR}).
$M^y_I$ is the upstream Mach number normal to the downstream wall. $\rho_I$ and $\vec v_I$ define a potential $\psi^I$ for the $I$ region:
$$\psi^I(\vec\xi)=-\pi(\rho_I)-\frac{|\vec v_I|^2}{2}+\vec v_I\cdot\vec\xi.$$

Given upstream data $\vec v_u=\vec v_I$ and $\rho_u=\rho_I$, Proposition \myref{prop:vdzero} defines a horizontal shock (\defm{R shock})
with $\vec v_R:=\vec v_d=0$. 
Let $\etat_R$ be its height; we choose $\etat_R$ as a parameter that determines $\vec v_I$, rather than
vice versa, so that we may regard $\etat_R$ defined as independent of $\gamma$ (of course $\vec v_I,M^y_I$ depend on $\gamma$ now).

Let $\epsilon\in(0,\overline\epsilon]$ (for some suitable small $\overline\epsilon>0$ which will be fixed later).
Of the two $L_d=\sqrt{1-\epsilon}$ points for this shock, as defined by Proposition \myref{prop:steady-shock-circle}, 
let $\vxit_R$ be the one farther from the origin (see Figure \myref{fig:frameR}). Set $\vec n_R=(0,-1)$.

Starting in the $R$ shock, Proposition \myref{prop:vdzero} yields a family of shocks with $v^y_d=0$. 
Each has two $L_d=\sqrt{1-\epsilon}$ points; let $\vxit_L$ be the one \emph{closer} to the origin.
We focus on choices $\etat_L\in(0,\etat_R]$. We call this shock the \defm{$L$ shock}.
Let $\vec n_L$ be the corresponding downstream normal.
$\vxit_L$ and $\vxit_R$ will be called the \defm{expected} corner locations (although they are most likely not the true locations, except for
$\epsilon=0$).

Let $c_C$, $\rho_C$, $\vec v_C$ be the downstream data of the $C$ shock ($C\in\{L,R\}$).
Note that $v^y_L=0$, but $v^x_L<0$ for $\etat_L<\etat_R$.

Define $\vec\xi_{BR}:=(c_R,0)$ and $\vec\xi_{BL}=(v^x_L-c_L,0)$.
Let $A=(\xi_{BL},\xi_{BR})\times\{0\}$; we will call $A$ the \defm{wall} (it is only the elliptic portion of the wall).
Let $P^*_R$ be a circular arc centered in $\vec v_R$ passing counterclockwise from $\vec\xi_{BR}$ to $\vxit_R$; 
let $P^*_L$ be the arc centered in $\vec v_L$ passing counterclockwise from $\vxit_L$ to $\vec\xi_{BL}$;
both \emph{excluding} the endpoints.
$P_C$ has radius $\sqrt{1-\epsilon}\cdot c_C$ (for $C=L,R$).

There is a $\delta_{nt}>0$ 
so that, for any unit vector $\vec t_S$ from $\vec n_R^\perp$ to $\vec n_L^\perp$ (counterclockwise) and any unit tangent $\vec t_P$ of 
$P^*_L$ or $P^*_R$, 
\begin{alignat}{1}
	|\vec t_S\times\vec t_P| &\geq \delta_{nt}>0. \myeqlabel{eq:deltant}
\end{alignat}
We choose extended arcs $\hat P_{L,R}$ that overshoot $\vxit_{L,R}$ by an angle $\delta_{\hat P}>0$, which we choose continuous in $\gamma,\etat_L$,
so that
\begin{alignat}{1}
	|\vec t_S\times\vec t_P| &\geq\delta_{nt}/2 \myeqlabel{eq:dntcond}
\end{alignat}
for the same $\vec t_S$, but unit tangents $\vec t_P$ of the \emph{extended} arcs $\hat P_{L,R}$. 
I.e.\ the possible shock tangents (restricted in \myeqref{eq:shocknormal} below) and arc tangents are uniformly not collinear.

$P^*_{L,R}$, $\hat P_{L,R}$, and later $P_{L,R}$, are called \defm{quasi-parabolic arcs} (or \defm{parabolic arcs}, by abuse of terminology, 
or short \defm{arcs}). (Of course these arcs are \emph{circular}; ``parabolic'' refers to the expected type of the PDE at these arcs.)

We use $P^{*(\epsilon)}_{L,R}$ and $\hat P^{(\epsilon)}_{L,R}$ to identify the arcs for a particular choice of $\epsilon$.

The Definitions \myref{def:Lambda}, \myref{def:b} and \myref{def:fusp} 
use many constants and other objects that will be fixed later on. 
In all of these cases, an upper (or lower) bound for each constant is found.
Whenever we say ``for sufficiently small constants'' (etc.), we mean that bounds for them are adjusted.
To rule out circularity, it is necessary to specify which bounds may depend on the values of which other bounds.
In the following list, bounds on a constant may only depend on bounds of other constants \emph{before} them. 
\begin{alignat}{1}
	& \delta_{\hat P},
	C_L,C_\eta,\delta_{SA},\delta_{Cc},\delta_{P\sigma},\delta_{Pn},\delta_d,\delta_\rho,\delta_{Lb},
	C_{Pt},C_{vx},C_{vL},C_{Sn},\delta_{vy},\notag\\
	& \qquad\delta_o,C_d,\epsilon,C_{\spC},r_I,\alpha,\beta. \myeqlabel{eq:constlist}
\end{alignat}
The constants $C_{\spC},r_I,\alpha,\beta$ may depend on $\epsilon$ itself, not just on an upper bound. 
$r_I$ may also depend on $\po$.
The reader may convince himself that the following discussion does respect this order.

The parameters $\gamma$, $\etat_L$ used in Leray-Schauder
degree arguments will be restricted to compact
sets below so that any constant that can be chosen continuous in them might as well be taken independent of them.
Dependence on other parameters like $\rho_I$ will not be pointed out explicitly. 

Constants $\delta_?$ as well as $\alpha,\beta-1,r_I,\epsilon$ are meant to be small and positive, constants $C_?$ are meant to be large 
and finite.

\begin{definition}
	\mylabel{def:Lambda}%
	For the purposes of degree theory we define a restricted parameter set (see Figure \myref{fig:etaL})
	$$\Lambda:=\{\lambda=(\gamma,\etat_L):\gamma\in[1,\overline\gamma],~\eta^*_L\in[\underline\eta^*_L(\gamma),\overline\eta^*_L(\gamma)]\}$$
	\sindex{Lambda}{$\Lambda$}%
	\sindex{lambda}{$\lambda$}%
	\sindex{gamma}{$\overline\gamma$}%
	\sindex{etaLstar}{$\underline\eta^*_L$}%
	\sindex{etaLstar}{$\overline\eta^*_L$}%
	with 
	$$\overline\gamma\in[1,\infty),\qquad 0<\underline\eta^*_L(\gamma)<\overline\eta^*_L(\gamma)\leq\etat_R$$
	where 
	\begin{alignat}{1}
		\overline\eta^*_L(\gamma) &= \begin{cases}
			\etat_R, & \gamma=1, \\
			\etat_R-C_\eta\epsilon^{1/2}, & \gamma>1
		\end{cases} \mylabel{eq:Ceta}
	\end{alignat}
	\sindex{Ceta}{$C_\eta$}%
	where $C_\eta$ (to be determined in Proposition \myref{prop:etaa-upperbound}) may depend on $\gamma$ but not on $\epsilon$.

	Moreover we restrict $\underline\eta^*_L(\gamma)$ so that \myeqref{eq:techcond} is satisfied.
\end{definition}

\begin{lemma}
	\mylabel{lemma:etax}%
	For sufficiently small $\epsilon$, with bound depending on $C_\eta$:

	There is an $\eta^x_L(\gamma)\in[0,\overline\eta^*_L(\gamma))$, continuous in $\gamma$,
	so that \myeqref{eq:techcond} is satisfied for all $\etat_L\in(\eta^x_L(\gamma),\etat_R]$,
	but never for $\etat_L\in(0,\eta^x_L(\gamma)]$. 
\end{lemma}
\begin{proof}
	\myeqref{eq:techcond} is satisfied for $\etat_L=\etat_R$ (see Figure \myref{fig:techcond}), by Proposition \myref{prop:steady-shock-circle}: in this
	case $L$ shock and $R$ shock coincide, and the $R$ shock never intersects the circle with center $\vec v_u=\vec v_I$ and radius $c_u=c_I$.

	Clearly the distance to that circle is strictly decreasing as $\etat_L$ decreases. By continuity there must be an $\eta^x_L$,
	depending continuously on $\gamma$, so that for $\etat_L=\eta^x_L$ the $L$ shock touches the circle. For all smaller $\etat_L>0$
	the circle is intersected, because $\etat_L,\etat_R>0$ does not allow the $L$ shock to pass below the circle.

	Since the $L$ shock tangent and location depends continuously on $\gamma$, $\eta^x_L$ must also be continuous in it.
\end{proof}

\if\dofigures%
\begin{figure}
\input{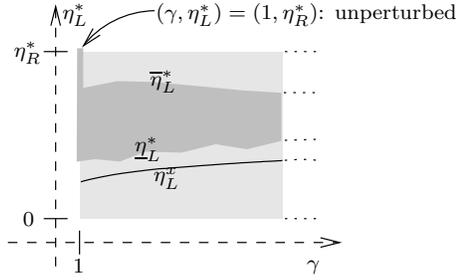}
\caption{$\eta^*_L$ below $\eta^x_L$ violate \myeqref{eq:techcond}. For $\gamma>1$ some distance from $\eta^*_R$ is needed for technical reasons.
The Leray-Schauder degree argument uses parameters in the dark shaded set $\Lambda$ which is path-connected for small $\epsilon>0$.}
\mylabel{fig:etaL}
\end{figure}
\fi%

\begin{lemma}
	\mylabel{lemma:Gamma-connected}%
	For sufficiently small $\epsilon$, with bound depending on $C_\eta$:

	For all $\underline\eta^*_L(\gamma)$ (continuous in $\gamma$) that satisfy
	\begin{alignat}{1}
		\eta^x_L(\gamma) &< \underline\eta^*_L(\gamma) < \overline\eta^*_L(\gamma), \myeqlabel{eq:ueLbd}
	\end{alignat}
	$\Lambda$ is path-connected and contains the point $(\gamma,\etat_L)=(1,\etat_R)$.
\end{lemma}
\begin{proof}
	The curve $\gamma\mapsto\frac{12}(\underline\eta^*_L+\overline\eta^*_L)$ is continuous and contained in $\Lambda$.
	It intersects every line $\{\gamma\}\times[\underline\eta^*_L,\overline\eta^*_L]$, including the one for $\gamma=1$
	which is $(\underline\eta^*_L,\eta^*_R]$. $\Lambda$ is the union of all these lines, so the proof is complete.
\end{proof}

\subsection{Function set and iteration}

\begin{definition}
	\mylabel{def:weighted-hoelder}%
	Let $U\subset\R^n$ open nonempty bounded with $\partial U$ uniformly Lipschitz.
	Let $F\subset\partial U$.
	For $k\in\N_0$, $\alpha\in[0,1]$ and $\beta\in(-\infty,k+\alpha]$ we define
	the \defm{weighted H\"older space} $\spC^{k,\alpha}_\beta(U,F)$ as the set of $u\in\spC^{k,\alpha}(\overline U-F)$ so that 
	$$\|u\|_{\spC^{k,\alpha}_\beta(U,F)}:=
	\sup_{r>0}r^{k+\alpha-\beta}\|u\|_{\spC^{k,\alpha}(\overline U-B_r(F))}$$
	is finite.
\end{definition}

\begin{definition}
	\mylabel{def:b}%
	For sufficiently small $\delta_{\hat P}>0$, 
	there is a function $b\in\spC^2(\R^2)$ with $b,|\nabla b|\leq 1$ so that $b=0$ on $\hat P_L^{(0)}$ and $\hat P_R^{(0)}$, $b>0$ elsewhere,
	$b$ even in $\eta$.
	From now on we fix a particular $b$.
	\sindex{b}{$b$}%
\end{definition}
\begin{proof}
	The construction is straightforward. $\delta_{\hat P}$ is taken so small that $\hat P_L^{(0)}$ and $\hat P_R^{(0)}$
	are well-separated. That is possible because $\xi_L^{*(0)}<\xi_R^{*(0)}$.
\end{proof}
\if\dofigures%
\begin{figure}
\input{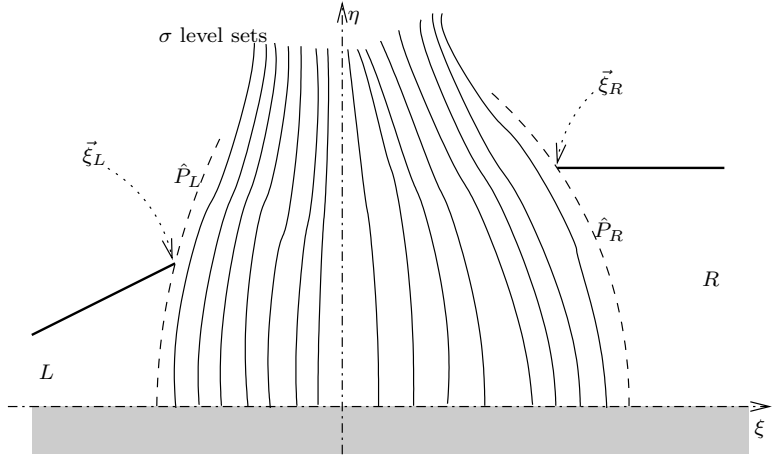}
\caption{Onion coordinates. $(\xi,\eta)$ plane shown; $\sigma$ level sets are green.}
\mylabel{fig:onion}
\end{figure}
\fi%
\begin{definition}
	\mylabel{def:fusp}%
	\ \par
	{\bf Onion coordinates}

	To define a function subset in a fixed Banach space, we need to map the domain with its free shock boundary to a fixed square.
	Define a $\spC^\infty$ change to coordinates $(\sigma,\eta)\in\R^2$ (see Figure \myref{fig:onion}) so that\sindex{sigma}{$\sigma$}
	\begin{enumerate}
	\item
		$\eta$ is preserved and $\sigma=\sigma(\xi,\eta)$ a $\spC^\infty$ function with a
		$\spC^\infty$ function $\xi=\xi(\sigma,\eta)$ inverting it,
	\item
		$\hat P_L^{(\epsilon)}$ maps to a subset of $\{\sigma=0\}$, 
	\item 
		$\hat P_R^{(\epsilon)}$ maps to a subset of $\{\sigma=1\}$,
	\item
		$\vec\xi_{BL}^{(\epsilon)}$ maps to $(0,0)$,
	\item
		$\vec\xi_{BR}^{(\epsilon)}$ maps to $(1,0)$,
	\item 
		and $A^{(\epsilon)}$ maps to $\{\sigma\in(0,1),~\eta=0\}$ precisely. 
	\item
		Let $T\subset S^1$ be the set of unit tangents of $\hat P_L^{(\epsilon)}$ and $\hat P_R^{(\epsilon)}$.\sindex{T}{$T$}
		For some constant $\delta_{P\sigma}>0$ depending only and continuously on $\lambda\in\Lambda$, 
		require\sindex{deltaPsigma}{$\delta_{P\sigma}$}
		that for every unit tangent $\vec t$ of a $\sigma$ level set, 
		\begin{alignat}{1}
			d(\vec t,T) &\leq \delta_{P\sigma}. \myeqlabel{eq:Tt}
		\end{alignat}
	\end{enumerate} 
	We require that the change of coordinates depends continuously (in $\spC^\infty$) on $\lambda\in\Lambda$.
	The construction is straightforward.

	Here and in what follows, we will use the weighted H\"older spaces $\spC^{2,\alpha}_\beta(\overline U)$,
	as in Definition \myref{def:weighted-hoelder}.
	The domain $U$ is either $[0,1]^2$ with $F=\{(0,1),(1,1)\}$,
	or $\overline\Omega$ with $F=\{\vxia_L,\vxia_R\}$ (to be defined).
	For the shock we use $U=[0,1]$ with $F=\{0,1\}$, or $U=[\xia_L,\xia_R]$ with $F=\{\xia_L,\xia_R\}$;
	for the arcs only the upper endpoints are in $F$ and for the wall $A$ we have $F=\emptyset$.
	We omit $F$ as it will be clear from the context.
	$C^{2,\alpha}_\beta$ are Banach spaces so that standard functional analysis applies.
	Moreover, 
	$C^{2,\alpha}_\beta(\overline\Omega)$ is continuously embedded in
	$C^1(\overline\Omega)$, so we have $C^1$ regularity in the corners as well, which is crucial.
	$\beta\in(1,2)$ and $\alpha\in(0,\beta-1]$ will be determined later.\sindex{alpha}{$\alpha$}\sindex{beta}{$\beta$}

{\bf Free boundary fit}

	Let $\cfusp$ be the set of functions $\po\in \spC^{2,\alpha}_\beta([0,1]^2)$ that satisfy all of the many conditions explained below.
	Require
	\begin{alignat}{1}
		\|\po\|_{\spC^{2,\alpha}_\beta([0,1]^2)} \leq C_\spC(\epsilon). \myeqlabel{eq:regularity}
	\end{alignat}

	For all $\sigma\in[0,1]$ define 
	\begin{alignat}{1}
		s(\sigma) &:= \frac{\po(\sigma,1)-\psi^I(0,0)}{v^y_I}; \myeqlabel{eq:sdef}
	\end{alignat}
	it satisfies $\po(\sigma,1)=\psi^I(\xi(\sigma,\eta),\eta)$ with $\eta=s(\sigma)$. 
	We define another coordinate transform by first
	mapping $(\sigma,\zeta)\in[0,1]$ to
	\sindex{zeta}{$\zeta$}%
	$(\sigma,\eta)$ with $\eta=\zeta s(\sigma)$, and then 
	mapping to $\vec\xi$ with the previous coordinate transform. 

	Let $\vxia_L$ resp.\ $\vxia_R$
	\sindex{xiLvector}{$\vxia_L$}%
	\sindex{xiRvector}{$\vxia_R$}%
	be the $\vec\xi$ coordinates for the $(\sigma,\zeta)$ plane points $(0,1)$ resp.\ $(1,1)$.
	Let $S$ be the $\vec\xi$ plane curve for $(0,1)\times\{1\}$. Define $P_L$, $P_R$ 
	\sindex{PL}{$P_L$}%
	\sindex{PR}{$P_R$}%
	to be the images of 
	$\{0\}\times(0,1)$ resp.\ $\{1\}\times(0,1)$. Finally, let $\Omega$ be the image of $(0,1)\times(0,1)$.

	Require shock-wall separation: 
	\begin{alignat}{1}
		\max_{\sigma\in[0,1]}s(\sigma) &\geq \delta_{SA}>0. \myeqlabel{eq:shockwall}
	\end{alignat}
	\sindex{deltaSA}{$\delta_{SA}$}%
	Require
	\begin{alignat}{1}
		& \text{$\xi=\xi(\sigma,\zeta)$ is strictly increasing in $\sigma$ for $\zeta=1$.} \myeqlabel{eq:s-welldef}
	\end{alignat}
	Require: corners close to target: 
	\begin{alignat}{1} 
		|\vec\xi_L-\vxit_L|,|\vec\xi_R-\vxit_R| \leq \epsilon^{1/2}, \myeqlabel{eq:cornerregion}
	\end{alignat}
	We require $\epsilon$ to be so small that $\vxia_C\in\hat P_C$ ($C=L,R$), 
	i.e.\ $\vxia_C$ may not be higher than the upper endpoint of $\hat P_C$.
	
	For later use we define $\eta^\pm_{L,R}:=\etat_{L,R}\pm\epsilon^{1/2}$ and let $\xi^\pm_{L,R}$ be so that
	$\vec\xi^\pm_{L,R}\in\hat P_{L,R}$.
	\sindex{xiplus}{$\xi^+_L$, $\xi^+_R$}%
	\sindex{etaplus}{$\eta^+_L$, $\eta^+_R$}%
	\sindex{ximinus}{$\xi^-_L$, $\xi^-_R$}%
	\sindex{etaminus}{$\eta^-_L$, $\eta^-_R$}%

	Corner cones: require\sindex{deltaCc}{$\delta_{Cc}$} 
	\begin{alignat}{1}
		\sup_{\vec\xi,\vec\xi'\in\overline\Omega}\measuredangle(\vec\xi-\vec\xi_C,\vec\xi'-\vec\xi_C)
		 &\leq \pi-\delta_{Cc} \qquad(C\in\{L,R\}).
		\myeqlabel{eq:cornercone} 
	\end{alignat}
	(As discussed in the introduction, $\measuredangle(\vec x,\vec y)$ is the counterclockwise angle from $\vec x$ to $\vec y$.)

	\myeqref{eq:shockwall} ensures that the map from $(\sigma,\zeta)$ to $\vec\xi$ is 
	a well-defined change of coordinates, uniformly nondegenerate (depending on $\delta_{SA}$ and $C_\spC$), 
	with $\spC^{2,\alpha}_\beta([0,1]^2)$ resp.\ $\spC^{2,\alpha}_\beta(\overline\Omega)$ regularity.
	Since the step from $(\sigma,\zeta)$ to $(\sigma,\eta)$ uses $\psi$, the entire coordinate change is as smooth as $\psi$.
	If we prove higher regularity for $\psi$ either in $(\sigma,\zeta)$ or $(\xi,\eta)$ coordinates, we immediately obtain
	the same higher regularity for the coordinate transform and in the respective other coordinates.

	\sindex{Omega}{$\Omega$}
	It is clear now that $\partial\Omega$ is the union of the disjoint sets $S$, $P_L$, $P_R$, $A$,
	$\{\vxia_R\}$, $\{\vxia_L\}$, $\{\vec\xi_{BL}\}$ and $\{\vec\xi_{BR}\}$. By \eqref{eq:shockwall}, $\Omega$ is a simply connected set.

	\myeqref{eq:s-welldef} ensures that $s$ can be defined as a function of $\xi$, which is the way we use it 
	from now on.

{\bf Iteration}

	Shock strength/density: require that 
	\begin{alignat}{1}
		-\xo-\frac{1}{2}|\nabla\xo|^2 >0, \myeqlabel{eq:rhoprep}
	\end{alignat}
	so that $\rho$ is well-defined, and require
	\begin{alignat}{1}
		\min_{\overline\Omega}\rho &\geq \rho_I+\delta_\rho. \myeqlabel{eq:rhomin}
	\end{alignat}
	\sindex{deltarho}{$\delta_\rho$}%

	Pseudo-Mach number bound: require 
	\begin{alignat}{1}
		L^2 \leq 1-\delta_{Lb}\cdot b \qquad\text{in $\overline\Omega$,} \myeqlabel{eq:ellip}
	\end{alignat}
	\sindex{deltaLb}{$\delta_{Lb}$}%
	(Note that $L$ is well-defined because by \myeqref{eq:rhomin} $\rho>0$, so $c>0$.)
	$b=0$ on $\hat P^{(0)}_{L,R}$ which have distance $\geq\frac\epsilon3$ (for sufficiently small $\epsilon$) from $\overline\Omega$, so 
	\myeqref{eq:ellip} implies
	\begin{alignat}{1}
		L^2\leq 1-\frac13|\nabla b|_{L^\infty}\delta_{Lb}\cdot\epsilon \leq 1-\frac13\delta_{Lb}\epsilon
		\qquad\text{in $\overline\Omega$,} \myeqlabel{eq:ellipC}
	\end{alignat}

	Require: there is\footnote{$\pn$ is the product of an iteration step with input $\po$.
		We will show/ensure in Proposition \myref{prop:pn-uqcont} that $\pn$ is unique and continuously 
		dependent on $\po$.} 
	a function $\pn\in \spC^{2,\alpha}_\beta(\overline\Omega)$ with the following properties:
	\begin{enumerate}
	\item
		$\po$ close to $\pn$: 
		\begin{alignat}{1}
			\|\po-\pn\|_{\spC^{2,\alpha}_\beta([0,1]^2)} &\leq r_I(\po) \myeqlabel{eq:oldnew}
		\end{alignat}
		where $r_I\in C(\cfusp;(0,\infty))$ is a continuous function to be determined later.
		Here and later we regard $\pn$ as defined on $[0,1]^2$ instead of $\overline\Omega$, via the coordinate
		transform from $\vec\xi$ to $(\sigma,\zeta)$ defined by $\po$ (see above).
		\sindex{rI}{$r_I$}%
	\item
		We require $r_I$ to be so small that
		\begin{alignat}{1}
			-\xn-\frac{1}{2}|\nabla\xn|^2 & >0, \myeqlabel{eq:tilderho} \\
			\nabla\pn &\neq \vec v_I, \myeqlabel{eq:faken}
		\end{alignat}
		so that in particular $\hat\rho$ is well-defined and positive.
	\item
		Moreover we require $r_I$ to be so small that
		\begin{alignat}{1}
			\big(c_0^2+(1-\gamma)(\xo+\frac{1}{2}|\nabla\xn|^2)\big)I-\nabla\xn^2 &>0, \myeqlabel{eq:tildeL}
		\end{alignat}
		i.e.\ is a (symmetric) positive definite matrix.
	\item
		Let $\BL=\BL(\po,\pn)$ be defined as
		\sindex{L}{$\BL$}%
		\begin{alignat}{1}
			& \Big(\big(c_0^2+(1-\gamma)(\xo+\frac{1}{2}|\nabla\xn|^2)\big)I-\nabla\xn^2\Big):\nabla^2\pn, \myeqlabel{eq:itn-inner} \\
			& \frac{|\nabla\xn|^2}{2} + \frac{(1-\epsilon)\big((\gamma-1)\xo+c_0^2\big)}{2+(1-\epsilon)(\gamma-1)}, \myeqlabel{eq:itn-parabolic} \\
			& \big(\hat\rho\nabla\xn-\rho_I\nabla\chi^I\big)\cdot\frac{\vec v_I-\nabla\pn}{|\vec v_I-\nabla\pn|}, \myeqlabel{eq:itn-shock} \\
			& \pn_\eta \Big). \myeqlabel{eq:itn-wall}
		\end{alignat}
		\myeqref{eq:itn-shock} is well-defined by
		\myeqref{eq:tilderho} and \myeqref{eq:faken}. The other components have no singularities.

		$\nabla\po\in\spC^{1,\alpha}_{\beta-1}$,
		so $|\nabla\xo|^2\in\spC^{1,\alpha}_{\beta-1}$, 
		so $\Big(\big(c_0^2+(1-\gamma)(\xo+\frac{1}{2}|\nabla\xn|^2)\big)I-\nabla\xn^2\Big)\in\spC^{1,\alpha}_{\beta-1}
			\contemb\spC^{0,\beta-1}\contemb\spC^{0,\alpha}$ (note $\alpha\leq\beta-1$ as required above), 
		and $\nabla^2\po\in\spC^{0,\alpha}_{\beta-2}$,
		so \myeqref{eq:itn-inner} is $\in\spC^{0,\alpha}_{\beta-2}$.
		In the same way we check that \myeqref{eq:itn-parabolic}, \myeqref{eq:itn-shock} and \myeqref{eq:itn-wall} are $\spC^{1,\alpha}_{\beta-1}$.
		Hence we may take the codomain of $\BL$ to be the Banach space
		$$Y:=\spC^{0,\alpha}_{\beta-2}(\overline\Omega)
		\times \spC^{1,\alpha}_{\beta-1}(\overline S)
		\times \spC^{1,\alpha}_{\beta-1}(\overline P_L)
		\times \spC^{1,\alpha}_{\beta-1}(\overline P_R)
		\times \spC^{1,\alpha}_{\beta-1}(\overline A).$$
		Alternatively, if we consider the pullback to $(\sigma,\zeta)$ coordinates, as defined by $\po$ above, we may consider
		$$\spC^{0,\alpha}_{\beta-2}([0,1]^2)
		\times \spC^{1,\alpha}_{\beta-1}[0,1]
		\times \spC^{1,\alpha}_{\beta-1}[0,1]
		\times \spC^{1,\alpha}_{\beta-1}[0,1]
		\times \spC^{1,\alpha}_{\beta-1}[0,1].$$
		In the same way we can discuss $\pn$ either in $\spC^{2,\alpha}_\beta(\overline\Omega)$ or in $\spC^{2,\alpha}_\beta([0,1]^2)$.

		With these topologies, clearly $\BL$ is a smooth function of $\po$ and $\pn$.

		Most importantly: require 
		\begin{alignat}{1}
			\BL(\po,\pn) &= 0. \myeqlabel{eq:pn}
		\end{alignat}
	\end{enumerate}

{\bf Other bounds}

	Require 
	\begin{alignat}{1}
		\|\po\|_{\spC^{0,1}(\overline\Omega)} \leq C_L \myeqlabel{eq:lip}
	\end{alignat}
	where $C_L$ may not depend on $\epsilon$. 
	\sindex{CL}{$C_L$}%

	$\chi_t$ and $\chi_n$ on parabolic arc: 
	\begin{alignat}{1}
		\max_{\overline{P_L}\cup\overline{P_R}}c^{-1}|\frac{\partial\xo}{\partial t}| &\leq C_{Pt}\epsilon^{1/2}, \myeqlabel{eq:partan} \\
		\max_{\overline{P_L}\cup\overline{P_R}}c^{-1}\frac{\partial\xo}{\partial n} &\leq -\delta_{Pn}. \myeqlabel{eq:parnor}
	\end{alignat}
	\sindex{CPt}{$C_{Pt}$}%
	\sindex{deltaPn}{$\delta_{Pn}$}%
	Here, $\delta_{Pt},\delta_{Pn}$ may depend \emph{only} on $\lambda$, but not on $\epsilon$ (or $\po$).

	Horizontal velocity: 
	\begin{alignat}{1} 
		\max_{\overline\Omega}\po_\xi &\leq C_{vx}\epsilon^{1/2}. \myeqlabel{eq:horvel}
	\end{alignat}
	\sindex{Cvx}{$C_{vx}$}%
	Vertical velocity:
	\begin{alignat}{1} 
		\po_\eta &\geq v^y_I+\delta_{vy}\qquad\text{in $\overline\Omega$.} \myeqlabel{eq:vertvel}
	\end{alignat}
	\sindex{deltavy}{$\delta_{vy}$}%
	Left corner shock tangential velocity: 
	\begin{alignat}{1}
		\max_{\overline\Omega}\nabla\po\times\vec n_L \geq \vec v_L\times\vec n_L+C_{vL}\epsilon^{1/2} \myeqlabel{eq:leftvel}
	\end{alignat}
	\sindex{CvL}{$C_{vL}$}%
	Shock normal: 
	Let $N\subset S^1$ (unit circle) be the set from $\vec n_R$ to $\vec n_L$ (counterclockwise).
	\begin{alignat}{1}
		\sup_Sd(\vec n,N) &\leq C_{Sn}\epsilon^{1/2}. \myeqlabel{eq:shocknormal}
	\end{alignat}
	\sindex{deltaSn}{$\delta_{Sn}$}%

	Set $\Sigma^1:=P_L$, $\Sigma_2:=S$, $\Sigma_3:=P_R$ and $\Sigma_4:=A$. 
	\sindex{Sigma}{$\Sigma^i$}%
	Write the components \myeqref{eq:itn-parabolic}, \myeqref{eq:itn-wall}, \myeqref{eq:itn-shock} of $\BL$
	as 
	$$g^i(\vec\xi,\xn(\vec\xi),\subeq{\nabla\xn(\vec\xi)}{=:\vec p})\qquad (i=1,\dotsc,4),$$ 
	\sindex{gi}{$g^i$}%
	where the $\vec\xi$ dependence includes the dependence on 
	$\xo(\vec\xi)$ and $\nabla\xo(\vec\xi)$. 

	$g^2$ has some singularities, but not on the set of $\vec\xi,\chi,\nabla\chi$ so that \myeqref{eq:vertvel}
	and \myeqref{eq:rhomin} (resp.\ \myeqref{eq:tilderho} and \myeqref{eq:faken}) are satisfied. That set is simply connected,
	so we can modify $g^2$ on its complement and extend it smoothly to $\overline\Omega\times\R\times\R^2$.
	The modification is chosen to depend smoothly on $\lambda$. 

	Require uniform obliqueness\footnote{In various articles 
	Lieberman uses the term ``uniformly oblique'', 
	probably with ``oblique'' in the sense of ``not tangential''. We adopt this terminology instead of the more common but less useful 
	sense ``not normal''
	(see e.g.\ \cite{gilbarg-trudinger} or \cite{palagachev}).}: 
	\begin{alignat}{1}
		|g^i_{\vec p}\cdot\vec n| &\geq \delta_o|g^i_{\vec p}|\qquad\forall\vec\xi\in\Sigma^i. \myeqlabel{eq:ndobb}
	\end{alignat}
	\sindex{deltao}{$\delta_o$}%
	\cucon{$\delta_o$ does not deteriorate with $\epsilon$. on $P_L,P_R$ the $\chi_n$ lower bound is achieved by $\epsilon=1/2$ 
	already. on $S$ we need only uniform shock strength (see corresponding prop). On $A$ trivial.}

	Functional independence in upper corners: for $i,j=1,2$ and for $i,j=2,3$ set
	$$G:=\begin{bmatrix}g^i_{p^1} & g^j_{p^1}\\g^i_{p^2} & g^j_{p^2}\end{bmatrix},$$
	regard it as a function of $\vec\xi$ (including the dependence on $\nabla\xn(\xi)$) 
	and require\sindex{deltad}{$\delta_d$}\sindex{Cd}{$C_d$}%
	\begin{alignat}{1}
		\|G\|,\|G^{-1}\| \leq C_d \qquad\text{in $B_{\delta_d}(\vec\xi_C)\cap\overline\Omega$, $C=L,R$.} \myeqlabel{eq:Gb}
	\end{alignat}

	Let $\cfusp$ be the set\footnote{The notation $\cfusp$ does not necessarily imply that $\cfusp$ is the closure of $\fusp$.} 
	of admissible functions so that all of these conditions are satisfied.
	\sindex{F}{$\fusp$, $\fusp_\lambda$, $\cfusp$, $\cfusp_\lambda$}%
	Define
	$\fusp$ to be the set of admissible functions such that all of these conditions are satisfied
	with \emph{strict} inequalities, i.e.\ replace $\leq,\geq$ by $<,>$, ``increasing'' by ``strictly increasing'' etc.

[This is the end of Definition \myref{def:fusp}.]
\end{definition}

\begin{remark}
	\mylabel{rem:fp}%
	If $\pn=\po$, then \myeqref{eq:itn-inner}, \myeqref{eq:itn-shock}, \myeqref{eq:itn-parabolic}, \myeqref{eq:itn-wall}
	and the definition of $S$ yield
	\begin{alignat}{1}
		(c^2I-\nabla\chi^2):\nabla^2\psi &= 0\qquad\text{in $\overline\Omega$,} \notag\\
		\chi_\eta &= 0 \qquad\text{on $\overline A$,} \notag\\
		\chi^I &= \chi \qquad\text{and} \notag\\
		(\rho\nabla\chi-\rho_I\nabla\chi^I)\cdot\vec n &= 0\qquad\text{on $\overline S$,} \notag\\
		L &= \sqrt{1-\epsilon}\qquad\text{on $\overline P_L\cup\overline P_R$}\notag
	\end{alignat}
	(we may take closures by regularity \myeqref{eq:regularity}).
\end{remark}

\begin{remark}
\mylabel{rem:reflection}%
In any point on $\overline{A}$, we can use even reflection (see Remark \myref{rem:symmetries}) of $\po$ across $A$ to obtain a point in the interior,
or (in the bottom corners) a point at a quasi-parabolic arc with the interior equation applying inside. 
$\po_\eta=0$ on $A$, for even reflection of $\po$, implies that the solution is $C^1$ across $\overline A$; then
necessarily it is also $C^{2,\alpha}$ (away from the shock-arc corners).

For $\po$ standard regularity theory immediately yields that the solution is analytic in the bigger domain near $\overline{A}$.
The same technique applied to $\pn$ and to solutions $\pnl$ of linearized equations 
(here $\po$, $\pn$ and $\pnl$ are reflected) yields $C^{2,\alpha}$ regularity (away from the shock-arc corners).
\end{remark}

\begin{proposition}
	\mylabel{prop:Lxn-iso}%
	For sufficiently small $\epsilon$ (with bound depending only on $C_{Pt}$) and $r_I$ (depending continuously and only on $\po,\delta_{vy}$):

	for all $\po\in\cfusp$, $\BL(\po,\pn')$ is well-defined for $\pn'$ near $\po$, and the Fr\'echet derivative
	$\partial\BL/\partial\pn'(\po,\po)$ (of $\BL$ with respect to its second argument $\pn'$,
	evaluated at $\pn'=\po$)
	is a linear isomorphism of $\spC^{2,\alpha}_\beta$ onto $Y$. 
\end{proposition}
\begin{proof}
	\myeqref{eq:itn-shock} is the only part of $\BL$ with a singularity. \myeqref{eq:rhomin} and \myeqref{eq:faken} guarantee
	that it is well-defined and smooth for $\pn=\po$. For $\pn$ in a sufficiently small neighbourhood of $\po$ (i.e.\ take $r_I$
	small), it stays well-defined.

	Let $\pnl\in \spC^{2,\alpha}_\beta(\overline\Omega)$; it is meant to be the first variation
	of $\pn$ (and, at the same time, $\xn$). 
	$\partial\BL/\partial\pn(\po,\po)\pnl$ is a tuple of functions in $Y$.
	Its first component (see \myeqref{eq:itn-inner}) is of type
	\begin{alignat}{1}
		(c^2I-\nabla\xo^2)
		:\nabla^2\pnl 
		+ \vec b\cdot\nabla\pnl 
		& \myeqlabel{eq:eqlin}
	\end{alignat}
	where $\vec b$ is some vector field.

	We linearize \myeqref{eq:itn-shock}: here we can use that we linearize at $\pn=\po$, so 
	$$|\vec v_I-\nabla\po|^{-1}(\vec v_I-\nabla\po)=\vec n$$
	because $\po=\psi^I$ defines the shock (see \myeqref{eq:sdef} in Definition \myref{def:fusp}).
	Result:
	\begin{alignat}{1}
		&\rho(1-c^{-2}\xo_n^2)\pnl_n-\rho c^{-2}\xo_n\xo_t\pnl_t-\rho c^{-2}\pnl \notag\\
		&\quad- (\rho\nabla\xo-\rho_I\nabla\chi^I)|\vec v_I-\nabla\po|^{-1}
			\subeq{(1-\Big(\subeq{\frac{\vec v_I-\nabla\po}{|\vec v_I-\nabla\po|}}{=\vec n}\Big)^2)}{=\vec t\vec t^T}\nabla\pnl.
		\myeqlabel{eq:shocklin}
	\end{alignat}
	The coefficient $\rho(1-c^{-2}\xo_n^2)$ of $\pnl_n$, positive by \myeqref{eq:rhomin} and \myeqref{eq:ellipC},
	has the opposite sign to the nonzero coefficient $-\rho c^{-2}$ of $\pnl$, negative by \myeqref{eq:rhomin}. 

	The other components linearize to
	\begin{alignat}{1}
		\nabla\xo\cdot\nabla\pnl &, \qquad\text{(parabolic)} \myeqlabel{eq:parlin} \\
		\xnl_n &. \qquad\text{(wall)} \myeqlabel{eq:walllin}
	\end{alignat}
	Note that the coefficient vectors $g^i_{\vec p}$ of $\nabla\pnl$ are the same as in \myeqref{eq:ndobb} and \myeqref{eq:Gb}.

	To prove the Proposition, we apply \cite[Theorem 1.4]{lieberman-crelle-1988}. 
	Our $\spC^{k,\alpha}_\beta(\Omega;F)$ spaces correspond to
	his spaces $H^{(-\beta)}_{k+\alpha}$ if $\alpha>0$ and $F\subset\partial\Omega$ is finite and contains
	all points where $\partial\Omega$ is not $C^2$ --- as is the case here. 
	We check the preconditions:
	\begin{enumerate}
	\item $\Sigma_i$ are $C^2$ curves (except perhaps for the endpoints), meeting in single points (corners).
		In each corner the two curves meet at an angle $0<\theta_{ij}\leq\theta_0$, with $\theta_0<\pi$.
		($0<\theta_{ij}$ is trivial from \myeqref{eq:shocknormal}).
		The lower bound is obvious; the upper bound is obvious for corners with $A$ and supplied by
		\myeqref{eq:cornercone} for corners with $S$.
	\item The equation is uniformly elliptic, by \myeqref{eq:ellipC}.
	\item All boundary operators are uniformly oblique, by \myeqref{eq:ndobb}. 
	\item Condition (1.17) in loc.cit.\ with $\beta_i\neq\beta_j$ is equivalent to \myeqref{eq:Gb}.
	\item Finally, the \emph{only} appearance of $\pnl$ is in \myeqref{eq:shocklin}, where its coefficient his nonzero
		and has the opposite sign as the coefficient of $\pnl_n$ (note that $\vec n$ is the \emph{inward} normal here).
		(This observation allows to apply maximum principles to obtain uniqueness.)
		This means (1.19) in loc.cit.\ is satisfied.
	\item The bottom corners are not a concern: by Remark \myref{rem:reflection} we can use even reflection across $A$.
	\item The other preconditions are technical and easy to verify.
	\end{enumerate}

	Theorem 1.4 in loc.cit.\ yields that $\partial\BL/\partial\xn$ is an isomorphism on $\spC^{2,\alpha}_\beta$ onto $Y$,
	if we choose $\alpha\in(0,1)$ and $\beta\in(1,2)$ sufficiently small, depending on the constants $\delta_o$, $C_d$, $\delta_{Lb}$, $\delta_d$
	and $\delta_{Cc}$.
\end{proof}

\begin{proposition}
	\mylabel{prop:pn-uqcont}%
	$r_I$ can be chosen so that $\pn$ is unique and depends continuously on $\lambda$ and on $\po\in\cfusp$ 
	(both $\po$ ad $\pn$ in the $\spC^{2,\alpha}_\beta([0,1]^2)$ topology).
\end{proposition}
\begin{proof}
	We use the subscript $\lambda$ for $\BL$, $\cfusp$ here to indicate their dependence on it.

	Take $r_I:=1$ first.
	$\BL_\lambda(\po,\pn')$ is well-defined for all $\po\in\cfusp_\lambda$ and $\pn'=\po$, as well as sufficiently small perturbations of $\pn'$,  
	by \myeqref{eq:rhoprep} and \myeqref{eq:ellipC}. It is easy to check that
	there is an $r_2>0$, depending continuously on $\po$ and $\lambda$, so that $\pn'\in B_{r_2}(\po)\mapsto\BL_\lambda(\po,\pn')$ is well-defined 
	and $C^1$. Take $r_I\leftarrow\min\{r_I,r_2\}$. This may shrink $\cfusp_\lambda$, but the properties of $\BL$ are not changed.

	Consider a particular $\lambda\in\Lambda$ and a corresponding $\po\in\cfusp_\lambda$.
	By Proposition \myref{prop:Lxn-iso} and the inverse/implicit function theorem for Banach spaces, there is an $r\in(0,r_I]$ so that 
	$\pn'\in B_r(\po)\mapsto\BL_{\lambda'}(\po',\pn')$ is a diffeomorphism
	for every $\po'\in B_r(\po)$ and $\lambda'\in\Lambda\cap B_r(\lambda)$.

	Let $r_3$ be the supremum of all $r$ with this property. $(\po,\lambda)\mapsto r_3(\po,\lambda)$ is continuous: for any $\po''\in B_r(\po)$
	and $\lambda''\in B_r(\lambda)$, set $r'':=r-\max\{|\po-\po''|,|\lambda-\lambda''|\}$. Then
	$B_{r''}(\po'')\subset B_r(\po)$ and $B_{r''}(\lambda'')\subset B_r(\lambda)$, so
	$\pn'\in B_{r''}(\po'')\mapsto\BL_{\lambda'}(\po',\pn')$ is a diffeomorphism for all $\po'\in B_{r''}(\po'')$
	and $\lambda'\in\Lambda\cap B_{r''}(\lambda'')$. Therefore 
	$$r_3(\po'',\lambda'')\geq r_3(\po,\lambda)-\max\{|\po-\po''|,|\lambda-\lambda''|\}.$$

	On the other hand, if $\max\{|\po-\po''|,|\lambda-\lambda''|\}<\frac12r_3(\po,\lambda)$, then $r_3(\po'',\lambda'')\geq r_3(\po,\lambda)/2$,
	so $\po\in B_{r_3(\po'',\lambda'')}(\po'')$
	and $\lambda\in B_{r_3(\po'',\lambda'')}(\lambda'')$, so we may apply the same argument with roles reversed to obtain
	$$r_3(\po,\lambda)\geq r_3(\po'',\lambda'')-\max\{|\po-\po''|,|\lambda-\lambda''|\}.$$
	Clearly $r_3$ is continuous.

	For $r=r_3$, the property need not hold, but we take $r_I\leftarrow\min\{r_I,\frac12r_3\}$. 
	This may shrink $\cfusp$ more, but again the properties of $\BL$ and $r_2$
	above are not changed. With this choice, $\pn$ from Definition \myref{def:fusp} must be unique (determined by \myeqref{eq:oldnew}
	and \myeqref{eq:pn}). It is also clear from the properties above that $(\po,\lambda)\mapsto\pn$ is a continuous (in fact $C^1$) map.
\end{proof}

\begin{proposition}
	\mylabel{prop:fusp-topology}%
	For $\delta_{Tt}$, $\epsilon$ and $r_I$ sufficiently small: 
	For all continuous paths $t\in[0,1]\mapsto\lambda(t)$ in $\Lambda$, 
	$\bigcup_{t\in(0,1)}\big(\{t\}\times\fusp_{\lambda(t)}\big)$ is open and
	$\bigcup_{t\in[0,1]}\big(\{t\}\times\cfusp_{\lambda(t)}\big)$ is closed\footnote{We make no statement about $\cfusp$ being the closure
	of $\fusp$. It certainly contains the closure, but it could be bigger, for example if one of the 
	inequalities in Definition \myref{def:fusp} becomes nonstrict in the interior without being violated.}
	in $[0,1]\times C^{2,\alpha}_\beta([0,1]^2)$.
\end{proposition}
\begin{proof}
	All conditions on $\po$ in Definition \myref{def:fusp} are inequalities which can be made scalar by taking a suitable supremum or infimum.
	Then their sides are continuous under $\spC^{2,\alpha}_\beta([0,1]^2)$ 
	changes to $\po$, and therefore $\pn$. (Most inequalities need only $\spC^1([0,1]^2)$.) 
\begin{enumerate}
\item Closedness:
	consider sequences $(t_n,\po_n)$ in $\bigcup_{t\in[0,1]}\big(\{t\}\times\cfusp_{\lambda(t)}\big)$ that converge to a limit $(t,\po)$.

	Let $\pn_n$ be associated to $\po_n$ as in Definition \myref{def:fusp}. 
	By continuity (Proposition \myref{prop:pn-uqcont}), $(\pn_n)$
	converges to a limit $\pn$ as well. By continuity of $\BL$ in $\po$, $\pn$ and $\lambda$, we have $\BL_{\lambda(t)}(\po,\pn)=0$ as well.

	By \myeqref{eq:sdef} for $\so_n,\po_n$ instead of $\so,\po$, 
	$(\so_n)$ converges in $\spC^{2,\alpha}_\beta[0,1]$ as well, 
	to a limit $\so$ which satisfies \myeqref{eq:sdef} itself.

	Most conditions on $\po$ are nonstrict and continuous inequalities, so they are still satisfied by $\po$.
	We check the strict inequalities explicitly and in order:

	\myeqref{eq:s-welldef}: this is implied by \myeqref{eq:shocknormal} and the nondegeneracy of the $(\sigma,\zeta)\mapsto(\xi,\eta)$
	coordinate change.
	
	\myeqref{eq:rhoprep} is implied by \myeqref{eq:rhomin}.

	\myeqref{eq:tilderho} resp.\ \myeqref{eq:faken} resp.\ \myeqref{eq:tildeL} are implied by \myeqref{eq:oldnew} resp.\ \myeqref{eq:vertvel}
	resp.\ \myeqref{eq:ellipC}, by choosing $r_I$ sufficiently small.

	All inequalities are satisfied, so $\po\in\cfusp$. 

\item Openness:	same proof, using that all inequalities are strict now, by definition of $\fusp$, hence preserved by sufficiently small perturbations.
\end{enumerate}
\end{proof}

\begin{definition}
	\mylabel{def:it}%
	Define 
	$\IT:\cfusp\rightarrow \spC^{2,\alpha}_\beta([0,1]^2)$
	to map $\po$ into $\pn$ as given in Definition \myref{def:fusp}, but pulled back to $(\sigma,\zeta)$ coordinates
	and the $[0,1]^2$ domain (see Definition \myref{def:fusp}) with the coordinate transform defined by $\po$.
\end{definition}

\subsection{Regularity and compactness}

To obtain regularity at the free boundary, we need a kludge: a transformation to a fixed boundary problem. With some further advances in elliptic
theory this step should become obsolete.

\begin{remark}
	\mylabel{rem:mutrans}%
	Using the $(\sigma,\eta)$ coordinates from Definition \myref{def:fusp},
	define the coordinate transformation from $\vec\xi$ to $x=(\sigma,\eta)$ and then to $y=(\sigma,\mu)$ where 
	$$ \mu := \psi(\sigma,\eta) - \psi^I(\sigma,\eta) = \psi(\sigma,\eta) - \psi^I(0,0) - v^y_I\eta. $$
	$\po=\psi^I$ on the shock, so $\mu=0$ there; $\mu>0$ in $\Omega$ (by \myeqref{eq:vertvel}).
	
	We check that this transformation satisfies all conditions of Proposition \myref{prop:ztrans} and Remark \myref{rem:shmor}, where we take $u=\psi$.
	The $\vec\xi$ to $x$ part is trivial, since it is independent of $\psi$. We study the second part:

	$$ y_x = \frac{\partial(\sigma,\mu)}{\partial(\sigma,\eta)}=\begin{bmatrix} 1 & 0 \\ 0 & -v^y_I \end{bmatrix}, \qquad 
		y_u = \frac{\partial(\sigma,\mu)}{\partial\po} = \begin{bmatrix} 0 \\ 1 \end{bmatrix}, \qquad 
		u_x = \begin{bmatrix} \psi_\sigma & \psi_\eta \end{bmatrix}.$$
	Obviously $y_x$ is uniformly invertible, and
	$$ y_x + y_u u_x = \begin{bmatrix} 1 & 0 \\ \psi_\sigma & \psi_\eta - v^y_I \end{bmatrix}$$
	is uniformly invertible by \myeqref{eq:vertvel}. The norms are bounded by constants depending only on $\delta_{vy}$ (to bound $\psi_\eta-v^y_I$ 
	below away from zero) and $C_L$ (to bound $\psi,\nabla\psi$).

	All conditions of Proposition \myref{prop:ztrans} and Remark \myref{rem:shmor} 
	are satisfied, so we obtain the analogues of
	\myeqref{eq:ellipC}, \myeqref{eq:ndobb} and \myeqref{eq:Gb}: the coordinate transformation
	yields a uniformly elliptic equation with uniformly oblique boundary conditions that are uniformly functionally independent
	(as functions of $\nabla\psi$) near the corner. 
	The constant for each property grows at most by a factor depending continuously and only on $\delta_{vy},C_L$.
\end{remark}

\begin{proposition}
	\mylabel{prop:regularity}%
	\mylabel{prop:fp-regularity}%
	For sufficiently small $\beta\in(1,2)$ and $\alpha\in(0,\beta-1)$, depending only on 
	$C_d,\delta_{Lb}\cdot\epsilon,\delta_o,C_L,\delta_{vy}$:
	\begin{enumerate}
	\item
		\begin{alignat}{1}
			\|s\|_{C^{0,1}[\xia_L,\xia_R]} &\leq C_{sL}  \myeqlabel{eq:sLip}
		\end{alignat}
		and
		\begin{alignat}{1}
			\|s\|_{C^{2,\alpha}_\beta[\xia_L,\xia_R]} &\leq C_s \myeqlabel{eq:sregu}
		\end{alignat}
		for $C_{sL}=C_{sL}(C_L,\delta_{vy})$ and $C_s=C_s(C_{\spC},\delta_{vy})$.
	\item
		For a \emph{fixed point} $\po$ of $\IT$:
		\begin{enumerate}
		\item
			\myeqref{eq:lip} is strict for sufficiently large $C_L$.
		\item 
			\myeqref{eq:regularity} is strict
			for sufficiently large $C_{\spC}=C_{\spC}(C_d,\delta_{Lb}\cdot\epsilon,C_L,\delta_o,\delta_{vy},\delta_d)$.
		\item
			For $K\Subset\overline\Omega-\{\vxia_L,\vxia_R\}$ and all $k\geq 0$, $\alpha'\in(0,1)$,
			\begin{alignat}{1}
				\|\po\|_{\spC^{k,\alpha'}(K)} &\leq C_{\spC K}
				\myeqlabel{eq:reguint}
			\end{alignat}
			where $C_{\spC K}=C_{\spC K}(d,C_L,\delta_o,\delta_{vy})$ is decreasing in $d:=d(K,\hat P_L^{(0)}\cup\hat P_R^{(0)})$ 
			and \emph{not} dependent on $\epsilon$.
		\item 
			$\po$ is analytic in $\overline\Omega-\{\vec\xi_L,\vec\xi_R\}$.
		\end{enumerate}
	\item
		For sufficiently small $r_I>0$, depending continuously and only on $\po$, 
		there are $\delta_\alpha,\delta_\beta>0$
		so that for all $\po\in\fusp$, 
		\begin{alignat}{1}
			\|\pn\|_{\spC^{2,\alpha+\delta_\alpha}_{\beta+\delta_\beta}(\overline\Omega)} &\leq C_\IT 
			\myeqlabel{eq:regu2}
		\end{alignat}
		Here, $C_{\IT},\delta_\alpha,\delta_\beta$ depend only on $C_d,\delta_{Lb}\cdot\epsilon,\delta_o,C_L,\delta_{vy}$,
	\end{enumerate}
\end{proposition}
\begin{proof}
\begin{enumerate}
\item
	The shock is implicitly defined by
	$$\po(\xi,s(\xi))-\psi^I(\xi,s(\xi))=0\qquad(\xi\in[\xi_L,\xi_R])$$
	(compare Definition \myref{def:fusp}, \myeqref{eq:sdef}).
	The derivative with respect to $s$ of the left-hand side is $\po_\eta-v^y_I$. \myeqref{eq:vertvel} bounds this away from zero.
	The implicit function theorem yields the derivative part of \myeqref{eq:sLip};
	bounds on $s$ itself are supplied by integration and the $\vxit_R$ part of \myeqref{eq:cornerregion} (for example).
	The implicit function theorem also yields \myeqref{eq:sregu} from \myeqref{eq:regularity}. 
	Henceforth we use without further mention that estimates on $\xo$ automatically yield corresponding estimates on $s$.
\item
	Now consider a fixed point $\po=\IT(\po)$. 
\begin{enumerate}
\item
	\myeqref{eq:ellipC} implies $L\leq 1$ which can be rewritten 
	\begin{alignat}{1}
		|\nabla\xo|^2\leq\frac{2(c_0^2+(1-\gamma)\xo)}{\gamma+1}. \myeqlabel{eq:lipode}
	\end{alignat}
	$\xo(\vec\xi_R)=\chi^I(\vec\xi_R)$ which is bounded by \myeqref{eq:cornerregion}.
	Using the differential inequality \myeqref{eq:lipode} vertically downwards from $\vec\xi_R$ to $A$, then along $A$, 
	and finally along any vertical line upwards to $S$, we achieve uniform bounds on $\xo$ and $|\nabla\xo|$ in
	all of $\overline\Omega$.
	This is \myeqref{eq:lip} which is strict if we take $C_L$ large enough.
\item
	By \myeqref{eq:ellipC} the equation
	$$(c^2I-\nabla\xo^2):\nabla^2\po=0$$
	is uniformly elliptic. Standard de Giorgi-Nash-Moser and then Schauder theory converts \myeqref{eq:lip} into
	interior $\spC^{k,\alpha}$ estimates for any $k\geq 2$ and $\alpha\in(0,1)$. 

	At the shock and near the shock-parabolic corners we first apply the coordinate transform from Remark \myref{rem:mutrans}. 
	This is necessary because we have only a $C^{0,1}$ bound of $S$ at this point; using $C^{1,\beta}$ regularity would cause circularity. 
	The transform yields a new problem with fixed boundary (both arc and shock are mapped into straight line segments).
	The uniform ellipticity \myeqref{eq:ellipC}, the uniform obliqueness of the boundary 
	conditions \myeqref{eq:ndobb}, and their uniform functional independence in the corner \myeqref{eq:Gb} are preserved by the 
	transform. 
	
	It is also crucial that the boundary conditions themselves are smooth. For this purpose we have combined\footnote{If we had
		used \myeqref{eq:momjump} by itself, it would contain $\vec n$ for which we have only an $L^\infty$ bound at this point.} 
	the two shock conditions
	\myeqref{eq:chijump} and \myeqref{eq:momjump} so that the boundary operator in \myeqref{eq:itn-shock} is smooth in $\vec\xi$,
	$\xn(\xi)$ and $\nabla\xn(\xi)$. This also requires a bound of $\rho$ below away from zero, by \myeqref{eq:rhomin},
	as well as a bound of $\nabla\po$ away from $\vec v_I$, by \myeqref{eq:vertvel}.

	At the boundaries away from the corners, we use \cite{lieberman-degiorgi-nonlinearoblique}, 
	which yields $C^{2,\alpha}$ regularity for any 
	$\alpha\in(0,1)$.
	Near the corners we apply Proposition \myref{prop:corner}. 
	It yields $C^{2,\alpha}_\beta$ 
	regularity for some range of $\alpha\in(0,1)$ and $\beta\in(1,2)$; we fix $\alpha,\beta$ below.

	We obtain \myeqref{eq:regularity}, with 
	$C_{\spC}=C_{\spC}(C_d,\delta_{Lb}\cdot\epsilon,C_L,\delta_o,\delta_{vy},\delta_d)$. 
	$\delta_{vy}$ is due to the transform from Remark \myref{rem:mutrans}.
	Moreover $C_s=C_s(C_{\spC},\delta_{vy})$, as discussed above.
\item
	In every $K\Subset\Omega$ with positive distance to $P_L$ and $P_R$, 
	we use that \myeqref{eq:ellip} provides
	$$\sup_KL^2\leq 1-\delta_{Lb}\sup_Kb$$
	where the right-hand side is $<1$ and \emph{independent} of $\epsilon$ (by Definition \myref{def:b}). 
	Thus the equation is uniformly elliptic in $K$, with ellipticity constant depending on $d(K,\overline P_L^{(0)}\cup\overline P_R^{(0)})$,
	but independent of $\epsilon$.
	Since the obliqueness constant $\delta_o$ in \myeqref{eq:ndobb} was already independent of $\epsilon$, the same arguments as before yield
	uniform in $\epsilon$ regularity in each $K$, for any H\"older norm. This is \myeqref{eq:reguint}. 
\item 
	For elliptic problems with analytic cofficients, analyticity of the solution is classical (e.g.\ \cite[Theorem 6.7.6']{morrey-book}) in the
	case of fixed boundaries. To deal with a free boundary, we use the transform from Remark \myref{rem:mutrans} again.
	Loc.cit.\ yields analyticity for the new problem. The inverse coordinate transformation is defined in terms of the new
	coordinates and the solution of the new problem, so it is analytic as well.
	Then $\psi$ itself is analytic in $\overline\Omega-\{\vec\xi_L,\vec\xi_R\}$.
\end{enumerate}
\item
	For a general $\po\in\cfusp$, not necessarily a fixed point of $\IT$, 
	$\po$ and $\pn$ in $\BL$ are different. 
	The boundary conditions are oblique derivative\footnote{
		For a \emph{Dirichlet} boundary condition, $C^{1,\beta}$ data would yield $C^{1,\beta}$, but not
		$C^{1,\beta+\delta_\beta}$ regularity. But for a \emph{derivative} condition we can gain one order.}
	conditions:
	$$g^k(\vec\xi,\pn(\vec\xi),\nabla\pn(\vec\xi)) = 0.$$
	Each $g^k$ is $\spC^{2,\alpha}_\beta$ in $\vec\xi$ (from\footnote{Here it is 
		crucial that $\nabla\po$ does not appear \emph{anywhere} in \myeqref{eq:itn-parabolic}, 
		\myeqref{eq:itn-shock}, \myeqref{eq:itn-wall}; otherwise $g^k$ would only be $C^{1,\alpha}_\beta$ in $\vec\xi$ which is not enough 
		to gain regularity.}
	$\xo$ in \myeqref{eq:itn-parabolic}), in particular $\spC^{1,\beta-1}$.
	
	Moreover each $g^k$ is $C^\infty$ in $\pn(\vec\xi)$ and $\nabla\pn(\vec\xi)$. For \myeqref{eq:itn-shock} 
	this requires that $\hat\rho$ is uniformly bounded below away from $0$,
	and $\nabla\pn$ uniformly bounded away from $\vec v_I$. This is guaranteed by \myeqref{eq:rhomin} and \myeqref{eq:vertvel}
	(bounding $\rho$ resp.\ $\nabla\po$) combined with \myeqref{eq:oldnew} (bounding $|\hat\rho-\rho|$ resp.\ $|\nabla\po-\nabla\pn|$), 
	if $r_I$ is chosen small enough (depending only and continuously on $\po$ and $\delta_{vy}$).

	The boundaries\footnote{Here we can apply Proposition \myref{prop:corner} without the 
		coordinate transform from Remark \myref{rem:mutrans}, because the shock is already known to be $C^{1,\beta}$.} 
	$\Gamma^k$ are $A,P_L,P_R$, which are perfectly smooth, and $S$,
	has a $C^{2,\alpha}_\beta$ bound \myeqref{eq:sregu}, implying $C^{1,\beta}$. 

	Uniform ellipticity, obliqueness and functional independence still hold by \myeqref{eq:ellipC}, \myeqref{eq:ndobb} and \myeqref{eq:Gb},
	combined with \eqref{eq:oldnew}, for $r_I$ sufficiently small(er).
	
	All combined, Proposition \myref{prop:corner} 
	and \cite[Corollary 1.4]{lieberman-pacj-1988} (away from the corners) yield
	$\spC^{2,\kappa}_\lambda$ regularity.
	Here, $\kappa$ depends on $\beta$ only whereas $\lambda$ does not depend on $\alpha,\beta$ at all.
	Therefore we may pick $\beta=1+\lambda/2$ and then $\alpha=\kappa/2$ for the resulting $\kappa$.
	With $\delta_\beta=\lambda/2$ and $\delta_\alpha=\kappa/2$, \myeqref{eq:regu2} is satisfied for suitable $C_\IT$.
\end{enumerate}
\end{proof}

\begin{proposition}
	\mylabel{prop:it-continuous-compact}%
	For sufficiently small $r_I>0$ (depending continuously and only on $\po,\delta_{vy}$),
	$\IT$ is a continuous and compact function of $\po$ and $\lambda$
	on $\bigcup_{\lambda\in\Lambda}\cfusp_\lambda$.
\end{proposition}
\begin{proof}
	Continuity: we have already shown in Proposition \myref{prop:Lxn-iso} that $\pn$ is a continuous function
	of $\po$. Therefore $\IT$ depends continuously on $\lambda$ and $\po$.

	Moreover $\IT$ is compact: by \myeqref{eq:regu2} in Proposition \myref{prop:fp-regularity} 
	the range of $\pn$ is a bounded subset of $\spC^{2,\alpha+\delta_\alpha}_{\beta+\delta_\beta}(\overline\Omega)$ 
	which
	is pre-compact in $\spC^{2,\alpha}_\beta(\overline\Omega)$.
	Pullback to $\sigma,\zeta$ coordinates is continuous in the latter topology, so the image under it is still 
	\footnote{We may actually lose regularity because the pullback is defined by $\po$, hence only $C^{2,\alpha}_\beta$, but
	we do retain compactness.} pre-compact.
	Altogether $\IT$ is a compact map.
\end{proof}

\subsection{Pseudo-Mach number control}

\mylabel{section:L-control}

\begin{proposition}
	\mylabel{prop:Lbounds}%
	For $\epsilon$ and $\delta_{Lb}$ sufficiently small, with bounds depending
	only on $\delta_{\rho}$:
	if $\po\in\cfusp$ is a fixed point of $\IT$, 
	then \myeqref{eq:ellip} is strict and
	\begin{alignat}{1}
		L^2 &< 1-\epsilon \qquad\text{in $\overline\Omega-\overline P_L-\overline P_R$.} \myeqlabel{eq:Leps}
	\end{alignat}
\end{proposition}
\begin{proof}
	$$d(\overline\Omega,\hat P_L^{(0)}\cup\hat P_R^{(0)})\geq\frac13\cdot\epsilon,$$
	for $\epsilon$ small enough.
	Remember from Definition \myref{def:b} that $b=0$ on $\hat P_L^{(0)}\cup\hat P_R^{(0)}$ . Therefore, on $P_L\cup P_R$:
	$$L^2 = 1-\epsilon < 1-\subeq{\|b\|_{C^{0,1}}}{\leq 1}\cdot d(P_L\cup P_R,\hat P_L^{(0)}\cup\hat P_R^{(0)}) \leq 1-\delta_{Lb}\cdot b,$$
	e.g.\ for $\delta_{Lb}\leq1$.

	On the shock, we may use \myeqref{eq:rhomin} combined with Proposition \myref{prop:L-minmax}
	to rule out that $L^2+\delta_{Lb}\cdot b$ has a maximum in a point where $L<1$ and $L\geq 1-\delta_{LS}$. 
	Here $\delta_{Lb}$ has to be chosen so that $|\delta_{Lb}\nabla b|\leq\delta_{LS}$ is satisfied.
	(Now $\delta_{Lb}$ depends continuously on $\delta_\rho$ as well.)
	Then $L^2+\delta_{Lb}\cdot b<1$ as well if, again, $\delta_{Lb}$ is small enough.

	In addition we can choose $\delta_{Lb}$ so small that $\delta_{Lb}\cdot b$ satisfies the preconditions of 
	Theorem 1 and Theorem 2 in \cite{elling-liu-ellipticity-journal} (where it is called $b$). Note that $b$ is even in $\eta$
	by Definition \ref{def:b},
	so $\partial b/\partial n=0$ on the wall.
	Let $\delta_{L\Omega}$ be the $\delta$
	from those theorems (it depends only and continuously on $\lambda$). 
	Then $L^2+\delta_{Lb}\cdot b$ cannot have a maximum in a point where $L^2\geq 1-\delta_{L\Omega}$.
	If, again, $\delta_{Lb}$ is chosen sufficiently small (no new dependencies), then $L^2+\delta_{Lb}\cdot b<1$ in $\Omega\cup A$,
	hence in $\overline\Omega$. Therefore \myeqref{eq:ellip} is strict.

	\myeqref{eq:Leps} can be shown in the same manner, by taking $b=0$ instead, using the actual boundary
	condition $L=\sqrt{1-\epsilon}$ on $P_L,P_R$ and and considering $\epsilon<\delta_{LS},\delta_{L\Omega}$.
\end{proof}

\subsection{Second derivatives on arcs}

\mylabel{section:parcs}
\mylabel{section:c-pararc}%

\if\dofigures%
\begin{figure}
	\center{\input{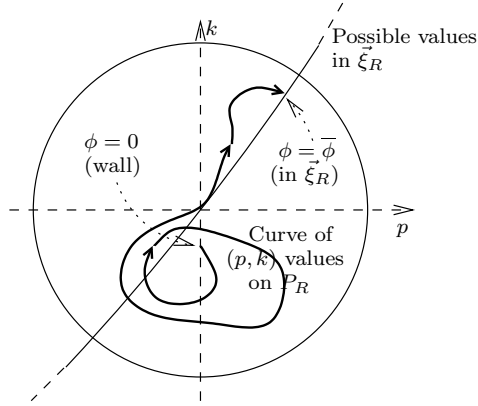}}
	\caption{%
	$\chi$, $\nabla\chi$ on $P_R$
	define a path $\phi\mapsto(p(\phi),k(\phi))$ (bold curve). 
	The path is constrained by \myeqref{eq:kphi} and \myeqref{eq:pphi}.
	The possible $k,p$ in $\vec\xi_R$ are on the thin solid curve which is parametrized by $\etaa_R$;
	one endpoint of the path is on this curve.
	The other endpoint is on the $k$ axis.}
	\mylabel{fig:pararc}
\end{figure}
\fi

Consider a fixed point $\chi\in\cfusp$ of $\IT$, so that 
$L^2=1-\epsilon$ is satisfied on $P_L,P_R$. This can be restated as
\begin{alignat}{1}
	\frac{1}{2}|\nabla\chi|^2 + \frac{(1-\epsilon)\big((\gamma-1)\chi-c_0^2\big)}{\gamma+1-\epsilon(\gamma-1)} &= 0.
	\myeqlabel{eq:Lsimple}
\end{alignat}
In what follows, $\vec n$ will be an outer normal to $P_L$ or $P_R$ and $\vec t=\vec n^\perp$; 
$\partial_n$ and $\partial_t$ are defined accordingly. When considering a particular point on an arc we also use
$\tau$ and $\nu$, which are Cartesian coordinates such that $\partial_\tau=\partial_t$ and $\partial_\nu=\partial_n$
in that point (only).

Take $\partial_t$ of \myeqref{eq:Lsimple} 
(using $\partial_t\partial_\tau=\partial^2_\tau$ and $\partial_t\partial_\nu=\partial_\tau\partial_\nu$):
\begin{alignat}{1}
	0 &= \chi_\tau\chi_{\tau\tau} + \chi_\nu\chi_{\nu\tau} 
	+ \frac{(\gamma-1)(1-\epsilon)}{\gamma+1-\epsilon(\gamma-1)}\chi_\tau \notag\\
	&= \chi_\tau\psi_{\tau\tau} + \chi_\nu\psi_{\nu\tau}
	- \frac{2}{\gamma+1-\epsilon(\gamma-1)}\chi_\tau. \myeqlabel{eq:Ltchi}
\end{alignat}

We have two equations (\myeqref{eq:Ltchi} and
the interior equation \myeqref{eq:chi}) for the three components of $D^2\chi$.
That is the case for other equation and boundary conditions as well; it is 
not really sufficient for any control. 
But here there is an additional tool:
\myeqref{eq:Leps} implies that $L$ attains its maximum in $\overline\Omega$ in \emph{every} point on $P_L\cup P_R$. Therefore
$(L^2)_\nu\geq 0$ on $P_L\cup P_R$, or equivalently
\begin{alignat}{1}
	\chi_\tau\psi_{\tau\nu} + \chi_\nu\psi_{\nu\nu} - \frac{2}{\gamma+1-\epsilon(\gamma-1)}\chi_\nu &\geq 0.
	\myeqlabel{eq:Lnchi}
\end{alignat}

From here on --- and in the next few sections --- we focus on $P_R$; all results have an analogous extension to $P_L$.

Combine \myeqref{eq:Lnchi}, \myeqref{eq:Ltchi}, \myeqref{eq:psi}:
\begin{alignat}{1}
	\begin{bmatrix}
		\chi_\tau & \chi_\nu & 0 \\
		0 & \chi_\tau & \chi_\nu \\
		c^2-\chi_\tau^2 & -2\chi_\tau\chi_\nu & c^2-\chi_\nu^2
	\end{bmatrix}
	\begin{bmatrix}
		\psi_{\tau\tau} \\ \psi_{\nu\tau} \\ \psi_{\nu\nu}
	\end{bmatrix}
	&=
	\frac{2}{\gamma+1-\epsilon(\gamma-1)}
	\begin{bmatrix}
		\chi_\tau \\ \chi_\nu \\ 0
	\end{bmatrix}
	+\begin{bmatrix}
		0 \\ a \\ 0
	\end{bmatrix}
\end{alignat}
where we use a ``slack variable'' $a\geq 0$.
Solutions (we use $\chi_\nu^2=(1-\epsilon)c^2-\chi_\tau^2$):
\begin{alignat}{1}
	\psi_{\tau\tau} &= 
	\frac{2}{\gamma+1-\epsilon(\gamma-1)} \Big(
		\frac{1+\epsilon}{1-\epsilon}\cdot\frac{\chi_\tau^2}{c^2}-\epsilon\Big)
		~\subeq{ -\frac{(c^2-\chi_\nu^2)\chi_\nu}{(1-\epsilon)c^4}\cdot a}{\geq 0}, 
	\myeqlabel{eq:psitautau}
\end{alignat}
The $a$ term is positive due to $\chi_n<0$ (by \myeqref{eq:parnor}).
\cucon{remarkable here: can sub by $\epsilon c^2-\chi_t^2$ in $a$ coefficient, so even if $a$ is
some bounded but \emph{nonzero} function, \emph{still} get tight control}

Now consider radial coordinates, centered in the origin, with $\phi=0$ corresponding to the positive $\xi$ axis. $P_R$ is at a fixed radius
$r=(1-\epsilon)^{1/2}c_R$ and covers $\phi\in[0,\overline\phi]$. In these coordinates:
\begin{alignat}{1}
	\chi_{\phi\phi} 
	&= r^2(\chi_{\tau\tau}-r^{-1}\chi_\nu) 
	= r^2(\psi_{\tau\tau}-1-r^{-1}\chi_\nu) \geq f(c^2,\chi_\phi), \notag\\
	f(h,p) &:= \frac{2}{\gamma+1-\epsilon(\gamma-1)}\big(
		\frac{1+\epsilon}{1-\epsilon}\cdot\frac{p^2}{h}-\epsilon r^2\big) 
	-r^2+\sqrt{r^2h(1-\epsilon)-p^2}. \myeqlabel{eq:refp}
\end{alignat}
On the other hand, $c^2=c_0^2+(1-\gamma)(\chi+|\nabla\chi|^2/2)$ and \myeqref{eq:Lsimple} yield
\begin{alignat}{1}
	(c^2)_\phi &= g(c^2,\chi_\phi), \qquad
	g(h,p) := \subeq{\frac{-2(\gamma-1)}{\gamma+1-\epsilon(\gamma-1)}}{=:\sigma_g}p. \myeqlabel{eq:cc-arc}
\end{alignat}
We seek stationary points of the ODE system $(p_\phi,h_\phi)=(f,g)$.
$g(h_0,p_0)=0$ obviously requires $p_0=0$.
\begin{alignat}{1}
	0 &= f(h_0,0) = -r^2\left(1+\frac{2\epsilon}{\gamma+1-\epsilon(\gamma-1)}\right)
	+r\sqrt{h_0}\sqrt{1-\epsilon} \notag\\
	\Rightarrow\qquad h_0 &= 
	\left(1+\frac{2\epsilon}{\gamma+1-\epsilon(\gamma-1)}\right)^2\frac{r^2}{1-\epsilon}
	\myeqlabel{eq:h0}
\end{alignat}
$g$ is already linear; we linearize $f$ around the stationary point:
\begin{alignat}{1}
	\partial_hf(h_0,0) 
	&= r\frac{\sqrt{1-\epsilon}}{2\sqrt{h_0}} 
	= \frac{1-\epsilon}{2(1+\frac{2\epsilon}{\gamma+1-\epsilon(\gamma-1)})} =: \sigma_f\notag
\end{alignat}

For easier treatment of the isentropic case, we change coordinates again (see Figure \myref{fig:pararc}): take
\begin{alignat}{1}
	k &:=(-\sigma_f/\sigma_g)^{1/2}(h-h_0).\myeqlabel{eq:refk}
\end{alignat}
From linearization we have
\begin{alignat}{1}
	p_\phi &\geq \sigma_f(h-h_0) + O((h-h_0)^2+p^2) = \subeq{(-\sigma_f\sigma_g)^{1/2}}{=:\sigma_\theta}\cdot k + O(k^2+p^2), \myeqlabel{eq:pphi} \\
	k_\phi &= (-\sigma_f/\sigma_g)^{1/2}\sigma_g p = -(-\sigma_f\sigma_g)^{1/2}p = -\sigma_\theta p. \myeqlabel{eq:kphi}
\end{alignat}

Here and later, $O(x^\alpha)$ is a term with absolute value $\leq Cx^\alpha$, for $|x|\leq R$, where $C$ and $R$ may
depend only on $C_L$, but none of the other constants.
While the elliptic equation degenerates as $\epsilon\downarrow 0$, all ODE we discuss here
are well-behaved for $\epsilon\downarrow0$. However, some $\gamma\downarrow 1$ are delicate, because $\sigma_g=\gamma-1+O(\epsilon)$,
so \myeqref{eq:refk} has a singularity; we make detailed comments in each case.

Represent $(p,k)$ by radial coordinates $(q,\theta)$, with $\theta=0$ corresponding to the positive $p$ axis and $\theta=\frac\pi2$ to
the positive $k$ axis. Then
\begin{alignat}{1}
	q_\phi 
	&= \partial_\phi\sqrt{p^2+k^2}
	= \frac{pp_\phi+kk_\phi}{q} 
	= \begin{cases}
		\geq O(q^2), & p \geq 0, \\
		\leq O(q^2), & p \leq 0.
	\end{cases} \myeqlabel{eq:qphi}
\end{alignat}
\begin{alignat}{1}
	\theta_\phi 
	&= \frac{pk_\phi-kp_\phi}{q^2} 
	= \begin{cases}
		\leq O(q)-\sigma_\theta , & k \geq 0, \\
		\geq O(q)-\sigma_\theta , & k \leq 0.
	\end{cases} \myeqlabel{eq:thetaphi}
\end{alignat}
\cucon{$|k|/q$ is uniform as $\gamma\downarrow 1$, so the $O$ terms are safe.}

\subsection{Arc control}

\begin{proposition}
	\mylabel{prop:pararc}%
	If $C_{Pt}<\infty$ is sufficiently large, if $\delta_{Pn}>0$ is sufficiently small,
	if $\epsilon$ is sufficiently small and $C_{Pv},C_{P\rho}$ sufficiently large,
	with bounds depending on $C_{Pt}$, 
	then for any fixed point $\chi$ of $\IT$, \myeqref{eq:partan} and \myeqref{eq:parnor}
	are strict, and
	\begin{alignat}{1}
		|\rho-\rho_C| &\leq C_{P\rho}\epsilon^{1/2} \qquad\text{and}  \myeqlabel{eq:rhoP} \\
		|\vec v-\vec v_C| & \leq C_{Pv}\epsilon^{1/2} \qquad\text{on $P_C$\quad($C=L,R$),}  \myeqlabel{eq:vP}
	\end{alignat}
	and in the case $\gamma>1$: 
	\begin{alignat}{1}
		\theta(\phi) &\not\in (\frac\pi2,\frac{3\pi}2-\sigma_\theta\overline\phi)+2\pi\Z \qquad\text{on $P_R$,} \myeqlabel{eq:thetasector} \\
		\theta(\phi) &\not\in (-\frac{\pi}2+\sigma_\theta\overline\phi,\frac\pi2)+2\pi\Z \qquad\text{on $P_L$}\notag
	\end{alignat}
	(if $q(\phi)\neq 0$).
\end{proposition}
\begin{proof}
	We focus on $P_R$ first.

	For $\vec\xi_R=\vec\xi_R^*$ we have $p(\overline\phi)=0$ and $c(\overline\phi)=c_R$, 
	by construction of the $R$ shock in Section \myref{section:parmset}.
	$p(\overline\phi),c(\overline\phi)$ depend smoothly on $\vec\xi_R$, so \myeqref{eq:cornerregion} yields
	\begin{alignat}{1}
		|p(\overline\phi)| &\leq C_{pk}\epsilon^{1/2}, \myeqlabel{eq:pcontrol}\\
		|c(\overline\phi)-c_R| &\leq C_{pk}\epsilon^{1/2} \myeqlabel{eq:ccontrol}
	\end{alignat}
	in $\vec\xi_R$, for some constant $C_{pk}$.
	$L^2=1-\epsilon$, combined with \myeqref{eq:partan}, as well as \myeqref{eq:partan} integrated in tangential direction, then implies
	\begin{alignat}{1}
		q(\phi)\leq C_q\epsilon^{1/2}\qquad\forall\phi\in[0,\overline\phi]. \myeqlabel{eq:qcontrol}
	\end{alignat}
	for $C_q=C_q(C_{Pt})$.
	($C^q$ may be $\geq C_{Pt}$, so this does not imply the sharp form of \myeqref{eq:partan} yet.)
\paragraph{Isothermal case}
	For $\gamma=1$ we need to consider only a single differential inequality, since $c$ is constant. 
	\myeqref{eq:refp} takes the form
	\begin{alignat}{1}
		p_\phi &\geq -(1-\epsilon)c^2\cdot\epsilon + O(p^2). \myeqlabel{eq:pphi-isen}
	\end{alignat}
	$\phi=0$ is the corner between $A,P_R$ where $\chi_\phi=r\chi_\eta=0$ by the boundary condition on $A$, so $p(\phi=0)=0$. 
	Integrating \myeqref{eq:pphi-isen} from $0$ to $\phi$ yields
	\begin{alignat}{1}	
		p(\phi) 
		&\geq O(\epsilon).
		\myeqlabel{eq:plower}
	\end{alignat}

	On the other hand $p(\overline\phi)$ is controlled by \myeqref{eq:pcontrol}.
	Integrating \myeqref{eq:pphi-isen} from $\overline\phi$ to $\phi$ yields
	\begin{alignat}{1}
		p(\phi) 
		&\leq O(\epsilon^{1/2}).
		\myeqlabel{eq:pupper}
	\end{alignat}
	\myeqref{eq:plower} and \myeqref{eq:pupper} combine to
	\begin{alignat}{1}
		c^{-1}r^{-1}\max_{\phi\in[0,\overline\phi]}p(\phi) 
			&< C_{Pt}\epsilon^{1/2}\notag
	\end{alignat}
	for $\epsilon$ sufficiently small and $C_{Pt}$ sufficiently large. This is the strict form of \myeqref{eq:partan}.
\paragraph{Isentropic case}
	For $\gamma>1$: first we show \myeqref{eq:thetasector}. 
	We fix $\theta\in[0,2\pi)$ here.

	Assume that $q(\phi')>0$ and 
	$\theta(\phi')\in(\pi,\frac{3\pi}2-\sigma_\theta\overline\phi)$ for some $\phi'\in[0,\overline\phi]$.
	Necessarily $\phi'>0$ because $p(\phi=0)=0$, so either $q(\phi=0)=0$ or $\theta(\phi=0)=\pi\pm\frac\pi2$.

	Let $\phi_0\in[0,\phi')$ be maximal so that
	$$\theta(\phi_0)\not\in(\pi,\frac{3\pi}2)\quad\text{or}\quad q(\phi_0)=0$$ 
	Such a $\phi_0$ must exist because $p(\phi=0)=0$.
	For $\phi\in(\phi_0,\phi')$, $k(\phi)<0$ and $p(\phi)<0$. 

	By the $p\leq 0$ part of \myeqref{eq:qphi} (in reverse direction), $q(\phi)>0$ for $\phi\in[\phi_0,\phi']$. 
	On the other hand \myeqref{eq:qcontrol} applies. Therefore $q(\phi_0)=0$ is not possible,
	so either $\theta(\phi_0)=\pi$ or $\theta(\phi_0)=\frac{3\pi}{2}$.

	\myeqref{eq:kphi} with $p<0$ (in reverse direction) shows that $k(\phi_0)\leq k(\phi')<0$, so $\theta(\phi_0)=\pi$ is not possible.

	For $\epsilon$ sufficiently small, with bound depending on $C_q$:
	the $k\leq 0$ part of \myeqref{eq:thetaphi} yields
	$$\theta(\phi_0)
	\leq\theta(\phi')+\sigma_\theta\subeq{(\phi'-\phi_0)}{\leq\overline\phi}+O(\epsilon^{1/2}),$$
	so since $\overline\phi<\frac\pi2$ and $\sigma_\theta<1$ (for $\epsilon$ small),
	$$\theta(\phi_0)\in(\pi,\frac{3\pi}2)$$
	Contradiction! So $\phi'$ cannot exist; $\theta(\phi)\not\in(\pi,\frac{3\pi}{2}-\sigma_\theta\overline\phi)$ for any
	$\phi\in[0,\overline\phi]$.
\if\dofigures%
\begin{figure}
\center{\input{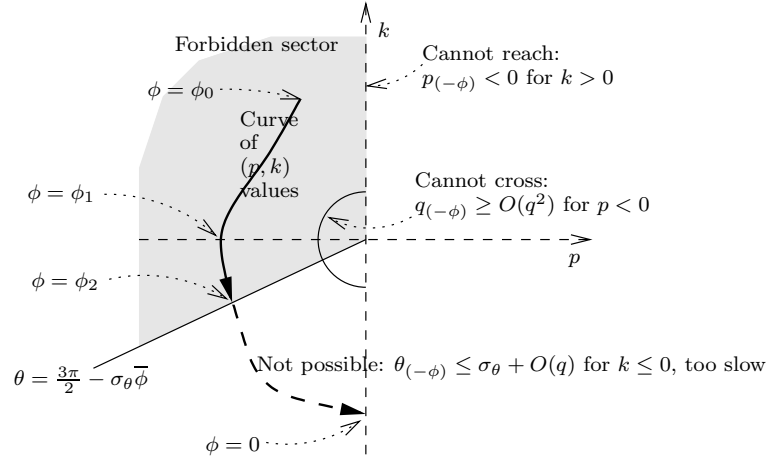}}
\caption{The second derivative inequality \myeqref{eq:thetaphi}, combined with the wall condition $p=0$, rules out a sector of $\chi_\phi,c^2$ values.}
\mylabel{fig:thetasector}
\end{figure}
\fi%
	
	Now assume $q(\phi_0)\neq 0$ and $\theta(\phi_0)\in(\frac\pi2,\pi]$ for some $\phi_0$
	(see Figure \myref{fig:thetasector}). Let $\phi_1\in[0,\phi_0)$ be maximal so that 
	$q(\phi_1)=0$ or $\theta(\phi_1)\not\in(\frac\pi2,\pi]$. 
	Again, $p(\phi=0)=0$, so such a $\phi_1$ must exist. 
	For $\phi\in(\phi_1,\phi_0]$, $k(\phi)\geq 0>p(\phi)$. 

	The $p<0$ part of \myeqref{eq:qphi} (in reverse direction) shows that $q(\phi)>0$ for all $\phi\in[\phi_1,\phi_0]$, so $q(\phi_1)=0$ is not possible.

	For sufficiently small $\epsilon$, using \myeqref{eq:qcontrol}, \myeqref{eq:pphi} yields $p_\phi(\phi)>0$ for those $\phi$.
	Thus $p(\phi_1)<p(\phi_0)$, so $\theta(\phi_1)=\frac\pi2$ is not possible either.

	Only $\theta(\phi_1)=\pi$ remains: but then $p(\phi_1)<0=k(\phi_1)$, so \myeqref{eq:kphi} yields $k_\phi(\phi_1)>0$.
	Therefore $\theta(\phi_1-\delta_1)=\pi+\delta_2$ for some small $\delta_1\in(0,\phi_1)$ and $\delta_2>0$. 
	This is in the sector we have already ruled out --- contradiction. The proof of \myeqref{eq:thetasector} is complete.

	Define
	$$Q:=\begin{cases}
		q, & p \geq 0, \\
		|k|, & p \leq 0.
	\end{cases}$$
	By \myeqref{eq:thetasector}, for $p<0$ necessarily $k\leq p\cot(\sigma_\theta\overline\phi)<0$, so
	$$Q_{(-\phi)}=-k_{(-\phi)}\overset{\myeqref{eq:kphi}}{=}-\sigma_\theta p\leq-\sigma_\theta k\tan(\sigma_\theta\overline\phi)=\sigma_\theta\tan(\sigma_\theta\overline\phi)Q.$$
	For $p\geq 0$, \myeqref{eq:qphi} yields $Q_{(-\phi)}\leq O(Q^2)$. Altogether $Q_{(-\phi)}\leq O(Q)$ (where $O$ is with respect
	to $|Q|\downarrow 0$), so integrating
	from $\overline\phi$ backwards yields 
	\begin{alignat}{1}
		Q(\phi) &\leq O(Q(\overline\phi)) \qquad(\phi\in[0,\overline\phi]).\notag
	\end{alignat}
	Using $|p|\leq|k|\tan(\sigma_\theta\overline\phi)$ for $p\leq 0$ again, we get
	\begin{alignat}{1}
		c^{-1}r^{-1}q(\phi) &\leq c^{-1}r^{-1}O(q(\overline\phi)) < C_{Pt}\epsilon^{1/2} \qquad(\phi\in[0,\overline\phi]),\notag
	\end{alignat}
	for sufficiently large $C_{Pt}$, and for $\epsilon$ sufficiently small with bound
	depending on $C_{Pt}$ only. This implies the strict form of \myeqref{eq:partan}.

	The strict form of \myeqref{eq:parnor} is immediate from $L^2=1-\epsilon$, in the form 
	$$\chi_n^2=-\sqrt{(1-\epsilon)c^2-\chi_t^2}$$
	(note that \myeqref{eq:parnor} (nonstrict) fixes the sign). Here $\delta_{Pn}$ is sufficiently small.
\paragraph{Density, velocity}
	Finally, we obtain \myeqref{eq:rhoP}: \myeqref{eq:cornerregion} combined with $\chi=\chi^I$ on the shock
	yields 
	$$|\chi(\vxia_R)-\chi^I(\vxit_R)|\leq C\epsilon^{1/2}$$
	for some constant $C$. Integrating \myeqref{eq:partan} along $P_R$
	we obtain
	$$\max_{\overline{P_R}}|\chi-\chi^I(\vxit_R)|\leq C'\epsilon^{1/2}$$
	for some other constant $C'$. Combined with \myeqref{eq:partan} and $L^2=1-\epsilon$ this implies \myeqref{eq:rhoP},
	for $C,C',C_{P\rho}$ depending only on $C_{Pt}$.

	\myeqref{eq:vP} is shown in the same manner.
\paragraph{Left arc}
	$P_L$ can be discussed in the same fashion, by noticing that in the $L$ picture (Figure \myref{fig:frameL}) it has
	the same properties as $P_R$, except for the wall not being horizontal which does not matter.
	Note that the mirror reflection in going back to the $R$ picture reverses the direction of $\phi$ and therefore
	changes $\theta$ to $-\theta$ (see \myeqref{eq:thetasector}).
\end{proof}

\subsection{Corners moving along arcs}

\mylabel{section:cornersmoving}

Since our shock is a free boundary, we cannot be sure where the shock-arc corner is located.
We study the behaviour of shock normal and downstream data when keeping the upstream data $\rho_u,\vec v_u$ fixed, 
imposing $L_d=\sqrt{1-\epsilon}$ and restricting the shock location to be on $P_R$.
(This is different from Proposition \myref{prop:vdzero}, where we imposed $v^y_d=0$ instead of the location.)

All the calculations in this section are done for the corner between $S$ and $P_R$, but each
has an analogous result for the left corner, using $L$ coordinates (Figure \myref{fig:frameL}).

We abbreviate $\omega:=\etaa_R$.
In this section $\partial_\omega$ refers to derivatives of upstream and downstream quantities as $\omega$ is varied, while
keeping $\vxia_R\in P_R$ and maintaining the shock conditions and the parabolic boundary condition.

We use dot notation $\pteq$ etc.\ on relations that hold only for $\omega=\etat_R$. No $\partial_\omega$ may be taken of such relations.
	
In addition we use the notation from Section \myref{section:shocks}; $\vec v_u=\vec v_I$, $\vec z=\nabla\chi$, $\vec v=\nabla\psi$ etc.

We required $\vxia_R\in\hat P^{(0)}_R$, so
\begin{alignat}{1}
	\xi &= \sqrt{(1-\epsilon)c_R^2-\eta^2}.\notag
\end{alignat}
Moreover
\begin{alignat}{1}
	\vec n &\pteq (0,-1), \quad\vec t\pteq(1,0) \notag\\
	z^x_u &=-\xi,\quad z^y_u = v^y_u-\eta, \quad z^n_u \pteq \eta-v^y_u, \quad z^t_u \pteq -\xi \notag\\
	z^x_d &\pteq -\xi, \quad z^y_d \pteq -\eta, \quad z^n_d \pteq \eta,\quad z^t_d \pteq -\xi \notag
\end{alignat}
Thus:
\begin{alignat}{1}
	\partial_\omega\xi &= \partial_\omega(\sqrt{(1-\epsilon)c_R^2-\eta^2}) =  -\eta/\xi, \notag\\
	\partial_\omega n^y &= -\partial_\omega\sqrt{1-(n^x)^2} = -\frac{n^x}{n^y}\partial_\omega n^x \pteq 0, \notag\\
	\partial_\omega z^x_u &= -\partial_\omega\xi = \eta/\xi,\qquad \partial_\omega z^y_u = -1. \notag
\end{alignat}
\begin{alignat}{1}
	\partial_\omega(|\vec z_u|^2) &= \partial_\omega(\xi^2+(\eta-v^y_u)^2) = 2\xi(-\eta/\xi)+2(\eta-v^y_u) = -2v^y_u.\notag
\end{alignat}

We use
\begin{alignat}{1}
	L_d^2 &= 1-\epsilon  \notag\\
	\Leftrightarrow\qquad |\vec z_d|^2 &= (1-\epsilon)c_d^2
	= (1-\epsilon)c_u^2 + \frac{(\gamma-1)(1-\epsilon)}{2}(|\vec z_u|^2-|\vec z_d|^2) \notag\\
	\Leftrightarrow\qquad |\vec z_d|^2 &= 
		\frac{2(1-\epsilon)}{\gamma+1+\epsilon(1-\gamma)}c_u^2
		+\frac{(\gamma-1)(1-\epsilon)}{\gamma+1+\epsilon(1-\gamma)}|\vec z_u|^2 
	\myeqlabel{eq:zdzuLone}
\end{alignat}

	\myeqref{eq:zdzuLone} yields
	\begin{alignat}{1}
		\partial_\omega(|\vec z_d|^2) &= \frac{\gamma-1}{\gamma+1}\partial_\omega(|\vec z_u|^2) + O(\epsilon)
		= \frac{-2(\gamma-1)}{\gamma+1}v^y_u + O(\epsilon). \myeqlabel{eq:zdeta}
	\end{alignat}
	On the other hand,
	\begin{alignat}{1}
		\partial_\omega(|\vec z_d|^2) &
		= 2z^x_d\partial_\omega(z^x_d)+2z^y_d\partial_\omega(z^y_d) \pteq -2\xi\partial_\omega(z^x_d)-2\eta\partial_\omega(z^y_d),\notag
	\end{alignat}
	so
	\begin{alignat}{1}
		\xi\partial_\omega(z^x_d)+\eta\partial_\omega(z^y_d) &\pteq \frac{\gamma-1}{\gamma+1}v^y_u + O(\epsilon). \myeqlabel{eq:cp1}
	\end{alignat}
	\myeqref{eq:momjump} can be restated
	\begin{alignat}{1}
		0 &= \rho_uz^n_u - \rho_dz^n_d  = \rho_u(z^x_un^x+z^y_un^y) 
		- \pi^{-1}\Big(\pi(\rho_u)+\frac{|\vec z_u|^2-|\vec z_d|^2}{2}\Big)(z^x_dn^x+z^y_dn^y) \notag
	\end{alignat}
	Take $\partial_\omega$:
	\begin{alignat}{1}
		\Rightarrow\qquad 0 &= 
		\rho_u(n^x\partial_\omega z^x_u+z^x_u\partial_\omega n^x
			+n^y\partial_\omega z^y_u+z^y_u\partial_\omega n^y)	
		-\frac{\rho_d}{2c_d^2}\big(\partial_\omega(|\vec z_u|^2)-\partial_\omega(|\vec z_d|^2)\big)z^n_d  \notag \\
		&-\rho_d(n^x\partial_\omega z^x_d+z^x_d\partial_\omega n^x
			+n^y\partial_\omega z^y_d+z^y_d\partial_\omega n^y)	\notag\\
		&\underset{\myeqref{eq:zdeta}}\pteq \rho_u(-\xi\partial_\omega n^x+1) 
		+ \frac{2\rho_dv^y_uz^n_d}{c_d^2(\gamma+1)}
		- \rho_d(-\xi\partial_\omega n^x-\partial_\omega z^y_d) \notag 
	\end{alignat}
	\begin{alignat}{1}
		\Rightarrow\qquad 
		\xi(1-\frac{\rho_u}{\rho_d})\partial_\omega n^x+\partial_\omega z^y_d
		&\pteq -\frac{\rho_u}{\rho_d}-\frac{2v^y_u\eta}{c_d^2(\gamma+1)} \notag\\
		\Rightarrow\qquad
		\frac{\xi v^y_u}{v^y_u-\eta}\partial_\omega n^x+\partial_\omega z^y_d
		&\pteq \frac{-\eta}{\eta-v^y_u}-\frac{2v^y_u\eta}{c_d^2(\gamma+1)} \myeqlabel{eq:cp2}
	\end{alignat}
	using $\rho_u/\rho_d=z^n_d/z^n_u\pteq\eta/(\eta-v^y_u)$. 
	Finally, $z^t_u=z^t_d$ yields
	\begin{alignat}{1}
		0 &= n^x(z^y_d-z^y_u)-n^y(z^x_d-z^x_u), \notag
	\end{alignat}
	so take $\partial_\omega$:
	\begin{alignat}{1}
		0 &= \partial_\omega n^x(z^y_d-z^y_u)+n^x(\partial_\omega z^y_d-\partial_\omega z^y_u)
		-\partial_\omega n^y(z^x_d-z^x_u)-n^y(\partial_\omega z^x_d-\partial_\omega z^x_u) \notag\\
		&\pteq -v^y_u\partial_\omega n^x+\partial_\omega z^x_d-\eta/\xi \notag
	\end{alignat}
	\begin{alignat}{1}
		\Rightarrow\qquad -v^y_u\partial_\omega n^x+\partial_\omega z^x_d &\pteq \eta/\xi \myeqlabel{eq:cp3}
	\end{alignat}
	\myeqref{eq:cp1}, \myeqref{eq:cp2} and \myeqref{eq:cp3} form a linear (nondegenerate) system for the three derivatives. Solution:
	\cucon{formulas checked in maple/deta.}
	\begin{alignat}{1}
		\partial_\omega z^y_d &= \frac{(\gamma-1)v^y_u-2(\gamma+1)\eta-2c_d^{-2}v^y_u\eta(\eta-v^y_u)}{(\gamma+1)(2\eta-v^y_u)} + O(\epsilon)
			\myeqlabel{eq:zydpre}
	\end{alignat}
	Using $v^y_d=z^y_d+\eta$: 
	\begin{alignat}{1}
		\partial_\omega v^y_d 
		&\pteq 2v^y_u\frac{\eta(v^y_u-\eta)c_d^{-2}-1}{(\gamma+1)(2\eta-v^y_u)} + O(\epsilon).
		\myeqlabel{eq:vyd-eta}
	\end{alignat}
	Note that
	\begin{alignat}{1}
		\partial_\omega v^y_d &\ptgtr 0
		\myeqlabel{eq:vyd-eta-positive}
	\end{alignat}
	(for sufficiently small $\epsilon$)
	because $v^y_u=v^y_I<0$, $\eta-v^y_u\pteq z^n_u>0$.

	\cucon{checks in maple/petaketa.txt and maple/pkcalc.txt}
	Transform \myeqref{eq:cp1} to
	\begin{alignat}{1}
		\partial_\omega(z^x_d) &\pteq -\frac{\eta}{\xi}\partial_\omega(z^y_d)+\frac{\gamma-1}{\gamma+1}\cdot\frac{v^y_u}{\xi} 
		+ O(\epsilon) \myeqlabel{eq:zxdpre}
	\end{alignat}
	(no need to evaluate further; $\partial_\omega(n^x)$ is not needed).
	Finally: the counterclockwise unit tangent for $P_R$ is 
	$$\vec t_P=\frac{\vec\xi^\perp}{|\vec\xi|}=r^{-1}\vec\xi^\perp,$$
	so
	$$p=\chi_\phi=r\chi_t=r\vec z\cdot\vec t_P=\vec\xi^\perp\cdot\vec z=\xi z^y_d-\eta z^x_d.$$
	Thus
	\begin{alignat}{1}
		\partial_\omega p &= \partial_\omega(\xi z^y_d-\eta z^x_d) = -\eta z^y_d/\xi+\xi\partial_\omega z^y_d-z^x_d-\eta\partial_\omega z^x_d \notag\\
		&\pteq \eta^2/\xi+\xi + \xi\partial_\omega z^y_d-\eta\partial_\omega z^x_d 
		= \subeq{\frac{\eta^2+\xi^2}{\xi}}{=(1-\epsilon)c_R^2/\xi} + \xi\partial_\omega z^y_d-\eta\partial_\omega z^x_d \notag\\
		&\overset{\text{\myeqref{eq:zxdpre}}}{\underset{\text{\myeqref{eq:zydpre}}}{\pteq}}
		\frac{\eta\big((\gamma+1)v^y_u-2\gamma\eta\big)-2c_d^2}{(\gamma+1)(2\eta-v^y_u)}\cdot\frac{v^y_u}{\xi} + O(\epsilon)\notag
	\end{alignat}
	Using 
	$\eta\pteq z^n_d=:\sigma c_u$, $v^y_u=z^y_u+\eta\pteq \eta-z^n_u = (\sigma-L^n_u)c_u$, as well as
	$$c_d^2=c_u^2+\frac{\gamma-1}{2}\big((z^n_u)^2-(z^n_d)^2\big)=\Big(1+\frac{\gamma-1}{2}\big((L^n_u)^2-\sigma^2\big)\Big)c_u^2,$$
	we obtain a more convenient formula:
	\begin{alignat}{1}
		p_\omega &\pteq 
		\frac{2+L^n_u\big((\gamma+1)\sigma+(\gamma-1)L^n_u\big)}{L^n_u+\sigma} \cdot \frac{-v^y_uc_u}{(\gamma+1)\xi} + O(\epsilon)
		\myeqlabel{eq:pc-zphi}
	\end{alignat}
	\cucon{verified that this follows from previous $p_\omega$}
	Since $\sigma=z^n_d/c_u<z^n_u/c_u=L^n_u$ for any admissible shock, we can argue that
	\begin{alignat}{1}
		p_\omega &\ptgeq \frac{2+(\gamma-1)L^n_u(\sigma+L^n_u)}{L^n_u+\sigma} \cdot \frac{-v^y_uc_u}{(\gamma+1)\xi} + O(\epsilon) \notag\\
		&\geq \Big(\frac{2}{L^n_u+\sigma}+(\gamma-1)L^n_u\Big) \cdot \frac{-v^y_uc_u}{(\gamma+1)\xi} + O(\epsilon) \notag\\
		&\geq \Big(\delta_{p\eta}+(\gamma-1)\sigma\Big) \cdot \frac{-v^y_uc_u}{(\gamma+1)\xi} + O(\epsilon) \myeqlabel{eq:petaineq}
	\end{alignat}
	for some $\delta_{p\eta}>0$. (Note: $L^n_u+\sigma$ is uniformly bounded
	because the set of possible shock locations is bounded.)
	Also,
	\begin{alignat}{1}
		(c_d^2-h_0)_\omega = (c_d^2)_\omega &\overset{L_d^2=1-\epsilon}{=} (1-\epsilon)\partial_\omega(|\vec z_d|^2) 
		\overset{\text{\myeqref{eq:zdeta}}}= \frac{-2(\gamma-1)}{\gamma+1}v^y_u+O(\epsilon) > 0 \myeqlabel{eq:pc-csq}
	\end{alignat}
	for $\epsilon$ sufficiently small.
	\begin{alignat}{1}
		k_\omega &\underset{\text{\myeqref{eq:pc-csq}}}{\overset{\text{\myeqref{eq:refk}}}{=}}
		(-\sigma_f/\sigma_g)^{1/2}\cdot\frac{-2(\gamma-1)}{\gamma+1}v^y_u 
		=
		\sqrt{\frac{\gamma-1}{\gamma+1}}(-v^y_u) + O(\epsilon)
		\myeqlabel{eq:keta}
	\end{alignat}
	(The $O$ term is uniform in $\gamma\downarrow 1$ because $(-\sigma_f/\sigma_g)^{1/2}$ has a $(\gamma-1)^{-1/2}$ singularity
	which is cancelled by the $\gamma-1$ numerator.)

\paragraph{Extreme corner locations}

	Now we study the behaviour of $p$, $k$, $q$ and $\theta$ (as introduced in Section \myref{section:c-pararc})
	for $\eta=\eta_R^\pm$. We use a superscript $\pm$ to indicate quantities evaluated for $\eta=\eta_R^\pm$;
	a superscript $*$ indicates $\etat_R$. We omit superscripts if the choice is unimportant (e.g.\ if the difference
	incurs an $O(\epsilon^{1/2})$ term which is dominated by something else). This is the case for $p_\omega,k_\omega$
	and other derivatives; we may conveniently evaluate them at $\eta^*_R$ (note $\eta^+_R-\eta^*_R=O(\epsilon^{1/2})$).

	$p^*=0$ and $k^*=O(\epsilon)$ by \myeqref{eq:h0} (with $r=\sqrt{1-\epsilon}c_R$ and $c_*=c_R$). 
	Then
	\begin{alignat}{1}
		p^+ &= p_\omega(\eta^+_R-\etat_R)+O(\epsilon), \mylabel{eq:pplus} \\
		k^+ &= k^*+k_\omega(\eta^+_R-\etat_R)+O(\epsilon) 
			\overset{\myeqref{eq:h0}}{=} 
		k_\omega(\eta^+_R-\etat_R)+O(\epsilon), \mylabel{eq:kplus}
	\end{alignat}
	where $p_\omega,k_\omega$ are the values at $\eta=\etat_R$.
	Therefore
	\begin{alignat}{1}
		q^+ &= \sqrt{p_\omega^2+k_\omega^2}(\eta^+_R-\etat_R)+O(\epsilon). \myeqlabel{eq:qetaB}
	\end{alignat}
	We estimate this:
	\begin{alignat}{1}
		&1-\left(\frac{\xi\sqrt{p_\omega^2+k_\omega^2}}{c_dv^y_u}\right)^2 \notag\\
		&\underset{\text{\myeqref{eq:pc-zphi}}}{\overset{\text{\myeqref{eq:keta}}}{=}}
		1
		- c_d^{-2}(v^y_u)^{-2}\left(\frac{\gamma-1}{\gamma+1}(v^y_u)^2 
			+ \Big(\frac{\frac{2}{\gamma+1}+L^n_u(\sigma+\frac{\gamma-1}{\gamma+1}L^n_u)}{(L^n_u+\sigma)\xi/c_u}v^y_u\Big)^2
		\right)\xi^2 + O(\epsilon)\notag \\
		&=
		\frac{c_u^2}{c_d^2}\left(\frac{c_d^2}{c_u^2}
		- \frac{\gamma-1}{\gamma+1}\frac{\xi^2}{c_u^2} 
		- \Big(\frac{\frac{2}{\gamma+1}+L^n_u(\sigma+\frac{\gamma-1}{\gamma+1}L^n_u)}{(L^n_u+\sigma)}\Big)^2 \right)
		 \notag + O(\epsilon) \notag\\
		&=
		\frac{c_u^2}{c_d^2}\left(\frac{c_d^2}{c_u^2}
		- \frac{\gamma-1}{\gamma+1}\frac{c_d^2-\eta^2}{c_u^2} 
		- \Big(\frac{\frac{2}{\gamma+1}+L^n_u(\sigma+\frac{\gamma-1}{\gamma+1}L^n_u)}{(L^n_u+\sigma)}\Big)^2 \right)
		 \notag + O(\epsilon) \notag\\
		&=
		\frac{c_u^2}{c_d^2}\left(\frac{2}{\gamma+1}\frac{c_d^2}{c_u^2}
		+ \frac{\gamma-1}{\gamma+1}\sigma^2
		- \Big(\frac{\frac{2}{\gamma+1}+L^n_u(\sigma+\frac{\gamma-1}{\gamma+1}L^n_u)}{(L^n_u+\sigma)}\Big)^2 \right)
		 \notag + O(\epsilon) \notag\\
		&=
		\frac{c_u^2}{c_d^2}\left(\frac{2}{\gamma+1}\frac{c_u^2+\frac{\gamma-1}{2}(|\vec z_u|^2-|\vec z_d|^2)}{c_u^2}
		+ \frac{\gamma-1}{\gamma+1}\sigma^2
		- \Big(\frac{\frac{2}{\gamma+1}+L^n_u(\sigma+\frac{\gamma-1}{\gamma+1}L^n_u)}{(L^n_u+\sigma)}\Big)^2 \right)
		 \notag + O(\epsilon) \notag\\
		&=
		\frac{c_u^2}{c_d^2}\left(\frac{2}{\gamma+1}\left(1+\frac{\gamma-1}{2}((L^n_u)^2-\sigma^2)\right)
		+ \frac{\gamma-1}{\gamma+1}\sigma^2
		- \Big(\frac{\frac{2}{\gamma+1}+L^n_u(\sigma+\frac{\gamma-1}{\gamma+1}L^n_u)}{(L^n_u+\sigma)}\Big)^2 \right)
		 \notag + O(\epsilon) \notag\\
		&= 
		\frac{2c_u^2\big((L^n_u)^2-1\big)}{(\gamma+1)^2c_d^2(L^n_u+\sigma)^2}
		\Big(2-(\gamma+1)\sigma^2+(\gamma-1)(L^n_u)^2\Big) + O(\epsilon)
		\myeqlabel{eq:qqq} 
	\end{alignat}
	The last factor is positive: \myeqref{eq:zndznucomp} yields
	$$\sigma\leq\frac{\gamma-1}{\gamma+1}L^n_u+\frac{2}{\gamma+1},$$
	so
	\begin{alignat}{1}
		2-(\gamma+1)\sigma^2+(\gamma-1)(L^n_u)^2 
		&\geq \frac{2(\gamma-1)}{\gamma+1}(L^n_u-1)^2.\notag
	\end{alignat}
	$L^n_u-1$ is uniformly positive since the corner shocks allowed by \myeqref{eq:cornerregion} are uniformly not vanishing, 
	for $\epsilon$ sufficiently small.
	All other factors are trivially positive (note $L^n_u>1,\sigma$). The right-hand side of \myeqref{eq:qqq} is positive, so
	\begin{alignat}{1}
		\sqrt{p_\omega^2+k_\omega^2} &< \frac{-c_Rv^y_I}{\xi} \myeqlabel{eq:qqeta}
	\end{alignat}
	and therefore
	\begin{alignat}{1}
		q^+ &\overset{\myeqref{eq:qetaB}}\leq \frac{-c_Rv^y_I}{\xi_R}(\eta^+_R-\etat_R) + O(\epsilon) \myeqlabel{eq:qeta}
	\end{alignat}
\paragraph{Extreme $\theta$ value}
	\begin{alignat}{1}
		\tan(\frac{\pi}{2}-\theta^+) &= \frac{p^+}{k^+} 
		\overset{\myeqref{eq:kplus}}{\underset{\myeqref{eq:pplus}}{=}}
		\frac{p_\omega}{k_\omega} + O(\epsilon) 
		\overset{\myeqref{eq:keta}}{\underset{\myeqref{eq:pc-zphi}}{=}} \sqrt{\frac{\gamma+1}{\gamma-1}}\cdot
		\frac{\frac{2}{\gamma+1}+L^n_u(\sigma+\frac{\gamma-1}{\gamma+1}L^n_u)}{(L^n_u+\sigma)\xi/c_u}  + O(\epsilon)\notag
	\end{alignat}
	On the other hand: by convexity of $\tan$ on $[0,\pi/2)$, with
	$\sigma_\theta\overline\phi<\overline\phi<\pi/2$,
	\begin{alignat}{1}
		\tan(\sigma_\theta\overline\phi)
		&\leq \sigma_\theta\tan\overline\phi 
		= \sigma_\theta\frac{\eta^+_R}{\xi^+_R}+O(\epsilon)
		=
		\sqrt\frac{\gamma-1}{\gamma+1}\cdot\frac{\sigma}{\xi/c_u}  + O(\epsilon)\notag
	\end{alignat}

	Then
	\begin{alignat}{1}
		\frac{ \tan(\sigma_\theta\overline\phi) }{ \tan(\frac{\pi}{2}-\theta^+) }
		&\leq \frac{\gamma-1}{\gamma+1}\cdot
			\frac{\sigma(L^n_u+\sigma)}{\frac{2}{\gamma+1}+L^n_u(\sigma+\frac{\gamma-1}{\gamma+1}L^n_u)} + O(\epsilon)\notag
	\end{alignat}
	The right hand side is $<1$, for sufficiently small $\epsilon$, if and only if
	\begin{alignat}{1}
		(\gamma-1)\sigma(L^n_u+\sigma) &< 2+L^n_u\big((\gamma+1)\sigma+(\gamma-1)L^n_u\big).\notag
	\end{alignat}
	Since $\sigma=z^n_d/c_u<z^n_u/c_u=L^n_u$ and $L^n_u>1$, this is always true.
	Therefore 
	\begin{alignat}{1}
		\theta^+ \in (0,\frac{\pi}{2}-\sigma_\theta\cdot\overline\phi)+2\pi\Z. \myeqlabel{eq:theta-phibar-plus} 
	\end{alignat}
	The result for $\vec\xi_R^-$ follows from symmetry:
	\begin{alignat}{1}
		\theta^- \in (\pi,\frac{3\pi}{2}-\sigma_\theta\cdot\overline\phi)+2\pi\Z. \myeqlabel{eq:theta-phibar-minus}
	\end{alignat}

\subsection{Corner bounds}

\mylabel{section:lowerbounds}

\begin{proposition}
	\mylabel{prop:etaa-lowerbound}%
	For $\epsilon$ sufficiently small:

	for any fixed point $\po\in\cfusp$ of $\IT$, 
	the lower bounds in \myeqref{eq:cornerregion} are strict: 
	$$\etaa_L>\eta_L^-,\qquad\etaa_R>\eta_R^-.$$
\end{proposition}
\begin{proof}
	For $\gamma=1$: we may borrow \myeqref{eq:plower} which contradicts
	\begin{alignat*}{1} 
		p^-
		&= \subeq{p^*}{=0} + p_\omega (\eta^-_R-\etat_R) + O(\epsilon) 
		= -p_\omega \epsilon^{1/2} + O(\epsilon)
	\end{alignat*}
	if $\epsilon$ is sufficiently small, because
	$p_\omega>0$ by \myeqref{eq:pc-zphi}.
	
	For $\gamma>1$: \myeqref{eq:thetasector} contradicts \myeqref{eq:theta-phibar-minus},
	for $\epsilon>0$ sufficiently small.

	Again the proof for the left corner is analogous: in $L$ coordinates (Figure \myref{fig:frameL}) $P_L$ and $\vxia_L$, $\vxit_L$
	have the same properties as $P_R$, $\vxia_R$, $\vxit_R$, except that $\vec\xi_{BL}$ is not on the horizontal axis which
	is irrelevant.
\end{proof}

To prove that $\etaa_R=\eta^+_R$ is impossible, a more global argument is needed. 

\if\dofigures%
\begin{figure}
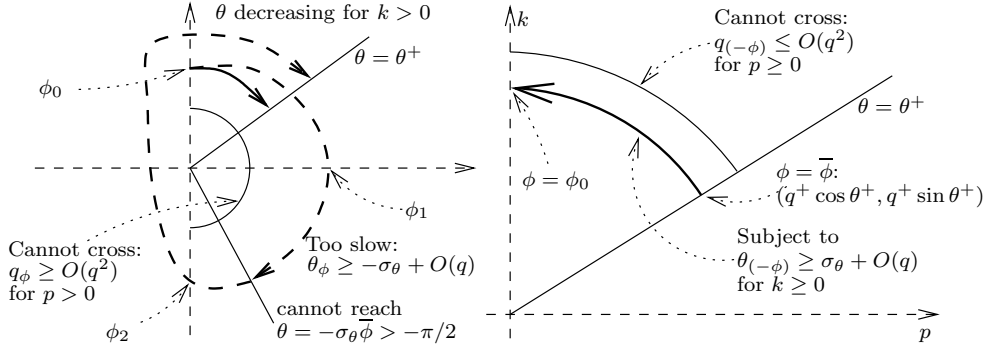

\input{pkright.pstex_t}
\input{cmax.pstex_t}
\caption{Left: $(p,k)$ value curve (bold) from a $k$ maximum on $P_R$ to the upper endpoint, for $\etaa_R=\eta^+_R$. The curve cannot go past $\theta^+$
	because \myeqref{eq:thetaphi} does not allow it to pass through the lower right quadrant ``fast'' enough.
	Right: $(p,k)$ curve (bold) in opposite direction; the constraints limit the value of $k$ maxima.}
\mylabel{fig:cmax}
\end{figure}
\fi%
\begin{proposition}
	\mylabel{prop:cbar}%
	Consider $\etaa_R=\eta^+_R$.
	There is a $\overline c$ so that for $\epsilon$ sufficiently small,
	$$c(\phi)<\overline c\qquad\text{for $\phi\in[0,\overline\phi]$.}$$
	For $\gamma>1$, $\overline c$ satisfies
	\begin{alignat}{1}
		\overline c &= \sqrt{h_0+C_{\overline c}\epsilon+(-\sigma_g/\sigma_f)^{1/2}q^+
		\sin(\theta^{+}+\sigma_\theta\overline\phi)}. \myeqlabel{eq:cbar}
	\end{alignat}
	where $C_{\overline c}$ is some constant, as in the $O$ terms;
	for $\gamma=1$ we may take any $\overline c>c$.
\end{proposition}
\begin{proof}
	Consider $\gamma>1$.
	For $k\leq 0$ the result is trivial.
	Assume that there is a $\phi_0\in[0,\overline\phi)$ so that $\phi\mapsto k(\phi)$ has a positive local maximum in $\phi=\phi_0$ 
	(see Figure \myref{fig:cmax}).
	For $\phi_0>0$ we need $k_\phi(\phi_0)=0$ which implies $p(\phi_0)=0$ by \myeqref{eq:kphi};
	for $\phi_0=0$ we have $p=0$ anyway. 
	By \myeqref{eq:pphi}, $k>0$ means $p_\phi>0$ (if $\epsilon$ is small enough, so that the $O(k^2)=O(\epsilon^{1/2}k)$ term is dominated
	by $\sigma_\theta k$).
	Therefore $p>0$ on $(\phi_0,\phi_2]$ for some $\phi_2>\phi_0$. Take $\phi_2\in(\phi_0,\overline\phi]$
	maximal with this property. 

	Assume that $\phi_2<\overline\phi$. Then $p(\phi_2)=0$ necessarily.

	\myeqref{eq:qphi} for $p>0$ implies $q>0$ on $[\phi_0,\phi_2]$, so $\theta$ is well-defined.
	Moreover $k$ is strictly decreasing on $[\phi_0,\phi_2]$, by \myeqref{eq:kphi}.

	Assume there is a $\phi_1\in(\phi_0,\phi_2]$ with $k(\phi_1)=0$, hence $k\leq0$ on $[\phi_1,\phi_2]$. 
	\myeqref{eq:thetaphi} (for $k\leq0$) integrated from $\phi_1$ to $\phi\in[\phi_1,\phi_2]$ implies
	$$\theta(\phi) \geq \subeq{\theta(\phi_1)}{=0} - \subeq{\sigma_\theta}{<1}\subeq{\overline\phi}{<\pi/2} + O(\epsilon^{1/2}) > -\frac{\pi}{2}.$$
	In particular $0\geq\theta(\phi_2)>-\pi/2$ --- contradiction to $p(\phi_2)=0$. The assumption was wrong; necessarily $k>0$ on $[\phi_0,\phi_2]$.

	\myeqref{eq:thetaphi} (for $k\geq 0$) implies that $\theta$ is strictly decreasing on $[\phi_0,\phi_2]$
	(for $O(q)=O(\epsilon^{1/2})$, i.e.\ $\epsilon$, sufficiently small). 
	Then $\theta(\phi_2)\in(0,\frac\pi2)$ which contradicts $p(\phi_2)=0$. The assumption was
	wrong; therefore $\phi_2=\overline\phi$. 

	By integrating \myeqref{eq:thetaphi} (for $k\geq 0$) from $\overline\phi$ back to $\phi$:
	\begin{alignat*}{1}
		\theta(\phi) & \geq \theta(\overline\phi) + \sigma_\theta(\overline\phi-\phi) + O(\epsilon^{1/2})
		= \theta^+ + \sigma_\theta(\overline\phi-\phi) + O(\epsilon^{1/2})
		\qquad(\phi\in[\phi_0,\overline\phi]).
	\end{alignat*}
	Integrate \myeqref{eq:qphi} (for $p\geq 0$) backwards:
	\begin{alignat*}{1}
		q(\phi) & \leq q(\overline\phi) + O(\epsilon)
		= q^+ + O(\epsilon)
		\qquad(\phi\in[\phi_0,\overline\phi]).
	\end{alignat*}
	Then 
	\begin{alignat*}{1}
		k_\phi(\phi)
		&=-\sigma_\theta p(\phi)
		=-\sigma_\theta q(\phi)\cos\theta(\phi) \\
		&\geq -\sigma_\theta q^+\cos(\theta^{+}+\sigma_\theta(\overline\phi-\phi))
		+O(\epsilon)
		\qquad(\phi\in[\phi_0,\overline\phi])
	\end{alignat*}
	so
	\begin{alignat*}{1}
		k(\phi_0)&\leq k(\overline\phi)+\sigma_\theta q^+
		\int_{\phi_0}^{\overline\phi}\cos\big(\theta^++\sigma_\theta(\overline\phi-\phi)\big)d\phi
		+O(\epsilon) \\
		&= q^+\sin\theta^+ - q^+\Big(\sin\theta^+-\sin\big(\theta^++\sigma_\theta\subeq{(\overline\phi-\phi_0)}{\leq\overline\phi}\big)\Big)
		+O(\epsilon) \\
		&\leq q^+\sin\big(\theta^++\sigma_\theta\overline\phi\big)
		+O(\epsilon)
	\end{alignat*}
	Finally we obtain a bound for $c$: let $\phi_0$ be the \emph{global} maximum point of $k$ on $[0,\overline\phi]$, then 
	\begin{alignat*}{1}
		\sup_{[0,\overline\phi]}c(\phi)^2 
		&= h_0+\sqrt\frac{-\sigma_g}{\sigma_f}k(\phi) 
		\leq h_0+\sqrt\frac{-\sigma_g}{\sigma_f}k(\phi_0) \\
		&\leq h_0+\sqrt\frac{-\sigma_g}{\sigma_f}q^+\sin\big(\theta^++\sigma_\theta\overline\phi\big)+C_{\overline c}\epsilon
	\end{alignat*}
	for $\epsilon>0$ small enough.
	This is exactly the statement.
\end{proof}

\begin{proposition}
	\mylabel{prop:psi-axi}%
	Again consider the case $\eta_R=\eta^+_R$.
	For $\gamma>1$ define
	\begin{alignat}{1}
		\tilde a(\phi) 
		&:= \sqrt{1-\epsilon}\cdot c_R
		\left(\Big[\left(\frac{\overline c}{c_R}\right)^2-\sin^2\phi\Big]^{1/2}-\cos\phi\right), \\
		a &:= \tilde a(0) = \sqrt{1-\epsilon}(\overline c-c_R) \myeqlabel{eq:a}
	\end{alignat}
	where $\overline c$ is as in Proposition \myref{prop:cbar}; for $\gamma=1$ take $a=\tilde a=0$.
	For $\epsilon$ sufficiently small,
	$$a=\max_{\phi\in[0,\overline\phi]}\tilde a(\phi),$$
	and $\psi+a\xi$ cannot have a local minimum (with respect to $\overline\Omega$) on $P_R\cup\{\vec\xi_{BR}\}$.
\end{proposition}
\begin{proof}
	For $\gamma=1$: $L^2=1-\epsilon$ implies $\chi_r\geq-\sqrt{1-\epsilon}\cdot c=-r$,
	so $\psi_r\geq 0$. By the Hopf lemma, this does not allow a local minimum of $\psi$ at $P_R$.
	In $\vec\xi_{BR}$ we argue that by Remark \myref{rem:reflection}, we may consider the even reflection of $\psi$ across $\overline A$
	which still satisfies the same equation, so the Hopf lemma also rules out a local minimum of $\psi$ in $\vec\xi_{BR}$.

	For $\gamma>1$:
	$$\frac{\tilde a}{\sqrt{1-\epsilon}\cdot c_R} \overset{\myeqref{eq:cbar}}{=}
	\Big( \frac{h_0}{c_R^2}-1+\frac{C_{\overline c}}{c_R^2}\epsilon+\cos^2\phi+\sqrt\frac{-\sigma_g}{\sigma_f}\frac{q^+}{c_R^2}
	\sin\big(\theta^+ + \sigma_\theta(\overline\phi-\phi)\big)\Big)^{1/2}
	-\cos\phi.$$
	Call the right-hand side $f$ and take $\partial_\phi$ of it:
	$$
	\frac{2(f-\cos(\phi))\sin\phi-\sqrt\frac{-\sigma_g}{\sigma_f}\frac{q^+}{c_R^2}\sigma_\theta
		\cos\big(\theta^+ + \sigma_\theta(\overline\phi-\phi)\big)}{2f}.
	$$
	$f>0$ by \myeqref{eq:h0} and $r=\sqrt{1-\epsilon}\cdot c_R$, so the denominator is positive.
	By \myeqref{eq:theta-phibar-plus}, $\cos(\theta^++\sigma_\theta(\overline\phi-\phi))>0$, so
	the numerator is negative unless $f\geq\cos\phi$. In that case $\phi=\frac{\pi}{2}+O(\epsilon^{1/2})$,
	because $f$ is $O(\epsilon^{1/2})$ due to $q^+=O(\epsilon^{1/2})$ and $h_0-c_R^2=O(\epsilon)$ (by \myeqref{eq:h0}).
	But $0\leq\phi\leq\overline\phi=\arctan\frac{\eta^+_R}{\xi^+_R}\leq\frac{\pi}{2}-\delta$,
	with $\delta>0$ uniformly in $\epsilon\downarrow0$, 
	so for sufficiently small $\epsilon$ there is a contradiction. 
	Thus $f$ is decreasing in $\phi$, so it attains its maximum in $\phi=0$.

	For a local minimum (with respect to $\overline\Omega$) of $\psi+a\xi$ on $P_R$ we need
	$$0=(\psi+a\xi)_\phi=\psi_\phi+a\xi_\phi=\psi_\phi-a r\sin\phi,$$
	so $\psi_t=r^{-1}\psi_\phi=a\sin\phi$ there. In $\vec\xi_{BR}$ this still holds (minimum or not)
	because we have $\psi_\phi=0=a\sin\phi$.
	Another minimum condition is
	\begin{alignat}{1}
		0&\geq(\psi+a\xi)_r=\psi_r+a\xi_r\overset{L^2=1-\epsilon}{=}r-\sqrt{c^2(1-\epsilon)-\psi_t^2}+a\cos\phi\notag\\
		&=\sqrt{1-\epsilon}\cdot c_R-\sqrt{c^2(1-\epsilon)-a^2\sin^2\phi}+a\cos\phi.\notag
	\end{alignat}
	This is equivalent (by squaring to eliminate the root and solving
	a quadratic inequality for $a$) to
	$$a\leq \sqrt{1-\epsilon}\cdot c_R\left(\sqrt{\left(\frac{c}{c_R}\right)^2-\sin^2\phi}-\cos\phi\right).$$
	But $a\geq\tilde a(\phi)$ which, by Proposition \myref{prop:cbar}, is greater than the right-hand side. Contradiction!
\end{proof}

\begin{lemma}
	\begin{alignat}{1}
		a &\leq \left(\frac{-v^y_I}{\xi_R}-\delta_a\right)(\eta^+_R-\etat_R). \myeqlabel{eq:aexp}
	\end{alignat}
	for some $\delta_a>0$ (depending continuously on $\lambda$), and $\epsilon>0$ sufficiently small.
\end{lemma}
\begin{proof}
	For $\gamma=1$ this is trivial since $a=0$ and $v^y_I<0<\xi_R$. 

	For $\gamma>1$:
	\myeqref{eq:h0} (with $r^2=c_R^2(1-\epsilon)$) means $h_0=c_R^2(1+O(\epsilon))$.
	Taylor expand the square root in \myeqref{eq:cbar} around $h_0$:
	\begin{alignat}{1}
		\overline c &= 
		\sqrt{h_0}+O(\epsilon)+\frac{1}{2\sqrt{h_0}}q^+(-\sigma_g/\sigma_f)^{1/2}\sin\big(\theta^++\sigma_\theta\overline\phi)
		+O(\epsilon) \notag\\
		&\overset{\myeqref{eq:h0}}{=} c_R+\frac{q^+}{2c_R}(-\sigma_g/\sigma_f)^{1/2}\sin\big(\theta^++\sigma_\theta\overline\phi)+O(\epsilon) \notag\\
		&\overset{\myeqref{eq:qeta}}{\leq}
		c_R+(\eta^+_R-\etat_R)\frac{-v^y_I}{2\xi_R}(-\sigma_g/\sigma_f)^{1/2}\sin\big(\theta^++\sigma_\theta\overline\phi)+O(\epsilon) \notag\\
		&\leq c_R+(\eta^+_R-\etat_R)\left(\frac{-v^y_I}{\xi_R}-\delta_a\right)\notag
	\end{alignat}
	for some $\delta_a>0$ because 
	$(-\sigma_g/\sigma_f)^{1/2}=2\sqrt{\frac{\gamma-1}{\gamma+1}}+O(\epsilon)<2$ 
	for $\epsilon>0$ small enough 
	and because
	$\sin\big(\theta^++\sigma_\theta\overline\phi)\big)<1$ by \myeqref{eq:theta-phibar-plus}.
	Use \myeqref{eq:a} to get \myeqref{eq:aexp}.
\end{proof}

\begin{proposition}
	\mylabel{prop:vyd-crit}%
	Consider $a$ as in \myeqref{eq:a}.
	For $\epsilon$ sufficiently small, the shock through $\vec\xi^+_R$ with upstream data $\vec v_I$ and $\rho_I$
	and tangent $(1,\frac{a}{-v^y_I})$ has $v^y_d>0$.
\end{proposition}
\begin{proof}
	The shock through $\vec\xi^*_R$ with tangent $(1,0)$ is the R shock where $v^y_d=0$ by construction.
	$v^y_d$ for the new slope and location differs from $0$ by (1) moving the shock up to $\vec\xi^+_R$
	from $\vxit_R$, while keeping it horizontal, and (2) rotating it while holding it in $\vec\xi^+_R$ to
	make its slope the above. Both of these changes are $O(\epsilon^{1/2})$, so
	it is sufficient to consider a first-order expansion using known derivatives. As before,
	we use uniformity in $\epsilon\downarrow 0$.

	For (1), we use \myeqref{eq:DvndDsigma} (note $v^n_d=\vec v\cdot\vec n=-v^y_d$, $\sigma=\vec\xi\cdot\vec n=-\eta$):
	$v^y_d$ changes by
	$$\left(1-\frac{\partial z^n_d}{\partial z^n_u}\right)(\eta^+_R-\eta^*_R) + O(\epsilon).$$

	For (2), we use \myeqref{eq:DvdxDnva} with $\vec n=(0,-1)$:
	\begin{alignat*}{1}
		\partial_\beta v^y_d &= (\partial_\beta\vec v_d)\cdot\vec n\subeq{n^y}{=-1}+(\partial_\beta\vec v_d)\cdot\vec t\subeq{t^y}{=0} 
		\overset{\myeqref{eq:DvdxDnva}}{=} z^t\left(1-\frac{\partial z^n_d}{\partial z^n_u}\right)
	\end{alignat*}
	Here $\nva=\arctan\frac{a}{-v^y_I}=O(\epsilon^{1/2})$. So
	$v^y_d$ changes by 
	\begin{alignat*}{1}
		z^t\left(1-\frac{\partial z^n_d}{\partial z^n_u}\right)\cdot\arctan\frac{a}{-v^y_I}+O(\epsilon) 
		&=\left(1-\frac{\partial z^n_d}{\partial z^n_u}\right)\frac{z^t}{-v^y_I}a + O(\epsilon) \\
		&\overset{\myeqref{eq:aexp}}{\underset{z^t\pteq-\xi_R}{\geq}}
			\left(1-\frac{\partial z^n_d}{\partial z^n_u}\right)(\frac{\xi_R}{-v^y_I}\delta_a-1)(\eta^+_R-\etat_R) + O(\epsilon).
	\end{alignat*}

	(1) and (2) combined: the change is
	\begin{alignat*}{1}
		\geq\left(1-\frac{\partial z^n_d}{\partial z^n_u}\right)\frac{\xi_R}{-v^y_I}\delta_a\subeq{(\eta^+_R-\etat_R)}{=\epsilon^{1/2}} 
			+ O(\epsilon).
	\end{alignat*}
	By \myeqref{eq:DvndDvnu} the first factor is $\geq2/(\gamma+1)$. 
	Therefore $v^y_d>0$ for $\etaa_R=\eta^+_R$, for $\epsilon$ sufficiently small.
\end{proof}

\begin{proposition}
	\mylabel{prop:etaa-upperbound}%
	Let $\chi\in\overline{\fusp}$ be a fixed point of $\IT$. 
	For $C_\eta$ sufficiently large and for $\epsilon>0$ sufficiently small,
	the upper part of \myeqref{eq:cornerregion} is strict:
	$$\etaa_C < \eta_C^+\qquad(C=L,R).$$
\end{proposition}
\begin{proof}
	Let $a$ be defined as in \myeqref{eq:a}. As shown in Proposition \myref{prop:psi-axi}, 
	$\psi+a\xi$ cannot have a local minimum at $P_L\cup\{\vec\xi_{BR}\}$.
	For $\etaa_R>\etat_R$, we have $(\psi+a\xi)_2=\psi_2>0$ in $\vec\xi_R$ by \myeqref{eq:vyd-eta-positive} (for sufficiently small $\epsilon$), so the minimum cannot be in $\vec\xi_R$
	either (note that the domain locally contains the ray downward from the corner).

	On the shock: $\psi+a\xi=\psi^I+a\xi$, so
	$$\partial_t(\psi+a\xi)=\partial_t(\psi^I+a\xi)=\vec v_I\cdot\vec t+\frac{a}{(1+s_1^2)^{1/2}}=\frac{v^y_Is_1+a}{(1+s_1^2)^{1/2}}.$$ 
	For a local minimum at the shock we need $\partial_t(\psi+a\xi)=0$,
	so
	$$s_1=\frac{a}{-v^y_I}.$$
	A \emph{global} minimum, in particular\ $\leq \psi(\vec\xi_R)+a\xi_R$, additionally requires
	that $\vxia_R$ (as well as the rest of the shock) is on or below the tangent through the minimum point, because $\psi^I$ and thus $\psi^I+a\xi$
	are decreasing in $\eta$. 
	Proposition \myref{prop:vyd-crit} shows that the shock through $\vxia_R$ with that tangent has 
	$v^y_d>0$ for $\etaa_R=\eta^+_R$. In the minimum point the tangent has same slope but is at least as high, 
	so the shock speed is at least as high, so $v^y_d=\psi_2>0$ is at least as high, in particular $>0$ too.
	But that contradicts a minimum (the ray vertically downwards from any shock point is locally contained in $\overline\Omega$). 
	Hence $\psi+a\xi$ cannot have a global minimum on the shock.

	$\psi_n=0$ on $A$ contradicts a minimum on $A$ (by the Hopf lemma).

	The equation \myeqref{eq:psi} yields $$(c^2I-\nabla\chi^2):\nabla^2(\psi+a\xi)=0$$
	($a\xi$ is linear), so the classical strong maximum principle rules out a minimum in the interior
	(unless $\psi+a\xi$ is constant, which means we are looking at the unperturbed solution which has
	$\etaa_R=\etat_R<\eta^+_R$).

	On $\overline{P_L}$: for $\gamma=1$ we argue that $\psi_r\leq 0$ on $\overline{P_L}$ (in fact $\psi_r<0$ except in the unperturbed
	case $\eta^*_L=\eta^*_R$), as in in the first paragraph of the proof of Proposition \myref{prop:psi-axi}, so the Hopf lemma
	rules out a maximum at $P_L\cup\{\vec\xi_{BL}\}$. In $\vec\xi_L$: either $\eta_L\leq\eta^*_L\leq\eta^*_R<\eta^+_R$
	so that $\psi(\vec\xi_L)>\psi(\vec\xi_R)=\psi(\vec\xi_R^+)$ due to $\psi=\psi^I$ on $\overline S$ and $\psi^I_\xi=0>\psi^I_\eta$, 
	or $\eta_L>\eta^*_L$ so that $\psi_\eta>0$ in $\vec\xi_L$,
	by the same analysis as for $\vec\xi_R$, ruling out a minimum.

	For $\gamma>1$:
	$$(\psi+a\xi)_\xi=\psi_\xi+O(\epsilon^{1/2})\overset{\myeqref{eq:vP}}{=}v^x_L+O(\epsilon^{1/2}).$$ 
	For $C_\eta$ sufficiently large and $\epsilon$ sufficiently small\footnote{We need $C_\eta>0$; otherwise the
	upper bound on $\epsilon$ would become zero as $\etat_L\uparrow\etat_R$.} 
	the right-hand side is negative: $v^x_L=0$ for $\etat_L=\etat_R$,
	it is strictly decreasing in $\etat_L$ (Proposition \myref{prop:vdzero}), and $C_\eta$ bounds $\etat_L$ away from $\etat_R$
	(see \myeqref{eq:Ceta}).
	Again, no minimum is possible. 

	Conclusion: no $\psi$ minimum anywhere --- contradiction!

	The argument for $\etaa_L$ is similar, using the $L$ picture (Figure \myref{fig:frameL}).
	The wall still passes through the origin, so the wall boundary condition $\chi_n=0$ implies $\psi_n=0$.
	For $\gamma=1$ this implies that $\psi$ cannot have minima at the wall (Hopf lemma).
	For $\gamma>1$ 
	we have to modify the argument at the wall since it is no longer horizontal.
	A minimum of $\psi+a\xi$ requires $(\psi+a\xi)_n<0$ (where $\vec n$ points \emph{out} of $\Omega$),
	so $a\xi_n<0$, so $\xi_n<0$ (by $a>0$). But the wall slope is positive, so $\xi_n>0$ (for outward $\vec n$) --- contradiction.

	The remaining arguments are as in $R$ coordinates.
\end{proof}

\subsection{Density bounds and shock strength}
\mylabel{section:densitycontrol}

\if\dofigures%
\begin{figure}
\center{\input{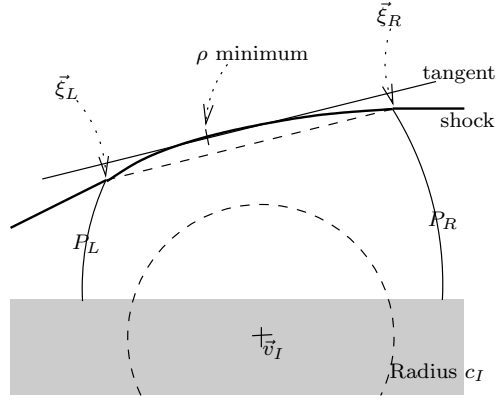}}
\caption{In a $\rho=\rho_-$ minimum at $S$, the shock tangent must be above $\vec\xi_L$ and $\vec\xi_R$, so by \myeqref{eq:techcond} the
minimum is separated from $B_{c_I}(\vec v_I)$. Therefore $\rho_-\gg\rho_I$.}
\mylabel{fig:shocktanarg}
\end{figure}
\fi%
\begin{proposition}
	\mylabel{prop:rho}%
	If $\epsilon$ and $\delta_\rho$ are sufficiently small (bounds depending only on $C_{Pt}$),
	then for any fixed point $\po\in\cfusp$ of $\IT$,
	the inequality \myeqref{eq:rhomin} is \emph{strict}.
\end{proposition}
\begin{proof}
	By Proposition \myref{prop:fp-regularity}, $\psi$ and hence $s$ are analytic. Thus we may use Proposition \myref{prop:c-principle} 
	which 
	rules out minima of $\rho$ in $\Omega$ and (using Remark \myref{rem:reflection}) at $A$,
	unless $\psi$ and $s$ are linear so that $\rho=\rho_R$ is constant and \eqref{eq:rhomin} is trivially strict. 
	However, $\rho\in C(\overline\Omega)$, so it must attain a global minimum somewhere.

	Case 1: $\rho$ attains its global minimum on $S$ (excluding $\vxia_{L,R}$). Then by Proposition \myref{prop:density-shock},
	the rest of the shock, including the corners $\vxia_L,\vxia_R$, must be below the tangent in that point
	(see Figure \myref{fig:shocktanarg}). 
	We assumed in Definition \myref{def:Lambda} that 
	the line through $\vxit_L$ and $\vxit_R$ does not touch or intersect the circle with radius $c_I$ centered in $\vec v_I$.
	Hence if $\epsilon$ is sufficiently small (depending on $\underline\eta^*_L$), then by \myeqref{eq:cornerregion}
	the shock tangent must have positive distance from the circle as well.
	Thus the global minimum of $\rho$ must be bounded below away from $\rho_I$.

	Case 2: $\rho$ attains its global minimum on $\overline{P_L}$ or $\overline{P_R}$
.
	By \myeqref{eq:rhoP} we know $\rho$ up to $O(\epsilon^{1/2})$.

	Combining both (nonexclusive) cases, we see that for sufficiently small $\delta_\rho$ and $\epsilon$, depending
	continuously on $C_{Pt}$, $\lambda$, $\underline\eta^*_L$, 
	\myeqref{eq:rhomin} is \emph{strict}.
\end{proof}

\subsection{Velocity and shock normal control}
\mylabel{section:v-control}

\begin{proposition}
	\mylabel{prop:ny}%
	\mylabel{prop:vx}%
	\mylabel{prop:vy}%
	If $\delta_{SA}$ is sufficiently small, 
	if $C_{vx},C_{vL}$ are sufficiently large (bounds depending only on $C_{Pt}$), 
	if $C_{Sn}$ is sufficiently large (bound depending only on $C_{vx},C_{vL}$), 
	if $\delta_{vy}$ and $\epsilon$ are sufficiently small ($\delta_{vy}$ bound depending only on $\delta_\rho,C_{Sn}$; $\epsilon$ bound
	depending only on $C_{Sn}$), and
	if $\delta_{Cc}$ is sufficiently small, then
	for any fixed point $\po\in\cfusp$ of $\IT$,
	the inequalities \myeqref{eq:shockwall}, \myeqref{eq:horvel}, \myeqref{eq:vertvel}, \myeqref{eq:leftvel},
	\myeqref{eq:shocknormal}, \myeqref{eq:cornercone} are \emph{strict}. Moreover 
	\begin{alignat}{1}
		|\chi_t| &\geq \delta_{\chi t} \qquad\text{(on $S\cap B_{\delta_d}(\vec\xi_C)$, $C=L,R$)} \myeqlabel{eq:chitshock}
	\end{alignat}
	for some constants $\delta_{\chi t},\delta_d>0$.
\end{proposition}
\begin{proof}
\begin{enumerate}
\item For horizontal velocity  \myeqref{eq:horvel}:
	assume that $v^x$ attains a positive global maximum (with respect to $\overline\Omega$) in a point $\vec\xi_0$ on $S$.
	Since $\vec v_I=(0,v^y_I)$ with $v^y_I<0$, this means $n^x\ptless0$ (because $n^y\ptless0$), i.e.\ $s_1(\xi_0)<0$. 

	$s_1(\xi_0)$ can be expressed as a continuous
	function of $v^x(\xi_0)$ and $\vec\xi_0$. The set of possible $\vec\xi_0$ is contained in the set of possible shock locations which
	is pre-compact. Therefore if $v^x$ has a maximum $=C_{vx}\epsilon^{1/2}$ in some $\vec\xi_0\in S$, then 
	\begin{alignat}{1}
		s_1(\xi_0) &\leq -C_{s1}\epsilon^{1/2} \mylabel{eq:s1x0}
	\end{alignat}
	where $C_{s1}=C_{s1}(C_{vx})>0$ is uniformly increasing in $C_{vx}$.

	Since $S$ and $\po$ are analytic (Proposition \myref{prop:fp-regularity}), we can apply
	Proposition \myref{prop:vshock} with $\vec w=(1,0)$. For a constant-state solution \myeqref{eq:horvel} is immediate. Otherwise
	\myeqref{eq:propvx1} and \myeqref{eq:propvx-s11} are satisfied.
	$n^x\ptless0$ means $w^n\ptless0$, so by \myeqref{eq:propvx-s11} $\kappa\ptless0$, i.e.\ $s_{11}\ptgtr0$.

	Now we can use a similar argument as for the density in Proposition \myref{prop:rho} and Proposition \myref{prop:density-shock}
	(see Figures \myref{fig:rhotangent} and \myref{fig:shocktanarg}):
	$s_{11}\ptgtr 0$ implies $s_1(\xi)<s_1(\xi_0)$ for $\xi<\xi_0$ near $\xi_0$. 
	On the other hand, for $\vxia_L=\vxit_L$ we have $s_1(\xia_L)\geq 0$ by construction of the $L$ shock in Section \myref{section:parmset},
	so for arbitrary $\vxia_L$ (satisfying \myeqref{eq:cornerregion}) 
	the continuous dependence of $s_1(\xia_L)$ on $\etaa_L$ and by \myeqref{eq:cornerregion} shows
	$s_1(\xia_L)\geq -C_2\epsilon^{1/2}$ for some constant $C_2$ independent of $\epsilon$. If we pick $C_{vx}$ so large that $C_{s1}>C_2$, 
	then $s_1(\xia_L)>s_1(\xi_0)$ by \myeqref{eq:s1x0} for any possible location of the left corner.

	Therefore we can pick $\xi_a\in(\xia_L,\xi_0)$ maximal so that $s_1(\xi_a)=s_1(\xi_0)$.
	Then $s_1(\xi)<s_1(\xi_0)$
	for $\xi\in(\xi_a,\xi_0)$, so by integration 
	$$s(\xi_a) > s(\xi_0)+s_1(\xi_0)\cdot(\xi_a-\xi_0).$$
	But that means the shock tangent in $\xi_a$ is parallel to the one in $\xi_0$
	but \emph{higher}, so $\sigma:=\vec\xi\cdot\vec n$ is smaller in $\xi_a$. 
	By \myeqref{eq:DvndDsigma}, that means $v^n_d$ is smaller in $\xi_a$, whereas
	$v^t$ is the same (parallel tangents). $n^x<0$, so $v^x_d$ is \emph{bigger} in $\xi_a$. Contradiction --- we assumed that
	we have a \emph{global} maximum of $v^x$ in $\xi_0$.

	Propositions \myref{prop:interior-velocity} and \myref{prop:v-wall} rule out local maxima of $v^x$ in $\Omega$ and on $A$,
	where we use that $\chi$ is analytic and that $\vec w$ is not vertical, i.e.\ not normal to the wall.

	On $\overline{P_L}\cup\overline{P_R}$ we can use \myeqref{eq:vP}, increasing $C_{vx}$ to $>C_{Pv}$ if necessary 
	(this makes $C_{vx}$ depend on $C_{Pt}$ as well). Now \myeqref{eq:horvel} is strict.
\item 
	In any point $\vec\xi\in S$, $s_1(\xi)$ is a function of $v^x_d$, with $\sign s_1=-\sign v^x$. $s_1(\xi)$ is continuous in $\vec\xi$
	and $v^x_d$, and the set of possible shock locations $\vec\xi$ is pre-compact, so \myeqref{eq:horvel} implies
	$$\sup\measuredangle(\vec n,\vec n_R)<C_{Sn}\epsilon^{1/2}$$
	where $C_{Sn}=C_{Sn}(C_{vx})$. 
\item The arguments for \myeqref{eq:leftvel} are analogous to those for \myeqref{eq:horvel}:
	the transformation from $R$ to $L$ coordinates (Figure \myref{fig:frameL}) turns $\vec n_L^\perp$ into $\vec n_R^\perp=(1,0)$ (and vice versa).
	The wall is never vertical in $L$ coordinates, so Proposition \myref{prop:v-wall} still applies to $v^x$.
	The other arguments are as before.

	Moreover \myeqref{eq:leftvel} implies 
	$$\sup\measuredangle(\vec n_L,\vec n)<C_{Sn}\epsilon^{1/2},$$
	where $C_{Sn}=C_{Sn}(C_{vx},C_{vL})$ now.
	\myeqref{eq:shocknormal} is strict with these choices.
\item 
	$v^y=0$ on $A$; $v^y=O(\epsilon^{1/2})$ on $P_L\cup P_R$ by \myeqref{eq:vP}. $v^y$ has no extrema in $\Omega$ (Proposition \myref{prop:vshock}); 
	to show \myeqref{eq:vertvel} it remains to discuss $S$.

	\myeqref{eq:rhomin} is in particular a lower bound for the shock strength, so $|v^n_d-v^n_u|$ is bounded away from $0$.
	\myeqref{eq:shocknormal} bounds the shock normals
	away from horizontal. Both combined imply \myeqref{eq:vertvel} is strict if $\delta_{vy}>0$ and $\epsilon>0$ are chosen small enough,
	with upper bound on $\delta_{vy}$ depending on $\delta_\rho$ and $C_{Sn}$, and upper bound on $\epsilon$ depending
	on $C_{Sn}$ only.
\item
	The shock normal bounds also imply \myeqref{eq:cornercone} is strict, for $\delta_{Cc}>0$ and $\epsilon>0$ sufficiently small(er),
	with $\epsilon$ bound depending only on $C_{Sn}$. Here we use \myeqref{eq:dntcond}: \myeqref{eq:shocknormal} shows 
	shock tangents are between $\vec n_L^\perp$ and $\vec n_R^\perp$ (up to $O(\epsilon^{1/2}$), and these 
	are bounded away from arc tangents.
\item
	Since the left corner is above the wall, the shock normal bounds imply \myeqref{eq:shockwall} is strict,
	for $\epsilon$ sufficiently small(er), with bound depending only on $C_{Sn}$, and for $\delta_{SA}$ sufficiently small.
\item
	Near each corner the shock normal bound bounds $\vec n$ away from the $\vec\xi$ direction, so 
	$|\chi^I_t|\geq\delta_{\chi t}$ and therefore \myeqref{eq:chitshock} for some $\delta_{\chi t}$. 
\end{enumerate}
\end{proof}

\subsection{Fixed points}

\begin{proposition}
	\mylabel{prop:oblique-corner}%
	For $\delta_o$ sufficiently small, with bounds depending only on $\delta_\rho$ and $C_L$,
	for $C_d$ resp.\ $\delta_d$ sufficiently large resp.\ small, with bounds depending only on $\delta_\rho$ and $C_L$,
	and for $\epsilon$ sufficiently small, with bounds depending only on $C_{Pt}$, $C_L$ and $\delta_\rho$: 

	If $\xo\in\cfusp$ is a fixed point of $\IT$, then \myeqref{eq:ndobb} and \myeqref{eq:Gb} are strict.
\end{proposition}
\begin{proof}
	First we check \myeqref{eq:ndobb}.
	\begin{enumerate}
	\item For the wall boundary operator \myeqref{eq:itn-wall}, the strict inequality \myeqref{eq:ndobb} is obvious.
	\item The parabolic boundary operator \myeqref{eq:itn-parabolic} has $\vec p$ derivative $g^i_{\vec p}=\nabla\xn$. 
		We use \myeqref{eq:partan} with sufficiently small $\epsilon$, depending on $C_{Pt}$,
		to obtain \myeqref{eq:ndobb} strictly (with $\delta_o=\frac12$ for example).
	\item For the shock boundary operator \myeqref{eq:itn-shock}: 
		\begin{alignat}{1}
			g^i_{\vec p}
			&= \hat\rho(1-\hat c^{-2}\nabla\xn^2)\frac{\vec v_I-\nabla\pn}{|\vec v_I-\nabla\pn|}
				-|\vec v_I-\nabla\pn|^{-1}
				\Big(1-\big(\frac{\vec v_I-\nabla\pn}{|\vec v_I-\nabla\pn|}\big)^2\Big)
			\cdot(\hat\rho\nabla\xn-\rho_I\nabla\chi^I) 
		\end{alignat}
		We exploit that for a \emph{fixed point} we have $\pn=\po=\psi^I$ at the shock, so
		$$\vec n=\frac{\vec v_I-\nabla\pn}{|\vec v_I-\nabla\pn|}=\frac{\nabla\chi^I-\nabla\xn}{|\nabla\chi^I-\nabla\xn|}$$
		where $\vec n$ is the downstream shock normal. 
		Note that 
		$$|\nabla\chi^I-\nabla\xn|=\chi^I_n-\xn_n$$
		because $\chi^I_t=\xn_t$ and $\chi^I_n>\xn_n$. 
		\begin{alignat}{1}
			g^i_{\vec p}
			&= \hat\rho(1-\hat c^{-2}\nabla\xn^2)\vec n
				-|\vec v_I-\nabla\pn|^{-1}\vec t~\vec t\cdot(\hat\rho\nabla\xn-\rho_I\nabla\chi^I) \notag\\
			&\overset{\xn_t=\chi^I_t}{=} \hat\rho(1-\hat c^{-2}\nabla\xn^2)\vec n
				-(\chi^I_n-\xn_n)^{-1}(\hat\rho-\rho_I)\xn_t\vec t \notag\\
			&\overset{\myeqref{eq:momjump}}{=} \hat\rho\Big((1-\hat c^{-2}\nabla\xn^2)\vec n
				-(\xn_n)^{-1}\xn_t\vec t\Big) \notag\\
			&= \hat\rho\Big((1-\hat c^{-2}\xn_n^2)\vec n-\xn_t\big((\xn_n)^{-1}+\hat c^{-2}\xn_n\big)\vec t\Big)
			\myeqlabel{eq:gpS}
		\end{alignat}
		\begin{alignat*}{1}
			\vec n\cdot g^i_{\vec p} &= \hat\rho\vec n^T(1-\hat c^{-2}\xn_n^2)\vec n-\subeq{\vec n\cdot\vec t}{=0}... \\
			&= \hat\rho(1-\hat c^{-2}\xn_n^2) \geq \delta_o'
		\end{alignat*}
		for some $\delta_{o'}$ depending only on $\delta_\rho$, because
		\myeqref{eq:rhomin} bounds downstream $\rho$ away from $\rho_I$, and that means $1-c^{-2}\xn_n^2=1-(L^n_d)^2$ is lower-bounded
		away from $0$, by the shock analysis in Section \myref{section:shocks}.
		$|g^i_{\vec p}|$ is easily bounded from above, using \myeqref{eq:lip},
		so \myeqref{eq:ndobb} is strict for a sufficiently small $\delta_o$, depending only on $\delta_\rho$ and $C_L$.
	\end{enumerate}

	Now we check \myeqref{eq:Gb}. For the parabolic-wall corners it is trivial: on $P_L$ or $P_R$, $g^i_{\vec p}=\nabla\xn$ is almost normal,
	by \myeqref{eq:partan}
	using a sufficiently small $\epsilon$ (bound depending only
	on $C_{Pt}$); on $A$ the vector $g^i_{\vec p}$ is normal; the corners enclose an angle exactly $\pi/2$, so the $g^i_{\vec p}$
	directions are independent (almost orthogonal).

	For shock-parabolic corners: on $P_L$ resp.\ $P_R$, again $g^i_{\vec p}=\nabla\xn$.
	On $S$, use \myeqref{eq:gpS}.
	We normalize both derivative vectors and consider their cross product:
	\begin{alignat}{1}
		&\frac{\nabla\chi}{|\nabla\chi|}\times 
		\frac{(1-c^{-2}\chi_n^2)\vec n-\chi_t\big((\chi^I_n)^{-1}+c^{-2}\chi_n\big)\vec t}{\sqrt{(1-c^{-2}\chi_n^2)^2
			+\chi_t^2\big((\chi^I_n)^{-1}+c^{-2}\chi_n\big)^2}} \notag\\
		&= 
		-\frac{\chi_t(1+\chi_n/\chi^I_n)}{|\nabla\chi|\sqrt{(1-c^{-2}\chi_n^2)^2
			+\chi_t^2\big((\chi^I_n)^{-1}+c^{-2}\chi_n\big)^2}}
	\end{alignat}
	Denominator: $|\nabla\chi|\geq\sqrt{1-\epsilon}\cdot c\geq c/2$ for $\epsilon\leq\frac12$.
	The square-root is lower-bounded by $1-c^{-2}\chi_n^2\geq\delta>0$ for some $\delta$ depending only on $\delta_\rho$
	(see above). 
	\myeqref{eq:chitshock} bounds the numerator away from $0$.

	Hence \myeqref{eq:G} is strict,
	for $C_d$ sufficiently large, $\delta_d$ and then $\epsilon$ sufficiently small, depending only and continuously on $\delta_\rho$. 
\end{proof}


\begin{proposition}
	\mylabel{prop:fp-boundary}%
	If the constants in \myeqref{eq:constlist} in Definition \myref{def:fusp} are chosen sufficiently small resp.\ large:

	for any $\lambda\in\Lambda$, 
	$\IT_{\lambda}$ cannot have fixed points on $\cfusp_{\lambda}-\fusp_{\lambda}$.
\end{proposition}
\begin{proof}
	Let $\xo\in\cfusp$ be a fixed point of $\IT$.
	We show that every inequality in the definition of $\cfusp$ is strict,
	so $\xo\in\fusp$.

	\myeqref{eq:lip} and \myeqref{eq:regularity} are strict by Proposition \myref{prop:regularity}.

	\myeqref{eq:shockwall} is strict by Proposition \myref{prop:ny}.

	\myeqref{eq:rhomin} is strict by Proposition \myref{prop:rho}.

	A fixed point satisfies $\po=\pn$, so $\|\po-\pn\|=r_I(\po)>0$ cannot be true.
	\myeqref{eq:oldnew} is strict.

	\myeqref{eq:ellip} strict is provided by Proposition \myref{prop:Lbounds}.

	Due to Proposition \myref{prop:Lbounds}, $L^2=1-\epsilon$ on each point of $\overline{P_L}\cup\overline{P_R}$,
	so we are in the situation of Section \myref{section:parcs} etc. 
	Proposition \myref{prop:pararc} shows that \myeqref{eq:partan} and \myeqref{eq:parnor} are strict.

	\myeqref{eq:horvel} is strict by Proposition \myref{prop:vx}.

	\myeqref{eq:vertvel} is strict by Proposition \myref{prop:vx}.

	\myeqref{eq:leftvel} is strict by Proposition \myref{prop:vx}.

	Propositions \myref{prop:etaa-lowerbound} and \myref{prop:etaa-upperbound} rule out
	$\etaa_L=\etat_L\pm\delta^{-1}\epsilon$ and
	$\etaa_R=\etat_R\pm\delta^{-1}\epsilon$ if $\delta$ is small enough,
	so \myeqref{eq:cornerregion} is strict.

	\myeqref{eq:cornercone} is strict by Proposition \myref{prop:ny}.

	\myeqref{eq:shocknormal} is strict by Proposition \myref{prop:ny}.

	Proposition \myref{prop:oblique-corner} shows that \myeqref{eq:ndobb} and \myeqref{eq:Gb} are strict.

	All inequalities are strict.
\end{proof}

\subsection{Leray-Schauder degree}
\mylabel{section:ls}

We determine the Leray-Schauder degree of $\IT$ on $\fusp$ for a particular choice of parameters $\lambda$: a straight horizontal shock 
($\etat_L=\etat_R$),
with $\gamma=1$ (see Figure \myref{fig:unperturbed}). This problem is simple enough to compute the degree precisely, although the discussion is
still difficult, especially due to the free boundary. 

Another option is to introduce a homotopy to an even
simpler problem (linear, fixed boundary, ...). But it would be necessary to prove a new set of a priori
estimates for a family of arbitrary, unphysical problems, hence a lot of work without useful sideeffects.

\newcommand{\xcon}{\overline\chi}%
\newcommand{\pcon}{\overline\psi}%
\newcommand{\scon}{\overline s}%
\newcommand{\Ocon}{\overline\Omega}%
\newcommand{\xoth}{\chi}%
\newcommand{\poth}{\psi}%
\newcommand{\soth}{s}%

\begin{proposition}
        \mylabel{prop:unperturbed-unique}%
	For sufficiently small $\epsilon$:

        For $\gamma=1$ and $\etat_L=\etat_R$, there are no fixed points of $\IT$ in $\fusp$ other than 
	the unperturbed solution.
\end{proposition}
\begin{remark}
	This result makes essential use of $\gamma=1$ as well as the considerable simplifications from using 
	the constant-state solution. 
	At the time of writing we do not 
	know a way of proving uniqueness for 
	$\gamma>1$ or $M_I<\infty$; 
	this is why we choose to argue by homotopy to $\gamma=1$, $\etat_L=\etat_R$.
\end{remark}
\begin{proof}[Proof of Proposition \myref{prop:unperturbed-unique}]
	Let $\pcon$ be the unperturbed solution and $\scon$ the corresponding shock ($\scon(\xi)=\etat_R$
	for all $\xi$).
	Consider another fixed point $\poth$ for the same case $\gamma=1$ and $\etat_L=\etat_R$.
	Let $\soth$ be its shock.

	Consider $\vec\xi$ coordinates as defined by $\poth=\poth(\sigma,\zeta)$ (see Definition \myref{def:fusp}). Then
	$$0=(c^2I-\nabla\chi^2):\nabla^2\poth=(c^2I-\nabla\chi^2):\nabla^2(\poth-\pcon)\qquad\text{(in $\Omega$)}$$
	because $\pcon$ is constant.
	The classical weak\footnote{The strong version applies, but is not needed} 
	maximum principle \cite[Theorem 10.1]{gilbarg-trudinger}
	implies that $\pcon-\poth$ must attain its
	global minimum and maximum on the boundary. This excludes the bottom boundary because
	we can make it interior by reflection (Remark \myref{rem:reflection}).

	Assume $\poth$ has an extremum (with respect to $\overline\Omega$) in $\vec\xi\in P_L\cup P_R$. Then $\poth_t=0$ there,
	so $\xoth_t=0$ because $\vec\xi\cdot\vec t=0$ on $P_R$ \emph{and} $P_L$ for $\etat_L=\etat_R$.
	So by\footnote{This argument breaks down for $\gamma>1$.}
	$|\nabla\chi|^2=(1-\epsilon)c^2$ we get $\chi_n=-(1-\epsilon)^{1/2}c$
	(the sign is fixed by \myeqref{eq:parnor}) and finally 
	$$\psi_n=\chi_n+\vec\xi\cdot\vec n=-c\sqrt{1-\epsilon}+c\sqrt{1-\epsilon}=0.$$
	Again there is a contradiction to the Hopf lemma. 
	A reflection argument (Remark \myref{rem:reflection}) also rules out extrema (with respect to $\overline\Omega$)
	in $\vec\xi_{BL}$ and $\vec\xi_{BR}$.\cucon{do need reflection here}

	Assume $\poth$ has a maximum $>\pcon$ on $\overline{S}$.
	By $\psi=\psi^I$ on $\overline S$, $\psi^I$ strictly decreasing in $\eta$,
	this corresponds to a minimum of $s$ smaller than $\scon=\etat_R$. 
	Then $s_\xi=0$ in that point, so $\graph s$ is horizontal there, like $\graph\scon$.
	Since it is lower than $\graph\scon$ the shock $S$ is weaker there, so by \myeqref{eq:DvndDsigma}
	we have $\psi_2<0$.
	But this is incompatible with a maximum. 

	Analogously a minimum $<\pcon$ is ruled out.

	Assume $\poth$ has a maximum $>\pcon$ in $\vec\xi_R$. Then $\psi=\psi^I$ on $S$, $\psi^I$ decreasing in $\eta$ means
	$\soth(\xi_R)<\etat_R$, so $\psi_2(\vec\xi_R)<0$ by \myeqref{eq:vyd-eta-positive}, for sufficiently small $\epsilon$.
	\cucon{\myeqref{eq:vyd-eta-positive} is only local, need to make sure we stay in range of validity}
	This contradicts a maximum of $\po$, because the negative vertical direction from $\vec\xi_R$ is contained in the domain.
	After reversing some signs and inequalities we rule out a minimum $<\pcon$ in the same ways. 
	Analogous arguments apply to $\vec\xi_L$.
	
	Altogether we must have $\poth=\pcon$. 
\end{proof}

\if\dofigures
\begin{figure}
\center{\input{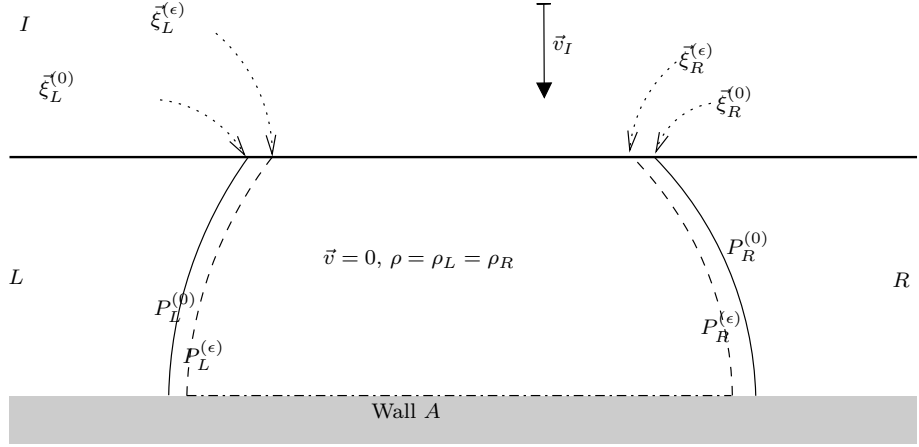}}
\caption{The unperturbed solution: a straight shock. This is the asymptotic limit for $M_I\uparrow\infty$ while holding
$M_I\sin\tau$ constant. In this limit, the wall corner moves to $\vec\xi=(-\infty,0)$.}
\mylabel{fig:unperturbed}
\end{figure}
\fi

\begin{proposition}
	\mylabel{prop:unperturbed-index}%
	The unperturbed solution with $\gamma=1$
	has index $\iota\neq 0$
	as a fixed-point of $\IT$ from Definition \myref{def:it}.
\end{proposition}
\begin{proof}
\newcommand{\FD}{\mathcal{D}}
\newcommand{\xfl}{\tilde\chi}
\newcommand{\pfl}{\tilde\psi}
\newcommand{\sfl}{\tilde s}
	We use \cite[Proposition 14.5]{zeidler-one}. Since $\IT$ is compact (Proposition \myref{prop:it-continuous-compact}), 
	we have to show that $I-\partial\IT/\partial\po(\pcon)$ has trivial kernel, where $\partial\IT/\partial\po(\pcon)$
	is the Fr\'echet derivative $\IT$ at $\po=\pcon$.
	If this is true, then the index of $\pcon$ as a fixed point of $\IT$ is $\pm1$.

	We consider $\po=\pn=\pcon$ in $\vec\sigma=(\sigma,\zeta)\in[0,1]$ coordinates (from Definition \myref{def:fusp}).
	We consider first variations $\po'$ of $\po$ in these coordinates: i.e.\ consider $\po+t\po'$ for $t\in\R$
	and evaluate $\partial_t$ at $t=0$. Let $\IT(\po)'$ etc.\ denote the resulting first variations of other
	objects. 

	Assume that $\po'=\partial\IT/\partial\po(\pcon)\po'$ (which is $\IT(\po)'$). We have to show $\po'=0$.

	$\po'=\IT(\po')$ implies that the variations of the $\vec\sigma\mapsto\vec\xi$ transforms defined by $\po$ and $\IT(\po)$ 
	(see Definition \myref{def:fusp}) are identical as well. We may write $(\vec\xi)'$ without distinction.
	Moreover $\po'=\pn'$. 

	The following relations are all meant to hold for $\po=\pn=\pcon$ and $\po'=\pn'$ only.

	We emphasize here that the variation $\po'$ is taken in $\vec\sigma$ coordinates. 
	The variation of $\xo=\po-\frac12|\vec\xi|^2$ is \emph{not} (necessarily) $\po'-\frac12|\vec\xi|^2$ because $\vec\xi$ varies as well.
	We deliberately vary $\po$, not $\xo$, because $\nabla\pcon=0$ in \emph{all} coordinates, allowing considerable simplifications.

	In particular:
	\begin{alignat}{1}
		(\nabla_{\vec\xi}\po)' 
		&= \big(\nabla_{\vec\xi}^T\vec\sigma\nabla_{\vec\sigma}\po\big)' \notag\\
		&= \big(\nabla_{\vec\xi}^T\vec\sigma\big)'\subeq{\nabla_{\vec\sigma}\po}{=0}
		+\nabla_{\vec\xi}^T\vec\sigma(\nabla_{\vec\sigma}\po)' 
		=\nabla_{\vec\xi}^T\vec\sigma\nabla_{\vec\sigma}\po' = \nabla_{\vec\xi}\po'.\notag
	\end{alignat}
	\begin{alignat}{1}
		(\nabla^2_{\vec\xi}\po)'
		&= \Big(\sum_k\frac{\partial\po}{\partial\sigma^k}\nabla_{\vec\xi}^2\sigma^k
		+\nabla_{\vec\xi}^T\vec\sigma\nabla_{\vec\sigma}^2\po\nabla_{\vec\xi}\vec\sigma\Big)' \notag\\
		&= \Big(\sum_k\frac{\partial\po}{\partial\sigma^k}\Big)'\nabla_{\vec\xi}^2\sigma^k
		+ \sum_k\subeq{\frac{\partial\po}{\partial\sigma^k}}{=0}\big(\nabla_{\vec\xi}^2\sigma^k\big)' \notag\\
		&+ \big(\nabla_{\vec\xi}^T\vec\sigma\big)'\subeq{\nabla_{\vec\sigma}^2\po}{=0}\nabla_{\vec\xi}\vec\sigma
		+\nabla_{\vec\xi}^T\vec\sigma\big(\nabla_{\vec\sigma}^2\po\big)'\nabla_{\vec\xi}\vec\sigma
		+\nabla_{\vec\xi}^T\vec\sigma\subeq{\nabla_{\vec\sigma}^2\po}{=0}(\nabla_{\vec\xi}\vec\sigma)' \notag\\
		&= \sum_k\frac{\partial\po'}{\partial\sigma^k}\nabla_{\vec\xi}^2\sigma^k
		+\nabla_{\vec\xi}^T\vec\sigma\nabla_{\vec\sigma}^2\po'\nabla_{\vec\xi}\vec\sigma = \nabla_{\vec\xi}^2\po'\notag
	\end{alignat}
	The same relations hold with $\pn$ instead of $\po$.

	Fr\'echet derivative of the interior equation:
	\begin{alignat}{1}
		0 &= \big(c^2I-(\nabla_{\vec\xi}\xn)^2\big)':\subeq{\nabla_{\vec\xi}^2\pn}{=0} + 
		\big(c^2I-(\nabla_{\vec\xi}\xn)^2\big):(\nabla_{\vec\xi}^2\pn)' \notag\\
		&= \big(c^2I-(\nabla_{\vec\xi}\xo)^2\big):\nabla_{\vec\xi}^2\po'\notag
	\end{alignat}
	The resulting right-hand side is a linear elliptic operator without zeroth-order term,
	applied to $\po'$. The classical maximum principle shows that $\po'$ cannot have a minimum in the interior.

	On the parabolic arcs,
	$$c^2=|\nabla_{\vec\xi}\xn|^2$$
	linearizes to
	\begin{alignat}{1}
		0 &= \nabla_{\vec\xi}\xo\cdot\big(\nabla_{\vec\xi}\xn\big)'
		= (\subeq{\nabla_{\vec\xi}\po}{=0}-\vec\xi)\cdot\big((\nabla_{\vec\xi}\pn)'-(\vec\xi)'\big) \notag\\
		&= \vec\xi\cdot(\vec\xi)'-\subeq{\vec\xi}{=|\vec\xi|\vec n}\cdot\nabla_{\vec\xi}\po' 
		= \subeq{\frac12(|\vec\xi|^2)'}{=(r^2)'=0} + |\vec\xi|\po'_n\notag\\
		\Rightarrow\qquad \po'_n &= 0 \mylabel{eq:paralin}
	\end{alignat}
	Here we use that variation of $\po$ may move $\vec\xi$ but keep it on $P_R$.
	\myeqref{eq:paralin} does not admit any extrema of $\pn'$, by the Hopf lemma.

	By reflection across $A$ we can also rule out extrema on $\overline{A}$, by applying the arguments
	for $P_L$ and $P_R$ resp.\ $\Omega$ there.

	Shock: $\po=\psi^I=\psi^I(0,0)+v^y_I\eta$, so 
	\begin{alignat}{1}
		\eta' &= (v^y_I)^{-1}\po' \qquad\text{on shock} \myeqlabel{eq:fdeta}
	\end{alignat}
	Moreover 
	\begin{alignat}{1}
		(\frac{\vec v_I-\nabla_{\vec\xi}\pn}{|\vec v_I-\nabla_{\vec\xi}\pn|})'
		&= \frac{-1}{|\vec v_I-\nabla_{\vec\xi}\pn|}
		\subeq{\left(1-\big(\subeq{\frac{\vec v_I-\nabla_{\vec\xi}\pn}{|\vec v_I-\nabla_{\vec\xi}\pn|}}{=\vec n}\big)^2\right)}{=(\vec t)^2}\nabla_{\vec\xi}\pn'
		=\frac{-(\pn')_t}{|\vec v_I-\nabla_{\vec\xi}\pn|}\vec t \myeqlabel{eq:fdnor}
	\end{alignat}

	Take the Fr\'echet derivative of \myeqref{eq:itn-shock}:
	\begin{alignat}{1}
		0 
		&= \big((\hat\rho\nabla_{\vec\xi}\xn-\rho_I\nabla_{\vec\xi}\chi^I)\cdot\frac{\vec v_I-\nabla_{\vec\xi}\pn}{|\vec v_I-\nabla_{\vec\xi}\pn|}\big)' \notag\\
		&= \big((\hat\rho)'\nabla_{\vec\xi}\xn+\rho(\nabla_{\vec\xi}\xn)'-\rho_I(\nabla_{\vec\xi}\chi^I)'\big)\cdot\vec n
		+ (\rho\nabla_{\vec\xi}\xn-\rho_I\nabla_{\vec\xi}\chi^I)\cdot(\frac{\vec v_I-\nabla_{\vec\xi}\pn}{|\vec v_I-\nabla_{\vec\xi}\pn|})' \notag\\
		&= 
		\rho c^{-2}\big(-\xn'-\nabla_{\vec\xi}\xn\cdot(\nabla_{\vec\xi}\xn)'\big)\xn_n 
			+ \rho(\nabla_{\vec\xi}\pn-\vec\xi)'\cdot\vec n 
			+ \rho_I(\vec\xi)'\cdot\vec n
			+ (\rho\nabla_{\vec\xi}\xn-\rho_I\nabla_{\vec\xi}\xo^I)\cdot(\frac{\vec v_I-\nabla_{\vec\xi}\pn}{|\vec v_I-\nabla_{\vec\xi}\pn|})'
		\notag\\
		&\overset{\myeqref{eq:fdnor}}{=}
		\rho c^{-2}\big(-\pn'+\vec\xi\cdot(\vec\xi)'
			-\nabla_{\vec\xi}\xn\cdot(\nabla_{\vec\xi}\pn)'
			+\subeq{\nabla_{\vec\xi}\pn}{=0}\cdot(\vec\xi)'
			-\vec\xi\cdot(\vec\xi)'
		\big)\xn_n \notag\\
		&+\rho(\nabla_{\vec\xi}\pn)'\cdot\vec n-\rho(\vec\xi)'\cdot\vec n+\rho_I(\vec\xi)'\cdot\vec n
		-\frac{(\rho\nabla_{\vec\xi}\xo-\rho_I\nabla_{\vec\xi}\chi^I)\cdot\vec t}{|\vec v_I-\nabla_{\vec\xi}\po|}(\po')_t \notag\\ 
		&= 
		-\rho c^{-2}\po'+\rho(1-c^{-2}\xo_n^2)(\po')_n 
		-\rho c^{-2}\xo_n\xo_t(\po')_t
		+(\rho-\rho_I)\eta'
		-\frac{\rho-\rho_I}{|\vec v_I-\nabla_{\vec\xi}\po|}\xo_t(\po')_t \notag\\ 
		&= 
		\rho(1-c^{-2}\xo_n^2)(\po')_n 
		+\Big(\frac{\rho_I-\rho}{|\vec v_I-\nabla_{\vec\xi}\po|}-\rho c^{-2}\xo_n\Big)\xo_t(\po')_t
		+(\rho-\rho_I)\eta'-\rho c^{-2}(\po') \notag\\
		&\overset{\myeqref{eq:fdeta}}{=}
		\rho(1-c^{-2}\xo_n^2)(\pn')_n 
		+\Big(\frac{\rho_I-\rho}{|\vec v_I-\nabla_{\vec\xi}\po|}-\rho c^{-2}\xo_n\Big)\xo_t(\po')_t
		+\big(\frac{\rho-\rho_I}{v^y_I}-\rho c^{-2}\big)\po' \notag\\
		&= 
		\subeq{-\rho(1-c^{-2}\xo_2^2)}{<0}(\pn')_2
		+\Big(\subeq{\frac{\rho_I-\rho}{|\vec v_I-\nabla_{\vec\xi}\po|}}{<0}\subeq{+\rho c^{-2}\xo_2}{<0}\Big)\xo_1(\po')_1
		+\big(\subeq{\frac{\rho-\rho_I}{v^y_I}}{<0}-\subeq{\rho c^{-2}}{>0}\big)\po'
		\myeqlabel{eq:shockl}
	\end{alignat}

	Now consider the case of a positive maximum of $\pnl$ on the shock.
	Then $(\po')_t=(\po')_1=0$ which also implies $(\vec n)'=0$ by \myeqref{eq:fdnor}.
	(This is natural because then we perturb the shock in a way that keeps it horizontal in
	that point, so the normal does not change in first order.) The two remaining terms:
	\begin{alignat*}{1}
		0 &= 
		\subeq{-\rho(1-c^{-2}\xo_2^2)}{<0}(\po')_2
		+\subeq{\big(\frac{\rho-\rho_I}{v^y_I}-\rho c^{-2}\big)}{<0}\po'
	\end{alignat*}
	For a positive local maximum of $\po'$ we need $(\po')_2\geq 0$, but this is incompatible with the previous equation.
	In the same way a negative local minimum is ruled out.

	Finally consider a positive maximum of $\po'$ in $\vec\xi_R$.
	The parabolic condition \myeqref{eq:paralin} means that $\nabla_{\vec\xi}\po'$ must be tangential to $P_R$ in the corner;
	for a maximum we need that it points counterclockwise, hence into the upper left quadrant, unless it is zero. 
	On the other hand the vector consisting of the $(\po')_2$ and $(\po')_1$ coefficients in \myeqref{eq:shockl}
	points into the lower right quadrant (because $\xo_1<0$ in the right corner), 
	so the scalar product is $\leq 0$. But the coefficient of $\po'$
	is $<0$, so the right-hand side of \myeqref{eq:shockl} is $<0$ --- contradiction!

	The same argument, with certain signs and inequalities reversed, rules out a positive maximum
	in the left corner as well as a negative minimum in either corner.

	We have ruled out that $\po'$ has a positive maximum or negative minimum anywhere in the domain.
	Therefore $\po'=0$, which is precisely what we had to show.
\end{proof}

\subsection{Existence of fixed points}

\begin{proposition}
	\mylabel{prop:probell}%
	For sufficiently small resp.\ large constants in \myeqref{eq:constlist}:
	$\IT$ has a fixed point for all $\lambda\in\Lambda$.
\end{proposition}
\begin{proof}
	Let $t\in[0,1]\mapsto\lambda(t)$ be any path in $\Lambda$.
	We have shown in Proposition \myref{prop:fusp-topology} that
	$U:=\bigcup_{t\in[0,1]}\big(\{t\}\times\fusp_{\lambda(t)}\big)$ is open in $[0,1]\times C^{2,\alpha}_\beta([0,1]^2)$, that 
	its closure is contained in 
	$W:=\bigcup_{t\in[0,1]}\big(\{t\}\times\cfusp_{\lambda(t)}\big)$,
	so 
	$$\partial U\subset V:=W-U=\bigcup_{t\in[0,1]}\Big(\{t\}\times(\cfusp_{\lambda(t)}-\fusp_{\lambda(t)})\Big).$$
	By Proposition \myref{prop:it-continuous-compact}, $\IT$ is a continuous and compact map on $W$.
	Finally, by Proposition \myref{prop:fp-boundary} we know $\IT$ cannot have fixed points on $V$, hence on $\partial U$.
	$U$ is bounded. Thus we may apply the property (D4*) in \cite[Section 13.6]{zeidler-one} to argue that the Leray-Schauder index
	of $\IT_{\lambda(t)}$ on $\fusp_{\lambda(t)}$ is constant in $t\in[0,1]$. 

	For $\lambda(0)=(\gamma,\etat_L)=(1,\etat_R)$, Proposition \myref{prop:unperturbed-index} shows that
	the Leray-Schauder degree of $\IT_{\lambda}$ is $\iota\neq 0$.
	Since $\Lambda$ is path-connected (Lemma \ref{lemma:Gamma-connected}), there is a path to any other $\lambda=\lambda(1)$.
	So $\IT_{\lambda}$ has degree $\iota\neq 0$ 
	for \emph{all} $\lambda\in\Lambda$. Nonzero degree requires at least\footnote{We expect that there is 
		only one, but we do not need this fact.	For all we prove, there could be more, e.g.\ two with 
		index $\iota$ and one with index $-\iota$ (which would still yield total L-S degree $\iota$).} 
	one fixed point. 
\end{proof}

\subsection{Construction of the entire flow}
\mylabel{section:entireflow}

\begin{proof}[Proof of Theorem \myref{th:elling-liu}]
	For all $\rho_I,c_I\in(0,\infty)$, $M^y_I\in(-\infty,0)$,
	for each choice of $\overline\gamma$, $\underline\eta^*_L$ and for all sufficiently small $\epsilon$, a separate $\Lambda$ is defined.
	For sufficiently small constants in \myeqref{eq:constlist}, Proposition \myref{prop:probell} yields fixed points $\po$ for all $\lambda\in\Lambda$. 
	Note that there is no lower bound on $\epsilon$, except that $\alpha,\beta$ and some other constants may deteriorate as $\epsilon\downarrow 0$.
	By Definition \myref{def:fusp}, Remark \myref{rem:fp}, 
	Proposition \myref{prop:pararc} and \myeqref{eq:reguint}, the fixed points satisfy
	\begin{alignat}{3}
		(c^2I-\nabla\chi^2):\nabla^2\psi &= 0 && \qquad \text{in $\Omega^{(\epsilon)}$}, \myeqlabel{eq:interior-eps} \\
		|\psi-\psi^I(\vxit_C)| & = O(\epsilon^{1/2}) && \qquad \text{and} \myeqlabel{eq:para-chi-eps} \\
		|\rho-\rho_C| & = O(\epsilon^{1/2}) && \qquad\text{and} \myeqlabel{eq:para-rho-eps} \\
		|\nabla\psi-\vec v_C| & = O(\epsilon^{1/2}) && \qquad \text{on $P_C^{(\epsilon)}$\qquad($C=L,R$),} \myeqlabel{eq:para-nablachi-eps} \\
		\chi &= \chi^I && \qquad \text{and} \myeqlabel{eq:shock1-eps}  \\
		(\rho\nabla\chi-\rho_I\nabla\chi^I)\cdot\vec n &= 0 && \qquad \text{on $S$,} \myeqlabel{eq:shock2-eps}  \\
		\chi_n &= 0 && \qquad \text{on $A$,} \\
		|\vxia_C-\vec\xi^{*(\epsilon)}_C| &= O(\epsilon^{1/2}) && \qquad \text{($C=L,R$).} \myeqlabel{eq:cornerdist-eps}
	\end{alignat}
	where the $O$ constants are independent of $\epsilon$. For regularity we have
	\begin{alignat}{3}
		\|\psi\|_{C^{0,1}(\overline\Omega^{(\epsilon)})} &\leq C_1, \myeqlabel{eq:lip-eps} \\
		\|\psi\|_{C^{k,\alpha}(K\cap\overline\Omega^{(\epsilon)})},|S|_{C^{k,\alpha}(K\cap\overline S^{(\epsilon)})} &\leq C_2(d) \notag\\
		\qquad\text{where $d:=d(K,\hat P_L^{(\epsilon)}\cup\hat P_R^{(\epsilon)})>0$} \myeqlabel{eq:cka-eps}.
	\end{alignat}
	for constants $C_1$ and $C_2(d)$ independent of $\epsilon$.

	\begin{figure}
	\input{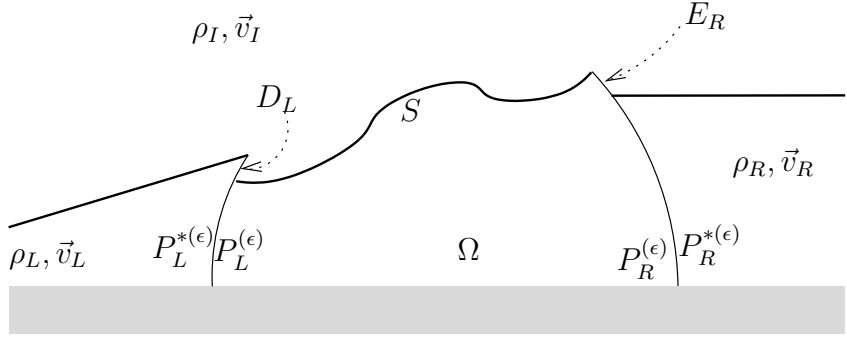}
	\caption{Extension from $\Omega$ to entire domain}
	\label{fig:epslimit}
	\end{figure}

	We extend each $\po$ to a function $\psi^{(\epsilon)}$ 
	defined on $\complement W$ as shown in Figure \ref{fig:epslimit}: we use $\rho_L,\vec v_L$ in the region enclosed by wall,
	$L$ shock and $P_L^{*(\epsilon)}$, we use $\rho_R,\vec v_R$ in the region enclosed by $R$ shock and $P_R^{*(\epsilon)}$,
	and $\rho_I,\vec v_I$ elsewhere. In each of the four regions, $\psi^{(\epsilon)}$ is a strong solution of
	self-similar potential flow, so we may multiply the divergence form \eqref{eq:chi-divform} 
	with any test function $\vartheta\in C_c^\infty(\complement W)$ and integrate over each of the four regions to obtain a sum
	of boundary integrals. Set
	$$D_C:=P_C^{*(\epsilon)}-P_C^{(\epsilon)},\qquad E_C:=P_C^{(\epsilon)}-P_C^{*(\epsilon)},\qquad(C=L,R).$$
	These curves have length $O(\epsilon^{1/2})$, so the integrals over them can be neglected because $\psi^{(\epsilon)}$ has
	$\epsilon$-uniformly bounded $C^{0,1}$ norm in each region. 
	The two integrals over $S^{(\epsilon)}$ cancel because of \myeqref{eq:shock2-eps}. The two integrals
	over $P_L^{(\epsilon)}\cap P_L^{*(\epsilon)}$ are $O(\epsilon^{1/2})$ due to \myeqref{eq:para-rho-eps} and \myeqref{eq:para-nablachi-eps},
	and same for $P_R$.
	The integrals over each piece of the wall vanish because $\chi_n=0$ in all cases.
	Therefore:
	\begin{alignat}{1}
		\int_{\complement W} \rho^{(\epsilon)}\nabla\chi^{(\epsilon)}\cdot\nabla\vartheta-2\rho^{(\epsilon)}\vartheta~d\vec\xi 
			&= O(\epsilon^{1/2}). \myeqlabel{eq:eps-weak}
	\end{alignat}
	
	By \myeqref{eq:cka-eps} with a diagonalization argument,
	for every compact $K\subset\complement W-\overline P_L^{*(0)}-\overline P_R^{*(0)}$
	we can find a sequence $(\epsilon_k)\downarrow 0$ so that $\psi^{(\epsilon_k)}$
	converges to $\psi^{(0)}$ in $C^{0,1}(K)$. Moreover, the convergence is bounded in $C^{0,1}(\complement W)$, so 
	we can take $\epsilon\downarrow 0$ in \myeqref{eq:eps-weak} to obtain
	\begin{alignat}{1}
		\int_{\complement W} \rho^{(0)}\nabla\chi^{(0)}\cdot\nabla\vartheta-2\rho^{(0)}\vartheta~d\vec\xi &= 0 .\myeqlabel{eq:zero-weak}
	\end{alignat}
	In addition, \myeqref{eq:shock1-eps}, \myeqref{eq:para-chi-eps} 
	and \myeqref{eq:lip-eps} show that 
	\begin{alignat}{1}
		\psi^{(0)} &\in C(\complement W) \myeqlabel{eq:zero-cont}
	\end{alignat}
	Finally, by construction of $\psi^{(\epsilon)}$,
	\begin{alignat}{1}
		\rho^{(0)}(s\vec\xi),\vec v^{(0)}(s\vec\xi) &\rightarrow \rho_I,\vec v_I\qquad(s\rightarrow\infty) \myeqlabel{eq:zero-cont-rhov}
	\end{alignat}
	for any $\vec\xi\in\complement\overline W-\{0\}$ (note that any ray from the origin is either in $\overline W$ or
	enters and stays in the $I$ region).	
	
	\myeqref{eq:zero-weak}, \myeqref{eq:zero-cont} and \myeqref{eq:zero-cont-rhov} show that $\phi(t,\vec x):=\psi^{(0)}(t^{-1}\vec x)$ defines
	a solution of \myeqref{eq:prob1}, \myeqref{eq:prob2}, \myeqref{eq:prob3}. 

	Taking $\overline\gamma\uparrow\infty$ 
	and $\underline\eta^*_L(\gamma)\downarrow\eta^x_L(\gamma)$ we obtain a solution for \emph{every} 
	$\gamma\in[1,\infty)$ and $\etat_L\in(\eta^x_L(\gamma),\overline\eta^*_R]$.
	Note that $\overline\eta^*_L\uparrow\overline\eta^*_R$ for $\epsilon\downarrow 0$,
	by Definition \myref{def:Lambda}.

	It remains to show that we have covered \emph{every} tip shock that satisfies the conditions of Theorem \myref{th:elling-liu}.
	Consider \emph{standard} coordinates (Figure \myref{fig:frameR}).
	By Proposition \myref{prop:vdzero}, the shocks with some upstream velocity $\vec v_I=(0,v^y_I)$, $v^y_I<0$, and density $\rho_I$ and
	downstream velocity parallel to the wall are \emph{uniquely} determined by $\etat_L$.
	The choices with $\etat_L\geq\etat_R$ cannot intersect the horizontal axis left of $\vxit_L$,
	as their shock slope is negative by Proposition \myref{prop:vdzero}.
	The choices with $\etat_L\leq 0$ are possible, but they are supersonic-subsonic at the intersection point (note that their other $L_d=1$ point
	must be above the axis, as their slope is positive and $\partial B_{c_d}(\vec v_d)$ is centered on the horizontal axis). 
	Therefore, all supersonic-supersonic tip shocks have $\etat_L\in(0,\etat_R]$ after passing from original to standard coordinates;
	there is only one such shock for each $\etat_L$ and we have constructed a solution for each that satisfies \myeqref{eq:techcond}.
\end{proof}

\begin{remark}
	\mylabel{rem:notvoid}%
	It remains to discuss whether there \emph{are} tip shocks that satisfy the conditions of Theorem \myref{th:elling-liu}. 
	Proposition \myref{prop:vdzero}, as discussed in the last paragraph of the proof of Theorem \myref{th:elling-liu}, 
	already settles that question. By Lemma \myref{lemma:etax}: for any $M^y_I<0$, 
	there is an interval $(\eta^x_L(\gamma),\etat_R]$ of $\etat_L$ so that \myeqref{eq:techcond} is satisfied.
	$\eta^x_L<\etat_R$, so there are some nontrivial solutions. For small $|M^y_I|$ we have $\eta^x_L>0$, so there are cases where
	\myeqref{eq:techcond} is violated. But if $|M^y_I|$ is sufficiently large, then $\eta^x_L=0$. 
	In particular if $M^y_I\leq-1$, then $\overline B_{c_I}(\vec v_I)$ is below the wall (in $R$ coordinates), 
	so \myeqref{eq:techcond} is always satisfied.
\end{remark}

\begin{remark}
	\mylabel{rem:structure}%
        In addition to mere existence we obtain some structural information in the proof:
	\begin{enumerate}
	\item The solution has the structure shown in Figure \myref{fig:frameorig}, with $L>1$ in the $I,L,R$ regions, $L<1$ in the elliptic 
		region.
	\item The solution has constant density and velocity in each of the $I,L,R$ regions. 
	\item The solution is analytic everywhere except perhaps at the $L,\Omega$ and $\Omega,R$ interfaces.
	\item The shock is Lipschitz; it is straight between the $I,L$ and $I,R$ regions; it is analytic between $I$ and the elliptic region 
		(away from the corners).
	\item Density and velocity are bounded. 
	\end{enumerate}
	It is expected that density and velocity are at least continuous. 
	However, the methods developed in this article yield boundedness everywhere, but 
	continuity
	only away from $\overline P^*_L\cup\overline P^*_R$.
	Note that $P_L^{(0)}$, $P_R^{(0)}$ can \emph{not} be classical shocks with smooth data on each side,  
	because the one-sided limit of $L$ on the hyperbolic side of the parabolic arcs is $=1$ everywhere ($>1$ is needed for positive shock strength).

	In some points, continuity can be obtained a posteriori. For example $v^x\leq 0$, as implied by \eqref{eq:horvel}, yields continuity
	in $\xi_{BR}$.

	Some other structural information: 
	\begin{enumerate}
	\item The possible (downstream) normals of the curved shock portion are between the R and L shock normals (counterclockwise). 
	\item The shock is admissible and does not vanish anywhere.	
	\item Therefore, the shock is above the line connecting the arc-shock corners.
	\item In the elliptic region, the velocity $v^x$ tangential to the wall is between $v^x_L$ to $v^x_R$. 
	\item In the elliptic region, the density $\rho$ is greater than $\rho_I$. 
	\end{enumerate}
	Additional information can be obtained from the inequalities in Definition \myref{def:fusp}.
\end{remark}

\section{Appendix}

\subsection{Regularity in 2D corners}
\mylabel{section:corner}

We obtain $\spC^{2,\alpha}_\beta$ bounds in a corner neighbourhood from merely $\spC^1$ (with a uniform $\spC^{0,1}$ bound).
Although our argument follows \cite[Theorem 2.1]{lieberman-pacj-1988} without major modification, we prefer to repeat it explicitly,
because we need a uniformly large neighbourhood.

We keep the proof simple by restricting ourselves to smooth coefficient functions and a priori $\spC^1$ rather than $\spC^{0,1}$ regularity.

\if\dofigures%
\begin{figure}
\center{\input{corn.pstex_t}}
\caption{Corner}
\mylabel{fig:corner}
\end{figure}
\fi%
\def\sugar{11}

\begin{proposition}
	\mylabel{prop:corner}%
	\begin{enumerate}
	\item
		Consider a point $x_0\in\R^2$ and polar coordinates $(r,\phi)$ with origin in that point. Let $R>0$.
		Consider two functions $\phi^1,\phi^2:[0,R]\rightarrow[-\frac{\pi}{2},\frac{\pi}{2}]$ so that $\phi^1<\phi^2$.
		For $j=1,2$ let $\Gamma^j$ be the curve $\{(r,\phi^j(r)):r\in(0,R)\}$. 
		Assume that $\Gamma^j$ are $C^{1,\sigma}$ curves ($\sigma\in(0,1]$), including the endpoints:
		\begin{alignat}{1}
			|\overline\Gamma^j|_{C^{1,\sigma}} &\leq C_\Gamma. \mylabel{eq:Gammaregu}
		\end{alignat}
		Let $I=\{(R,\phi):\phi\in(\phi^1(R),\phi^2(R))\}$. 
		Let $U=\{(r,\phi):r\in(0,R),~\phi\in(\phi^1(R),\phi^2(R))\}$.
		Assume there is a $\theta<\pi$ so that
		\begin{alignat}{1}
			\sup_{r\in(0,R)}\phi^2(r) &\leq\frac{\theta}{2},\qquad \inf_{r\in(0,R)}\phi^1(r) \geq -\frac{\theta}{2}. \myeqlabel{eq:phithetabd}
		\end{alignat}

		Consider a function $u\in \spC^3(U)\cap\spC^2(U\cup\Gamma^1\cup\Gamma^2)\cap\spC^1(\overline U)$ with
		\begin{alignat}{1}
			\|u\|_{C^1(\overline U)} &\leq C_u. \myeqlabel{eq:uLip}
		\end{alignat}
		Assume that $u$ satisfies
		\begin{alignat}{1}
			a^{ij}(x,u(x),Du(x))u_{ij}(x) 
			&= 0 \qquad \forall x\in U \myeqlabel{eq:u}
		\end{alignat}
		(we use Einstein convention in this section)
		as well as boundary conditions
		\begin{alignat}{1}
			g^k(x,u(x),Du(x)) &= 0 \qquad \forall x\in\Gamma^k,~k\in\{1,2\}. \myeqlabel{eq:Gammacond}
		\end{alignat}
		Here $a^{ij}$ and $g^k$, as functions of $x,u,\nabla u$, 
		are assumed\footnote{Note that $g^k$ is defined for all arguments. See \cite{lieberman-pacj-1988}
		for more general circumstances.} to satisfy
		\begin{alignat}{1}
			\|a^{ij}\|_{\spC^1(\overline U\times\R\times\R^2)},\|g^k\|_{\spC^1(\overline U\times\R\times\R^2)} &\leq C_c. \myeqlabel{eq:coeffnorm}
		\end{alignat}

		Assume uniform ellipticity: there is a $\delta_e>0$, independent of $x\in\overline U$, $z\in\R$, $p\in\R^2$ and
		$y\in\R^2$, so that
		\begin{alignat}{1}
			a^{ij}(\vec x,z,p)y^iy^j &\geq \delta_e|y|^2. \myeqlabel{eq:Ce}
		\end{alignat}

		Moreover, assume uniform non-degenerate obliqueness: there is a $\delta_o>0$,
		independent of $x\in\Gamma^k$, $z\in\R$ and $p\in\R^2$ so that 
		\begin{alignat}{1}
			g^k_p(x,z,p)\cdot n(x) &\geq \delta_o, \myeqlabel{eq:ndob}
		\end{alignat}
		where $n$ is the outer unit normal to $\Gamma^k$ ($k=1,2$).

		Finally let $C_d$ be independent of $x\in\overline U$, $z\in\R$ and $p\in\R^2$, so that
		\begin{alignat}{1}
			\|G^{-1}\| &\leq C_d, \myeqlabel{eq:G}
		\end{alignat}
		where $G$ is the $2\times 2$ matrix with $k$th column $g^k_p(x,z,p)$.

		Then 
		there are $R'\in(0,R)$, $\lambda>0$ and $C_r<\infty$ so that
		\begin{alignat}{1}
			|\nabla u(x)-\nabla u(x_0)| &\leq C_r|x-x_0|^\lambda \qquad\text{for $x\in U$}. \myeqlabel{eq:Colambda}
		\end{alignat}
		$R',\lambda,C_r$ depend only on $C_d$, $\delta_e$, $C_c$, $\delta_o$, $C_u$, $C_\Gamma$ and $R$,
		except $R'$ may also depend on $\sigma$.
	\item
		Assume in addition that 
		\begin{alignat}{1}
			\|G\|_{C^{1,\sigma}(\overline U\times\R\times\R^2)} &\leq C_\Gamma. \myeqlabel{eq:ghregu}
		\end{alignat}
		Then there are $\kappa>0$, $C_r<\infty$, $R'>0$ so that 
		\begin{alignat}{1}
			\|u\|_{\spC^{2,\kappa}_{1+\lambda}(B_{R'}(x_0)\cap\overline U,\{x_0\})}\leq C_r. \myeqlabel{eq:Ctabeta}
		\end{alignat}

		$\lambda$ is as before. $R'$ may have changed, but has the same dependencies.
		$C_r$ may have changed and may depend on $\sigma$ now.
		$\kappa$ depends continuously and only on $C_d$, $\delta_e$, $C_c$, $\delta_o$, $C_u$, $C_\Gamma$, $\sigma$, $R$.
	\end{enumerate}
\end{proposition}
\begin{proof}
	In this proof let $C$, $C_i$, $\delta$ and $\delta_i$ (for numbers $i$) represent constants that may depend continuously
	and only on $C_d$, $\delta_e$, $C_c$, $\delta_o$, $C_u$, $C_\Gamma$, $R$ (but not $\sigma$). $C,\delta$ may change from occurence to occurence.
	$\delta>0$ is meant to be small, $|C|<\infty$ large.
\if\sugar\par{\bf Constructing $w$: interior}\par\fi%
	Define 
	$$v(x):=g^1(x,u(x),Du(x))+Mw(x).$$ 
	Set
	$$w(r,\phi):=r^\lambda f(\phi) \qquad f(\phi):=1-\mu e^{-B\phi}$$
	with $\lambda\in(0,\frac12]$ and $\mu,B>0$ to be determined. We take $\mu$ so small that $f\geq\frac12$.
	In a given point of $U$, rotate coordinates to be angular and readial and let $a^{rr},a^{r\phi},a^{\phi\phi}$ be the corresponding components 
	of $a^{ij}$.
	\begin{alignat*}{1}
		&a^{ij}w_{ij} \\
		&= a^{rr}w_{rr}+2a^{r\phi}(r^{-1}w_{r\phi}-r^{-2}w_\phi)+a^{\phi\phi}(r^{-2}w_{\phi\phi}+r^{-1}w_r) \\
		&= \big(
			a^{rr}\lambda(\lambda-1)f
			+2a^{r\phi}(\lambda-1)f'
			+a^{\phi\phi}(f''+\lambda f)
		\big)r^{\lambda-2} \\
		&= \Big(
			\subeq{a^{rr}\lambda(\lambda-1)f}{\leq 0}
			+\subeq{2a^{r\phi}(\lambda-1)}{\leq2|a^{r\phi}|}\cdot \mu Be^{-B\phi} 
			+\subeq{a^{\phi\phi}}{\geq 0}\big(\lambda\subeq{f}{\leq 1}-\mu B^2e^{-B\phi}\big)
		\Big)r^{\lambda-2} \\
		&\leq \big(
		e^{-B\phi}\mu B(2|a^{r\phi}|-Ba^{\phi\phi})
		+\lambda a^{\phi\phi}
		\big) r^{\lambda-2}
	\end{alignat*}
	We pick\footnote{We need full ellipticity later, but here only $a^{rr},a^{\phi\phi}>0$ matter.}
	$$B=(1+2\sup|a^{r\phi}|)/\inf a^{\phi\phi}>0$$
	so that
	\begin{alignat*}{1}
		a^{ij}w_{ij}
		&\leq \big(
		-\mu Be^{-B\phi}
		+\lambda a^{\phi\phi}
		\big) r^{\lambda-2}
	\end{alignat*}
	For any $\mu>0$ we can pick a $\lambda=\lambda(\mu)>0$ so small that
	\begin{alignat}{1}
		a^{ij}w_{ij} &\leq -\frac{\mu B}{2}e^{-B\phi}r^{\lambda-2}. \myeqlabel{eq:w-interior}
	\end{alignat}

\if\sugar\par{\bf Boundary condition for $w$}\par\fi%

	Let $\nu,\tau$ be \emph{Cartesian} coordinates that are normal resp.\ tangential in a given point on $\Gamma^2$
	(then $\partial_\nu=\partial_n$ and $\partial_\tau=\partial_t$, but higher derivatives differ).
	Let $\nu$ be outer normal and $\tau$ pointing away from the corner (\emph{clockwise} from $\nu$ direction).

	The angle between radial and tangential direction is $\leq C(R')^\sigma$, because of \myeqref{eq:Gammaregu}. We can make $R'$ small to control it.
	Note that $C$ can change from line to line.
	\begin{alignat*}{1}
		w_\nu
		& \geq (1-C(R')^\sigma) r^{-1}w_\phi - C(R')^\sigma|w_r| \\
		&= (1-C(R')^\sigma)r^{\lambda-1}\mu Be^{-B\phi} - C(R')^\sigma\lambda r^{\lambda-1}(1-\mu e^{-B\phi}) \\
		&= \big(B\mu e^{-B\phi}-C(R')^\sigma\big)r^{\lambda-1} \\
		|w_\tau|
		&\leq (1-C(R')^\sigma))|w_r|+C(R')^\sigma|w_\phi| \leq (\lambda + C(R')^\sigma B\mu)r^{\lambda-1}
	\end{alignat*}	
	For any constant $T$,
	\begin{alignat}{1}
		w_\nu+Tw_\tau &\geq \Big(B\mu e^{-B\phi}-C(R')^\sigma-T\big(\lambda+C(R')^\sigma B\mu\big)\Big)r^{\lambda-1}. \myeqlabel{eq:wboundary}
	\end{alignat}

\if\sugar\par{\bf $D^2u$ and $D(g^1)$}\par\fi%

	$$(g^1)_i:=\partial_{x^i}\big(g^1(x,u,Du(x))\big)=g^1_{p^k}u_{ik} + C\qquad(i=1,2).$$
	($C$ is a remnant of terms containing only $u(x),Du(x)$ which can be estimated by $C_u$ via \myeqref{eq:uLip}.)
	Combined with \myeqref{eq:u} we have a system
	$$\begin{bmatrix}
		a^{11} & 2a^{12} & a^{22} \\
		g^1_{p^1} & g^1_{p^2} & 0 \\
		0 & g^1_{p^1} & g^1_{p^2}
	\end{bmatrix}\begin{bmatrix}
		u_{11} \\ u_{12} \\ u_{22}
	\end{bmatrix}=\begin{bmatrix}
		0 \\ (g^1)_1 \\ (g^1)_2.
	\end{bmatrix}+C.$$
	The system matrix has determinant
	$$=a^{11}(g^1_{p^2})^2-2a^{12}g^1_{p^1}g^1_{p^2}+a^{22}(g^1_{p^1})^2
	=\big((g^1_p)^\perp\big)^TA(g^1_p)^\perp \overset{\myeqref{eq:Ce}}{\geq} \delta_e|g^1_p|^2
	\overset{\myeqref{eq:coeffnorm}}{\geq} \delta_eC_c>0.$$
	Combined with an upper bound on the matrix (from \myeqref{eq:coeffnorm})
	we obtain that 
	\begin{alignat}{1}
		\begin{bmatrix}
			u_{11} \\ u_{12} \\ u_{22}
		\end{bmatrix}
		&= D(x)\nabla(g^1)+C \myeqlabel{eq:ddudg}
	\end{alignat}
	where $D(x)\in\R^{3,2}$ with $|D(x)|\leq C$.

\if\sugar\par{\bf Interior PD inequality}\par\fi%

	\begin{alignat}{1}
		a^{ij}v_{ij} &= g^1_{p^k}a^{ij}u_{kij} + (g^1_{p^kp^\ell}u_{j\ell}+C)u_{ik} + Ma^{ij}w_{ij} + C. \myeqlabel{eq:Av}
	\end{alignat}
	Again we use $C$ to abbreviate terms depending on $u,Du$ only, which are bounded by \myeqref{eq:uLip}.
	
	Take $\partial_k$ of \myeqref{eq:u}:
	\begin{alignat}{1}
		a^{ij}u_{ijk} + (a^{ij}_{p^\ell}u_{k\ell}+C)u_{ij} &= 0\notag
	\end{alignat}
	Multiply with $g^1_{p^k}$ and substitute into \myeqref{eq:Av}:
	\begin{alignat}{1}
		a^{ij}v_{ij} &= -(C+g^1_{p^k}a^{ij}_{p^\ell}u_{k\ell})u_{ij} + (g^1_{p^kp^\ell}u_{j\ell}+C)u_{ik} + Ma^{ij}w_{ij} + C. \notag
	\end{alignat}

	Apply \myeqref{eq:ddudg}:
	\begin{alignat}{1}
		a^{ij}v_{ij} &= Ma^{ij}w_{ij} + e^{ij}(g^1)_i(g^1)_j + e^i(g^1)_i + C \notag
	\end{alignat}
	for some coefficients $e^{ij}(x),e^i(x)=C$.
	Use $v_i = (g^1)_i + Mw_i$ to obtain
	\begin{alignat}{1}
		&a^{ij}v_{ij} + \subeq{\big ( - e^{ij}v_j + M(e^{ij}+e^{ji})w_j - e^i \big)}{=:q^i}\cdot v_i 
		= Ma^{ij}w_{ij} + M^2e^{ij}w_iw_j - Me^iw_i + C \notag \\
		&\overset{\myeqref{eq:w-interior}}{\leq} 
		- \frac{M\mu B}{2}e^{-B\phi}r^{\lambda-2}
		+ CM^2|\nabla w|^2+C \notag\\
		&\leq
		- \frac{M\mu B}{2}e^{-B\phi}r^{\lambda-2}
		+ CM^2(\lambda^2\subeq{f^2}{\leq 1}+\mu^2B^2e^{-2B\phi})r^{2\lambda-2}+C \notag
	\end{alignat}
	We take $M=M'(R')^{-\lambda}$ with $M'\geq 1$ to be determined:
	\begin{alignat}{1}
		a^{ij}v_{ij} + q^iv_i  
		&\leq
		\Big( \subeq{- \frac{M'\mu B}{2}e^{-B\phi}(r/R')^\lambda}{(I)}
			+ C(M')^2(r/R')^{2\lambda}(\subeq{\lambda^2}{(II)}+\subeq{\mu^2B^2e^{-2B\phi}}{(III)})\Big)r^{-2}
		+ C \notag
	\end{alignat}
	Given any $M'$ and $R'\in(0,1]$, we take $\mu$ (and $\lambda$, which depends on it) 
	so small(er) that the $(III)$ term is dominated by the $(I)$ term.
	Then we pick $\lambda$ even small(er) so that the $(II)$ term is also dominated by the $(I)$ term. 
	In both cases we use that $(r/R')^\lambda\geq(r/R')^{2\lambda}$ because $0<r\leq R'$. Now
	\begin{alignat}{1}
		a^{ij}v_{ij} + q^iv_i  
		&\leq
		- \frac{M'\mu B}{4}e^{-B\phi}\subeq{(r/R')^\lambda r^{-2}}{\geq(R')^{-2}}
		+ C. \notag
	\end{alignat}
	Finally we take $R'$ so small that the $C$ term is dominated. 
	We obtain
	\begin{alignat}{1}
		a^{ij}v_{ij} + q^iv_i  
		&< 0
		\qquad\text{in $U\cap B_{R'}(x_0)$.} \notag
	\end{alignat}
	Therefore $v$ cannot have minima in that set.
\if\sugar\par{\bf Boundary PD inequality}\par\fi%

	Take $\partial_\tau$ of the boundary condition \myeqref{eq:Gammacond} on $\Gamma^2$ to obtain
	\begin{alignat}{1}
		g^2_{p^\tau}u_{\tau\tau} + g^2_{p^\nu}u_{\nu\tau} + C &= 0. \notag 
	\end{alignat}
	We plan to take a linear combination of this with the interior equation \myeqref{eq:u} to obtain
	a new boundary condition of the form
	\begin{alignat}{1}
		C &= (g^1)_\nu + T(g^1)_\tau \myeqlabel{eq:gGammaf} \\
		&= g^1_{p^\nu}u_{\nu\nu}+(Tg^1_{p^\nu}+g^1_{p^\tau})u_{\tau\nu}+Tg^1_{p^\tau}u_{\tau\tau} \notag
	\end{alignat}
	This problem can be written
	$$\begin{bmatrix}y_1&y_2&-1\end{bmatrix}\begin{bmatrix}
		a^{\tau\tau} & 2a^{\tau\nu} & a^{\nu\nu} \\
		g^2_{p^\tau} & g^2_{p^\nu} & 0 \\
		Tg^1_{p^\tau} & Tg^1_{p^\nu}+g^1_{p^\tau} & g^1_{p^\nu}
	\end{bmatrix}\overset{!}=0.$$
	It is solvable if and only if the determinant is $0$, which is a linear equation for $T$: solution
	$$T = \frac{C}{a^{\nu\nu}\cdot(g^2_{p^\tau}g^1_{p^\nu}-g^2_{p^\nu}g^1_{p^\tau})} = \frac{C}{a^{\nu\nu}\Det G}.$$ 
	The denominator is bounded away from $0$ by \myeqref{eq:Ce} and \myeqref{eq:G}, so $|T|\leq C$. 
	The solutions are 
	\begin{alignat}{1}
		y_1 &= \frac{g^1_{p^\nu}}{a^{\nu\nu}}, \qquad
		y_2 = \frac{a^{\tau\tau}(g^1_{p^\nu})^2-2a^{\tau\nu}g^1_{p^\tau}g^1_{p^\nu}+a^{\nu\nu}(g^1_{p^\tau})^2}{a^{\nu\nu}\det G}; \notag
	\end{alignat}
	clearly $|y_1|,|y_2|\leq C$. Therefore we are justified in writing $C$ on the left-hand side of \myeqref{eq:gGammaf}. 

	Now:
	\begin{alignat*}{1}
		v_\nu+Tv_\tau 
		&=  M(w_\nu+Tw_\tau) + (g^1)_\nu+T(g^1)_\tau \\
		&\underset{\text{\myeqref{eq:gGammaf}}}{\overset{\text{\myeqref{eq:wboundary}}}{\geq}} 
		M\Big(\mu Be^{-B\phi} - C(R')^\sigma - T\big(\lambda+C(R')^\sigma B\mu\big)\Big)r^{\lambda-1} + C \\
		&= 
		M'(R')^{-\lambda}\Big(\mu Be^{-B\phi} - C(R')^\sigma - T\big(\lambda+C(R')^\sigma B\mu\big)\Big)r^{\lambda-1}+ C
	\end{alignat*}
	We can choose $\lambda$ and $R'$ so small that
	\begin{alignat*}{1}
		v_\nu+Tv_\tau 
		&\geq
		M'(R')^{-\lambda}\frac{\mu B}{2}r^{\lambda-1}e^{-B\phi} + C 
		\qquad\text{for all $r\in(0,R')$.} 
	\end{alignat*}
	The $C$ term is bounded, so we can pick $R'$ so small(er) that
	\begin{alignat*}{1}
		v_\nu+Tv_\tau &> 0
		\qquad\text{for all $r\in(0,R')$.} 
	\end{alignat*}
	This inequality implies that $v$ cannot have minima on $\Gamma^2\cap B_{R'}(\vec x)$ because otherwise $v_\nu\leq 0$ and $v_\tau=0$
	in the minimum point would cause a contradiction.
\if\sugar\par{\bf Conclusion of corner $\spC^{1,\lambda}$}\par\fi%
	Let 
	$$I':=\{(R',\phi):\phi\in(\phi^1(R'),\phi^2(R'))\}, \qquad U':=\{(r,\phi):r\in(0,R'),~\phi\in(\phi^1(r),\phi^2(r))\}.$$
	We are still free to choose $M'$ ($\geq 1$). Since $Mw\geq \frac{M'}{2}$ on $\overline{I'}$,
	whereas $g^1(x,u(x),Du(x))=C$,
	we can take $M'$ so large that $v\geq 0$ on $\overline{I'}$.

	$v$ must attain its minimum over $\overline{U'}$ on $\overline{I'}\cup\{x_0\}\cup\Gamma^1$.
	On $\overline{\Gamma^1}$ the boundary condition $g^1=0$ yields $v=w\geq 0$. Therefore $v$ attains its minimum
	$v=0$ in $x_0$. Hence $v=g^1+Mw\geq 0$ in $\overline U$, so
	$$g^1\geq -Mw\geq -\frac{M'}{2}\left(\frac{r}{R'}\right)^\lambda.$$

	Moreover we can apply the same arguments to $v:=g^1-Mw$, by switching inequalities and signs and replacing ``minimum'' with
	``maximum'' whereever appropriate. Then
	$$g^1\leq\frac{M'}{2}\left(\frac{r}{R'}\right)^\lambda.$$

	The same arguments, but with the roles of $\Gamma^1$ and $\Gamma^2$ reversed, yield a bound on $g^2$.
	Now we argue that $g^1$ and $g^2$, as functions of $\nabla u$ with $x$ and $u(x)$ held fixed, are uniformly functionally independent, 
	by \myeqref{eq:G}. Therefore
	$$|\nabla u(x)-\nabla u(x_0)|\leq Cr^\lambda.$$

\if\sugar\par{\bf Full $\spC^{2,\alpha}_\beta$ result}\par\fi%

	This concludes the proof of \myeqref{eq:Colambda}. 
	The $C^{2,\kappa}_{1+\lambda}$ result follows from \cite[Corollary 1.4]{lieberman-pacj-1988}, where $R$ is the distance from the corner,
	and $\osc_{\Omega_R}Du/R=R^{\lambda-1}$ by \myeqref{eq:Colambda}.
\end{proof}

\subsection{Free boundary transformation}

For a column vector $x$ and a column-vector-valued function $u$, let $u_x$ be the transpose of the gradient
and $u_{xx}$ the Hessian. For ``row'' instead of ``column'' let $u_x$ be the gradient.
Boundary normals are row vectors.

\begin{proposition}
	\mylabel{prop:ztrans}%
	Consider a coordinate transformation $y=y(x,u(x))$. 
	\begin{enumerate}
	\item
		It is nondegenerate if and only if
		$$y_x+y_uu_x$$
		is regular. Let $C_d$ be an upper bound for the norms of the matrix and its inverse (for some fixed matrix norm).
	\item
		A quasilinear equation 
		$$A(x,u(x),u_x(x)):u_{xx}(x)+b(x,u(x),u_x(x))=0$$
		transforms into another quasilinear equation.
		The first equation is elliptic if and only if the second one is,
		and the ellipticity constants are comparable, up to constant factors depending only and continuously on $C_d$.
	\item
		Consider a boundary condition 
		$$g(x,u(x),u_x(x))=0.$$
		As before we say it is oblique, with obliqueness constant $\delta_o>0$, if
		$$|ng_p|\geq\delta_o|g_p|,$$
		where $p$ represents $u_x$.
		If 
		\begin{alignat}{1}
			D:=|1-u_x(y_x+y_uu_x)^{-1}y_u|\geq\delta_D>0, \myeqlabel{eq:Dcond}
		\end{alignat}
		then the new
		boundary condition is oblique if and only if the old one is, and the obliqueness constants are comparable, 
		up to constant factors depending only and continuously on $\delta_D$ and $C_d$.
	\item
		Under the same assumptions, 
		consider two functions $g^1,g^2$ of $u_x$. The smaller angle between $g^1_p,g^2_p$ is nonzero if and only if
		the smaller angle between $g^1_q,g^2_q$ is nonzero. They are comparable, up to constant factors depending only
		and continuously on $C_d,\delta_D$. Here $q$ represents $u_y$.
	\end{enumerate}
\end{proposition}
\begin{proof}
	\begin{enumerate}
	\item Obvious.
	\item The new top-order coefficient matrix is
		$$ (y_x+y_uu_x)A(y_x+y_uu_x)^T $$
		It is still symmetric positive definite, and the ellipticity constant bound is obvious.
	\item We omit the $x$ and $u(x)$ arguments from the notation of the boundary conditions.
		$$u(x) = u(y(x,u(x))).$$
		Take the $x$ derivative:
		$$u_x = u_y(y_x+y_uu_x) \qquad\Rightarrow\qquad u_y=u_x(y_x+y_uu_x)^{-1}.$$
		If $h(u_y)=0$ is the boundary condition in $y$ coordinates, then
		$$g(u_x) = h\big(u_x(y_x+y_uu_x)^{-1}\big).$$
		Expressed with $p,q$:
		$$g(p) = h\big(p(y_x+y_up)^{-1}\big),$$
		so
		$$g_p=\Big((y_x+y_up)^{-1}-p(y_x+y_up)^{-1}y_u(y_x+y_up)^{-1}\Big)h_q = (1-p(y_x+y_up)^{-1}y_u)(y_x+y_up)^{-1}h_q.$$
		\cucon{we regard $g_p,h_q$ as column vectors, $u_x,u_y$ as row vectors}
		With $p=u_x$ we have
		$$g_p = (1-u_x(y_x+y_uu_x)^{-1}y_u)(y_x+y_uu_x)^{-1}h_q.$$
		Let $n$ be the normal in $x$ coordinates, $N$ the normal in $y$ coordinates. Then 
		$$n = \frac{N(y_x+y_uu_x)}{|N(y_x+y_uu_x)|}.$$
		\cucon{we regard normals as gradients of scalar functions whose zero set is the boundary, locally}
		So
		$$ng_p=\frac{1-u_x(y_x+y_uu_x)^{-1}y_u}{|N(y_x+y_uu_x)|}Nh_q.$$
		Clearly $ng_p\neq 0$ iff $Nh_q\neq 0$, as long as $D>0$. The constant comparison is obvious.
	\item 
		The two functions transform like the boundary condition above:
		$$g^k_p=(1-u_x(y_x+y_uu_x)^{-1}y_u)(y_x+y_uu_x)^{-1}h^k_q$$
		where $g^k(u_x)=h^k(u_y)$. Then clearly $g^1_p$ and $g^2_p$ are collinear if and only if $h^1_q$ and $h^2_q$ are,
		and the constants are obvious.
	\end{enumerate}
\end{proof}

\begin{remark}
	\mylabel{rem:shmor}%
	We use the Sherman-Morrison formula: for an invertible matrix $M$, the rank 1 perturbation $M+ab^T$ is invertible 
	if and only if $b^TM^{-1}a \neq -1$.
	Taking $M=y_x+y_uu_x$, $a=-y_u$ and $b=u_x^T$, we see that \myeqref{eq:Dcond} is precisely the condition for $y_x$ to
	be invertible as well.
\end{remark}

\bibliographystyle{amsalpha}
\bibliography{../elling}

\end{document}